\documentclass{emulateapj}

\usepackage{graphicx}
\usepackage{lscape}
\usepackage{times}
\makeatother

\newcommand{\npropto}{\ \raise -0.truept\hbox{\rlap{\hbox{$/$}}\raise0.truept
        \hbox{$\propto$}\ }}

\newcommand{\lsim}{\ \raise -2.truept\hbox{\rlap{\hbox{$\sim$}}\raise5.truept
        \hbox{$<$}\ }}
\newcommand{\gsim}{\ \raise -2.truept\hbox{\rlap{\hbox{$\sim$}}\raise5.truept
        \hbox{$>$}\ }}

\newcommand{\ngrbs}{154}
\newcommand{\nhosts}{46}
\newcommand{\nemi}{33}

\slugcomment{ApJ in press}
\shorttitle{GRB host galaxies}
\shortauthors{Savaglio, Glazebrook, \& Le Borgne}

\begin{document}

\title{The Galaxy Population Hosting Gamma-Ray Bursts}

\author{S.\ Savaglio\altaffilmark{1}, K.\
Glazebrook\altaffilmark{2}, \& D.\ Le Borgne\altaffilmark{3}}

\altaffiltext{1}{Max Planck Institute for Extraterrestrial Physics, Garching, Germany, savaglio@mpe.mpg.de}
\altaffiltext{2}{Centre for Astrophysics \& Supercomputing, Swinburne University of Technology, Hawthorne, Australia, 
karl@astro.swin.edu.au}
\altaffiltext{3}{CEA, Irfu, SAp, Centre de Saclay, F-911191 Gif-sur-Yvette, France; damien.leborgne@cea.fr}

\begin{abstract} 

We present the most extensive and complete study of the properties for the largest sample (\nhosts\ objects) of gamma-ray burst (GRB) host galaxies. The redshift interval and the mean redshift of the sample are $0<z<6.3$ and $z=0.96$ (look-back time: 7.2 Gyr), respectively; 89\% of the hosts are at $z \leq 1.6$.  Optical-near-infrared (NIR) photometry and spectroscopy are used to derive stellar masses, star formation rates (SFRs), dust extinctions and metallicities. The average stellar mass is $10^{9.3}$ M$_\odot$, with a 1$\sigma$ dispersion of 0.8 dex. The average metallicity for a subsample of 17 hosts is about 1/6 solar and the dust extinction in the visual band (for a subsample of 10 hosts) is $A_V=0.5$. We obtain new relations to derive SFR from [OII] or UV fluxes, when Balmer emission lines are not available. SFRs, corrected for dust extinction, aperture-slit loss and stellar Balmer absorption are in the range 0.01-36 M$_\odot$ yr$^{-1}$. The median SFR per unit stellar mass (specific SFR) is 0.8 Gyr$^{-1}$. Equivalently the inverse quantity, the median formation timescale is 1.3 Gyr.  Most GRBs are associated with the death of young massive stars, more common in star-forming galaxies. Therefore GRBs are an effective tool to detect star-forming galaxies in the universe.  Star-forming galaxies at $z<1.6$ are a faint and low-mass population, hard to detect by conventional optical-NIR surveys, unless a GRB event occurs. There is no compelling evidence that GRB hosts are peculiar galaxies. More data on the subclass of short GRB are necessary to establish the nature of their hosts.

\end{abstract}


\section{Introduction}

The first gamma-ray burst (GRB) ever discovered was in the year 1967 (Klebesadel et al.\ 1973), but it took 30 more years to finally identify these sources as extragalactic and
cosmologically distributed (Metzger et al.\ 1997). As of today (2008 September 22), the total
number of GRBs with known redshift\footnote{For the most complete list of GRB redshifts, see the URL: http://www.mpe.mpg.de/$\sim$jcg/grbgen.html, maintained by J. Greiner.} is \ngrbs, half of which are at
$z>1.4$, and 69\% were discovered by the dedicated space mission Swift (Gehrels et al.\  2004) after 2005 January.

Although the GRB population with redshift is rather small,  studies dedicated to the hosting galaxies are often deep and can cover the wavelength range from the radio, to mid-IR, to optical, and UV in imaging and spectroscopy 
(e.g., Bloom et al.\ 2002; Le Floc'h et al.\ 2006; Priddey et al.\ 2006; Prochaska et al.\ 2006; Berger et al.\ 2007b; Ovaldsen et al.\ 2007).  The typical nature of GRB hosts is of a faint star-forming galaxy, dominated by a young stellar
population (Christensen et al.\ 2004), detected at any redshift from 0 to 6.3 (Berger et al.\ 2007b).
Luminosities are generally low (Chary et al.\ 2002; Le Floc'h et al.\ 2003), indicating low masses and low metallicities (Gorosabel et al.\ 2005b; Wiersema et al. 2007a; Kewley et al. 2007). New cosmological simulations suggest that hosts associated with long GRBs are representative of the whole galaxy population (Nuza et al.\ 2007).

Many hosts are fainter  than the observational limits achieved today by the typical galaxy survey at high redshift (see for instance Cimatti et al.\ 2002; Abraham et al.\ 2004; Reddy et al.\ 2006; Noeske et al.\ 2007). In fact, the bright optical GRB afterglow often facilitates obtaining a spectroscopic redshift and the mere presence of a GRB encourages deeper photometric and spectroscopic observing campaigns than in conventional galaxy surveys. It is not clear yet whether this faint population is stand-alone and characterized by the association with a GRB event (Stanek et al.\ 2006; Fruchter et al.\ 2006), or whether we see many faint galaxies simply because these are the most common galaxies in the universe. Wolf \& Podsiadlowski (2007) showed that there is no dramatic difference between GRB hosts and what is expected for a galaxy population tracing star formation.

This paper is dedicated to the study of the largest possible sample of GRB host galaxies. In the past, many different tools applied to individual or a few GRB hosts led to very heterogeneous  results, not always easy to compare. We used a compilation of GRB hosts  
to measure a large number of galaxy parameters in a  robust and consistent way, and try to establish the role of GRB hosts in the cosmological scenario of galaxy formation and evolution.

\begin{figure}
%\centerline{\includegraphics[scale = .45]{f1.pdf}}
\centerline{\includegraphics[scale = .45]{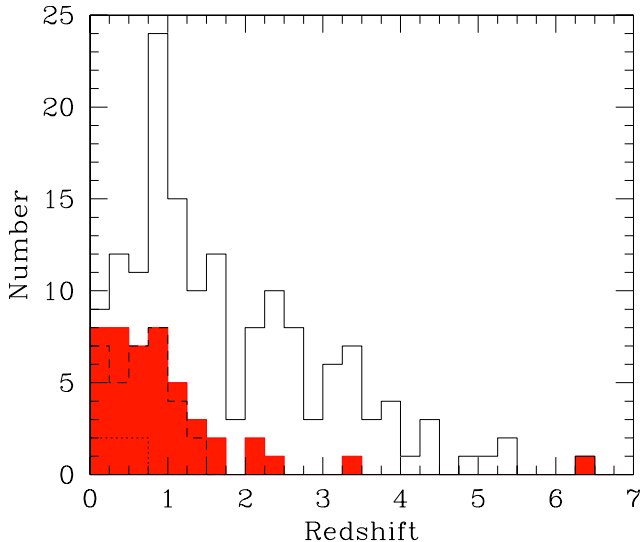}}
\caption {Histogram of the total sample of GRBs with measured redshift (\ngrbs\ objects, solid line), for
the GRB hosts studied in this work (\nhosts\ objects, filled histogram). The dashed line
represents the subsamples of hosts with detected emission lines (\nemi\ objects). The dotted line represents the subsample of hosts associated with short GRBs (six objects).}
\label{f1}
\end{figure}

The sample is selected by requiring that optical and/or near-IR (NIR)
photometry (available in the literature) have allowed the host identification. 
For a better stellar mass estimate, the important observable is the galaxy photometry redward of the
4000\AA\ break (Glazebrook et al.\ 2004), which restricts our search to (mainly) GRBs with $z\lsim3$. The total number of GRB hosts for which the stellar mass is estimated is \nhosts\ (Figure~\ref{f1}). For a subsample of \nemi, rest-frame optical emission-line fluxes from spectra are also available.  In the past, the stellar mass for several GRB hosts was derived by Chary et al.\ (2002), Castro Cer{\'o}n et al.\ (2006), and  Micha{\l}owski et al.\ (2008).

Throughout the paper, we distinguish between the subsamples of galaxies originating in short GRBs, which are associated with neutron star/black hole mergers possibly in evolved stellar populations, or long GRBs, associated with core-collapsed SNe and more abundant in young stellar populations (Woosley 1993). It has been argued by different authors that the hosts of the two classes are substantially different, the former in early- as well as late-type galaxies, the latter mainly in young star-forming galaxies (e.g., Prochaska et al.\ 2006; Berger et al.\ 2007c). Long GRB hosts are the vast majority (seven out of eight), as the long duration allows an easier identification of the afterglow, whereas the detection of short afterglows is very recent (Gehrels et al.\ 2005).

Our group created a public database, where many GRB-host observed parameters, available from the literature, are classified and stored. The database, the largest of its kind, is called GRB Host Studies (GHostS) and 
is accessible at the Web site www.grbhosts.info. GHostS is an interactive tool
offering many features. It also incorporates Virtual Observatory services, in particular Sloan Digital Sky Survey (SDSS) and Digital Sky Survey (DSS) sky viewing. 

The paper is organized as follows: in \S\ref{sample}, we describe the sample
selection; in \S\ref{SP}, multiband photometry and emission lines are used to derive galaxy parameters; results are given in \S\ref{gas} and \S\ref{metallicities}; the discussion and summary are presented in \S\ref{discussion} and \S\ref{summary}.  Throughout the paper we adopt an $h \equiv H_o/100=0.7$, $\Omega_M = 0.3$, $\Omega_\Lambda = 0.7$ cosmology (Spergel et al.\ 2003).

\section{The Sample}\label{sample}

The sample selection is based on the requirement that multiband photometry is available for the GRB host, mainly
optical and NIR photometry. Some mid-IR detections from Spitzer are also included in our analysis (Le Floc'h et al.\ 2006). The sample is listed in Table~\ref{tsample}, for which we report redshift,  the type of GRB (short or long), and the morphological classification, as given by Conselice et al.\ (2005) and Wainwright et al.\ (2007). The Galactic color excess $E_G(B-V)$ for each object (Table~\ref{tsample}) is estimated using the reddening maps by Schlegel et al.\ (1998), and is less than 0.15 for 83\% of the cases. All measurements reported in this work are corrected for the Galactic extinction (Cardelli et al.\ 1989). Figure~\ref{f1} shows the number of objects in the sample per redshift bin and the comparison with the
total sample of GRBs with measured redshift.

We compare observed parameters, with the same parameters  measured in field galaxies, observed at different redshifts. In particular, we use results from the $0.4<z<2$ Gemini Deep Deep Survey (GDDS; Abraham et al.\ 2004), the Lyman break galaxies (LBGs; Erb et al.\ 2006; Reddy et al.\ 2006), and the local dwarf galaxies (Lee et al.\ 2006). The GDDS is an ultra-deep NIR-selected survey ($K<20.6$, $I<24.5$) targeting galaxies in the ``redshift desert'' ($0.8<z<2$). GDDS is designed to find the most massive galaxies up to $z=2$. The limit in $K$ gives very high sensitivity to low-mass galaxies at $z=1$ (e.g., Savaglio et al.\ 2005). LBGs are UV-selected (observed ${\cal R}<25.5$) high-$z$ galaxies ($1.4<z<2.6$). Dwarfs in the local universe (distance $D<5$ Mpc) have absolute magnitudes at 4.5 $\mu$m in the interval $-19.9<M_{[4.5]}<-13.3$.

Measurements from GDDS and the GRB hosts are treated in the same way, in particular in the modeling analysis we use the initial mass function (IMF) proposed by Baldry \& Glazebrook (2003). A correction for the different IMFs, generally Salpeter is assumed, is necessary for the other samples.

The GRB-host sample studied here contains \nhosts\ objects (30\% of all GRBs with measured
redshift). Six of these are associated with short GRBs, all at redshift $z<0.7$, with a mean redshift $z=0.38$(Table~\ref{tsample}).
The mean and median redshift of the long-GRB sample is $z=1.05$ and $z=0.84$, respectively, with 88\% being at $z\leq1.60$, when the universe was 4 Gyr old (i.e., 29\% of the age today). 

We note that the GRB hosts that are considered have relatively low redshift, in comparison with the mean and median values for the total GRB population with known redshift ($z=1.79$ and $z=1.32$, respectively). This is because GRB hosts are faint and hard to detect for $z>2$, and also because redshifts in the pre-Swift era were mainly determined from strong optical emission lines, redshifted in the more difficult NIR regime for $z>1.5$. Most GRBs ($\sim70$\%) in our sample belong to the pre-Swift era.
This can potentially bias the results that we discuss in this paper. For instance, GRBs are associated with regions of star formation. As the star formation rate (SFR) density in the universe is a strong function of the galaxy stellar mass (Juneau et al.\ 2005), it might be that GRB hosts at low and high redshifts are associated with less- and more-massive galaxies, respectively.

From the multiband photometry, we derive a `pseudophotometry' to create a homogeneous sample, as described in \S\ref{phot} (Table~\ref{tcol}). Thirty-three out of \nhosts\ hosts have detected optical emission lines, indicating ongoing star formation (Table~\ref{temi}). This is partly an observational bias, as the redshift of a GRB is often determined from the emission lines in the host. The bluest of these lines is the [OII] at $\lambda=3727$ \AA. As many hosts are spectroscopically observed in the optical,
the highest redshift of the sample is marked by the  [OII]$\lambda3727$ line, at $z\sim1.6$ (Figure~\ref{f1}). Details are given in \S\ref{spe}. 

\begin{figure*}
%\centerline{\includegraphics[scale = .7]{f2.pdf}}
\centerline{\includegraphics[scale = .7]{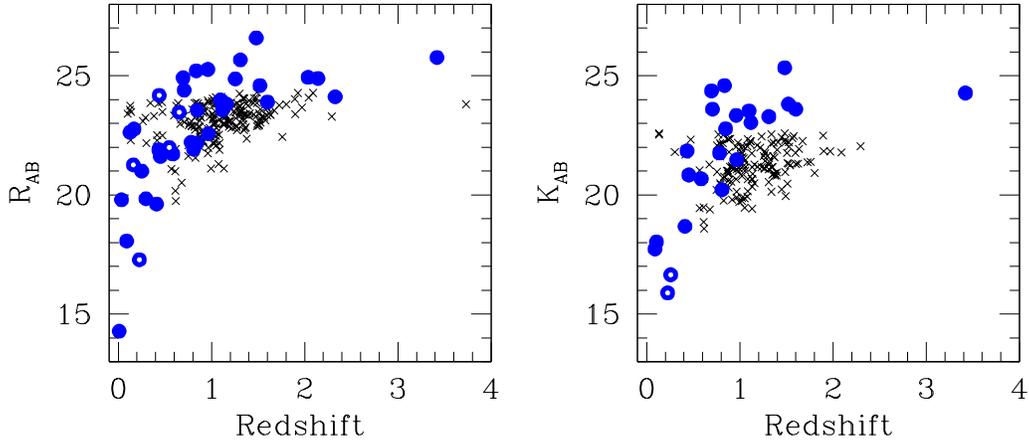}}
\caption {$R_{AB}$ (left plot) and $K_{AB}$ (right plot) observed magnitudes as
a function of redshift, for GRB hosts (filled circles) and GDDS field galaxies (crosses). The filled circles with white dots are short-GRB hosts. Errors for GRB-host magnitudes are generally below 0.2 mag.}
\label{mag}
\end{figure*}

\begin{figure*}
%\centerline{\includegraphics[scale = .92]{f3.pdf}}
\centerline{\includegraphics[scale = .92]{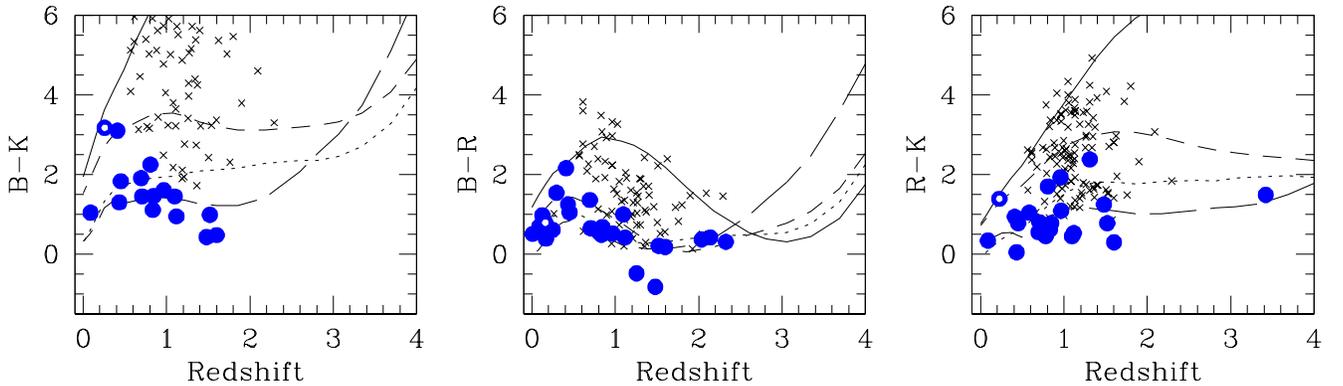}}
\caption {From left to right, $B-K$, $B-R$, and $R-K$ AB apparent colors as a function of 
redshift, for the GRB hosts (filled circles) and GDDS field galaxies
(crosses). The filled circles with white dots are short-GRB hosts. The curves are predicted colors as a function of redshift for galaxies with E (solid line), Sbc (short-dashed line), irregular (dotted), and starburst (long-dashed line) stellar populations (Le Borgne \& Rocca-Volmerange 2002).}
\label{col}
\end{figure*}

\subsection{The photometric sample}\label{phot}

The photometry of the sample is mainly covering the observed optical (including
the $U$ band) and NIR bands. For a subsample, mid-IR detections
or upper limits are also available (Le Floc'h et al.\ 2006; Berger et al.\ 2007b). Only GRB hosts detected in at least two bands are included in the sample, the reddest of which has to be above the 4000\AA\ break. This means mainly $z\lsim3.4$ hosts, the only exception being the host of GRB~050904 at $z=6.3$, for which only optical and mid-IR upper limits are available (Berger et al.\ 2007b). The total number of GRBs with $z\lsim3.4$ is 109, two-fifths of which are in our sample. The rest were not considered, because optical-NIR photometry is not available, either because observations were never attempted or not deep enough.

The detection of the host above the 4000\AA\ break is necessary to determine one of the galaxy key parameters, the total stellar mass. Parameters derived from only UV photometry are very uncertain, because  the mass-to-light ratio in the UV can vary by a very large factor in young stellar populations and due to the presence of dust.

The main difficulty of dealing with the photometric sample of GRB hosts is that it is very
heterogeneous and sparse.  For instance the data taken by different groups are treating Galactic dust
extinction or aperture corrections in different ways. Moreover, heterogeneous filters are often used. In a few cases the GRB host is observed in the same band with different telescopes, and results differ by more than the observational uncertainties.

To reduce the confusion, from the observed multiband photometry we have derived a ``pseudophotometry",  which is  a homogeneous photometry for a reduced set of filters. Magnitudes for the pseudophotometry in the AB system, corrected for Galactic extinction, are reported in Table~\ref{tcol}. 

The principle of ``pseudophotometry" is to first define a large set of commonly used filters including SDSS, Hubble Space Telescope (HST), Bessel, Johnson, and IRAC filters. Then, we reduce this set of filters to the filters which are necessary to represent all the photometric points involved in the available GRB-host observations. To do so, we use a criterion  $| \log \lambda_1 - \log \lambda_2 | < \log(1+300/5000)$, or maximum wavelength difference of 300 \AA\ at 5000 \AA: the effective wavelengths $\lambda_1$ of the filters involved in the real photometry must be close enough to one of the effective wavelengths $\lambda_2$ included in the reduced set of filters. We thereby build  a small set of 16 filters, ranging from $U$ band to IRAC 8$\mu$m band, which we use in practice for our spectral energy distribution (SED) fitting. All the GRB-host SEDs are therefore described with this set of filters. The effect of using the reduced set of filters instead of the exact bibliographic filters is small for our study: the offset in effective wavelengths is typically less than 200 \AA. For each GRB host, we checked with Monte Carlo simulations that random shifts of this order on the filters wavelengths produce small errors on the mass estimates. We note that not all observed magnitudes are in Table~\ref{tcol}. For more details on the observed photometry, see the GHostS database.

The $R_{AB}$ and $K_{AB}$ magnitudes for the sample are shown in Figure~\ref{mag}. The comparison with the GDDS $K_{AB}<22.5$ galaxies shows that GRB hosts are generally faint galaxies, with about half being fainter than $R_{AB}=23.5$ and $K_{AB}=22.5$. All short-GRB hosts are at $z<0.7$.  This could be a selection effect, as short afterglows are harder to detect than long afterglows (Gehrels et al.\ 2005). In Figure~\ref{col}, we show the apparent colors of the sample. The comparison with a complete sample of GDDS galaxies, and the predicted colors assuming different stellar populations (E/S0, Sbc, irregulars, and starbursts), indicates that GRB hosts are generally blue star-forming galaxies (see also Berger et al.\ 2007a). The short-GRB hosts are still too few to conclude anything about their stellar population from their colors.

\subsection{The Spectroscopic Sample}\label{spe}

Fluxes of optical emission lines from [OII]$\lambda3727$ to the [SII]$\lambda\lambda6716,6731$ doublet, originating in the star-forming (or HII) regions are measured for a subsample of \nemi\ GRB hosts and corrected for Galactic extinction (Table~\ref{temi}). Errors are reported when available. The [OII] and H$\alpha$ lines are detected up to $z=1.31$ and $z=0.45$, in 31 and 11 hosts, respectively. H$\beta$ and [OIII]$\lambda\lambda4959,5007$ are both detected in 19 hosts. The [NeIII]$\lambda3869$ line is present in 13 hosts; the [NeIII]-to-[OII] flux ratio is $>0.2$ in six hosts. Such relatively strong values indicate high temperature or high ionization, which is expected in  the presence of hot massive stars. Strong [NeIII] is common in HII regions and blue compact dwarf galaxies (Stasi{\'n}ska 2006). Contamination from Active Galactic Nucleus (AGN) is unlikely, as seen from emission-line ratios (Section 5.6). 

The true emission-line fluxes can be several times higher than those measured.
Some corrections are necessary in order to have a better estimate of the total SFR, metallicity, and dust extinction in the hosts. The slit-aperture flux loss is determined for each host individually. It depends on the size of the host with respect to the slit aperture used and the seeing conditions during observations. The aperture correction is reported in the last column of Table~\ref{temi}, but it is not applied for the fluxes given in this table. This is the factor we have to multiply the observed flux to obtain a more realistic total flux. It is `1' (no correction necessary, almost half of the hosts) either when it is negligible, or it is already applied by authors of the papers from where fluxes are taken, or  the redshift is high enough (galaxies get smaller) for the given slit (generally 1 arcsec). In all other cases it is $>1$ and derived by us, mainly comparing the observed multiband photometry with the flux-calibrated host spectrum, or it is estimated by considering the galaxy size in comparison with the slit aperture used and the seeing condition during observations. One arcsec corresponds to a physical size from redshift $z=1.31$ to $z=0.009$ (the highest and lowest redshift in the sample) of 8.4 to 1.9 kpc, respectively. The aperture correction is $>1$ and $\leq2$ for eight hosts. For the remaining 5, it is $>2$ and $\leq5$. Of course our aperture correction is not perfect, but it is the best that can be done, given the inhomogeneity of the sample.

Dust extinction in the host (see \S\ref{dust} for details) is determined from the Balmer decrement for the subsample with simultaneous H$\beta$ and H$\alpha$ detection (10 GRB hosts, Table~\ref{tbal}), and is calculated assuming a gas temperature  $T=10^4$ K.  The dust extinction is not applied for fluxes in Table~\ref{temi}, but is applied when estimating SFRs. 

The stellar absorption correction, generally more important in evolved stellar populations, is estimated only for the Balmer lines, as it is typically negligible for the metal lines. In local irregular and spiral galaxies, the typical stellar Balmer absorption equivalent width (EW) is 3 \AA, with 2 \AA\ uncertainty (Kobulnicky, Kennicutt, \& Pizagno 1999). It is lower in extragalactic HII regions,  EW$= 2$ \AA\ (McCall et al.\ 1985). 

To estimate the stellar Balmer absorption, we inspected the observed host spectrum. In some cases it is clearly negligible (e.g., the emission-line flux is high with respect to the stellar continuum) and no correction is applied. In all other cases it is generally estimated by adding 2 \AA\ to the measured, rest-frame equivalent width of the lines. In one case (the host of GRB~060505), the stellar correction is the one estimated by Th{\"o}ne et al.\ (2008).  

In Table~\ref{tbal}, we list the intrinsic-to-observed flux ratios (after and before stellar absorption correction) for Balmer emission lines. For H$\alpha$, the correction is up to 13\% of the observed emission flux (mean value 4\%). For H$\beta$ and H$\gamma$ it is clearly higher, up to 30\% and 56\% (mean 10\% and 25\%), respectively (Table~\ref{tbal}). The stellar absorption correction, although generally not very important, was never applied before in GRB-host studies. For the remaining of the paper, we will use emission fluxes corrected for stellar absorption and aperture-slit loss unless otherwise stated.

\begin{figure*}
%\centerline{\includegraphics[scale = .8]{f4a.pdf}}
\centerline{\includegraphics[scale = .8]{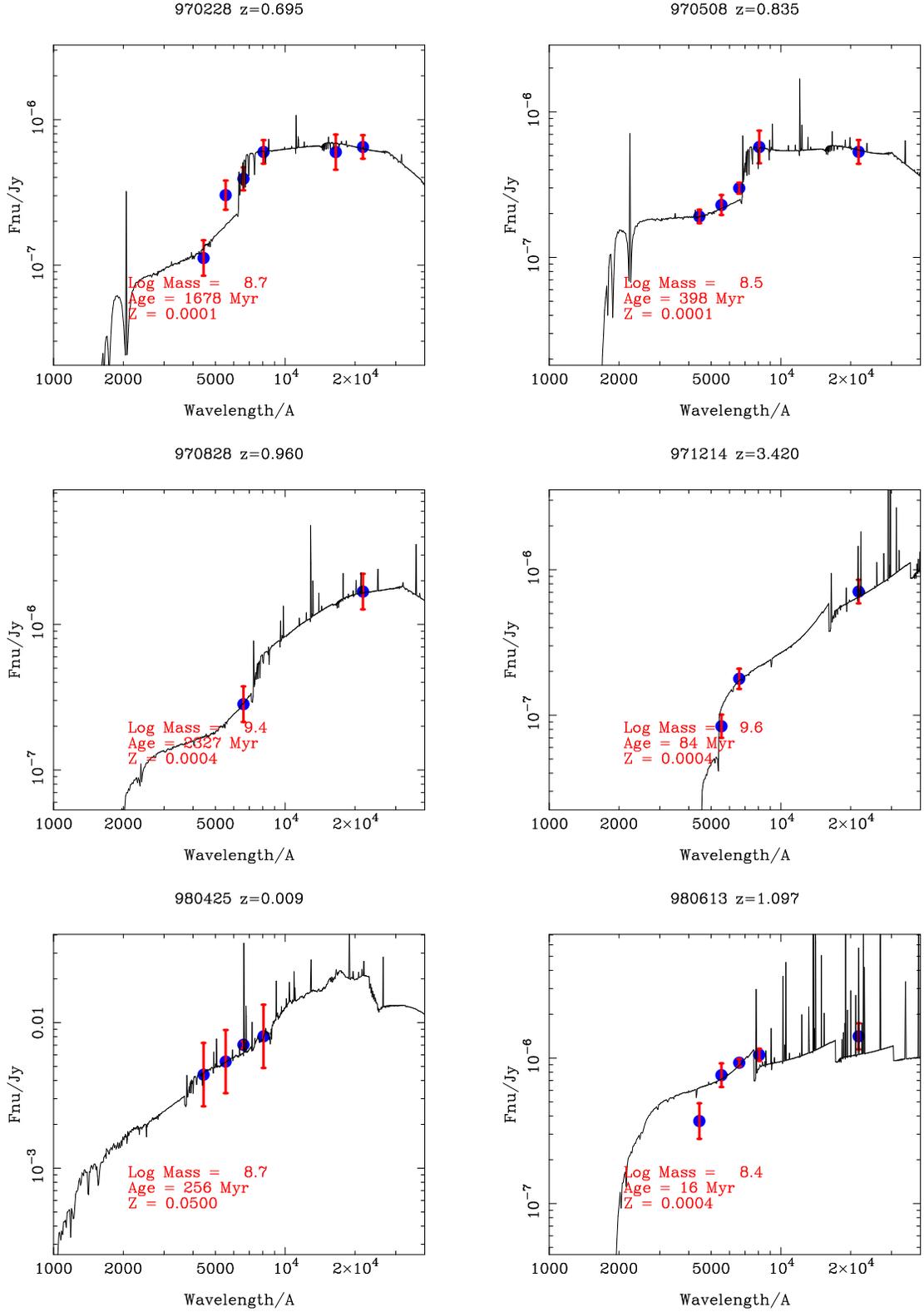}}
\caption {Observed SEDs of GRB hosts, and  synthetic spectra derived from the SED best fit, using P\'EGASE.  Reported best-fit parameters refer to an individual Monte Carlo realization, whereas those given in Table~\ref{tsed}  refer to an average from all Monte Carlo realizations. The metallicity, dust extinction and age derived from the best fit are very uncertain, due to the degeneracy between these parameters. On the other hand, the robustness of mass fitting, despite the age--metallicity degeneracy, is well known (e.g., Shapley et al.\ 2005).}
\label{fSED1}
\end{figure*}

\begin{figure*}
\setcounter{figure}{3}
%\centerline{\includegraphics[scale = .8]{f4b.pdf}}
\centerline{\includegraphics[scale = .8]{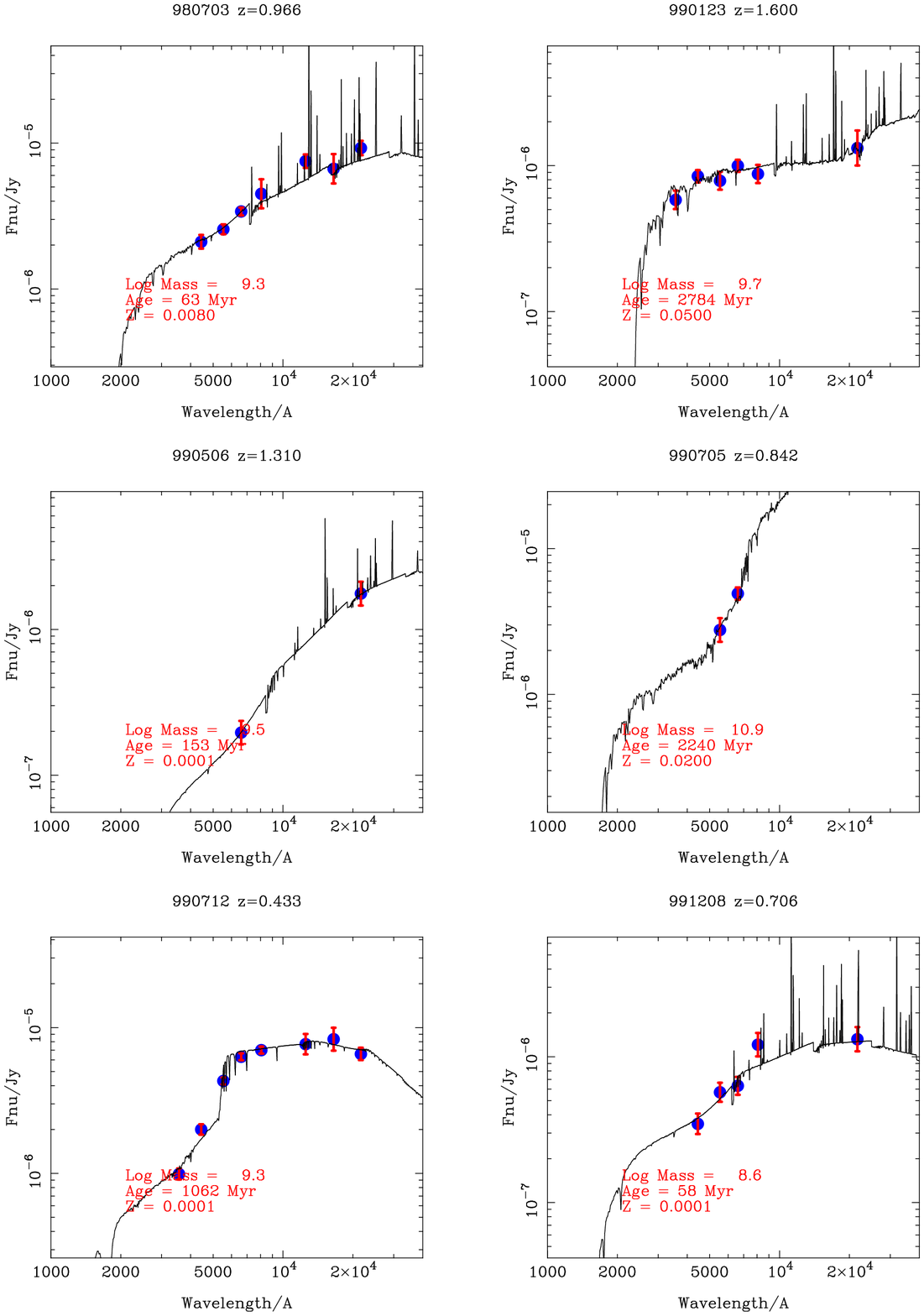}}
\caption {\it Continued.}
\label{fSED2}
\end{figure*}

\begin{figure*}
\setcounter{figure}{3}
%\centerline{\includegraphics[scale = .8]{f4c.pdf}}
\centerline{\includegraphics[scale = .8]{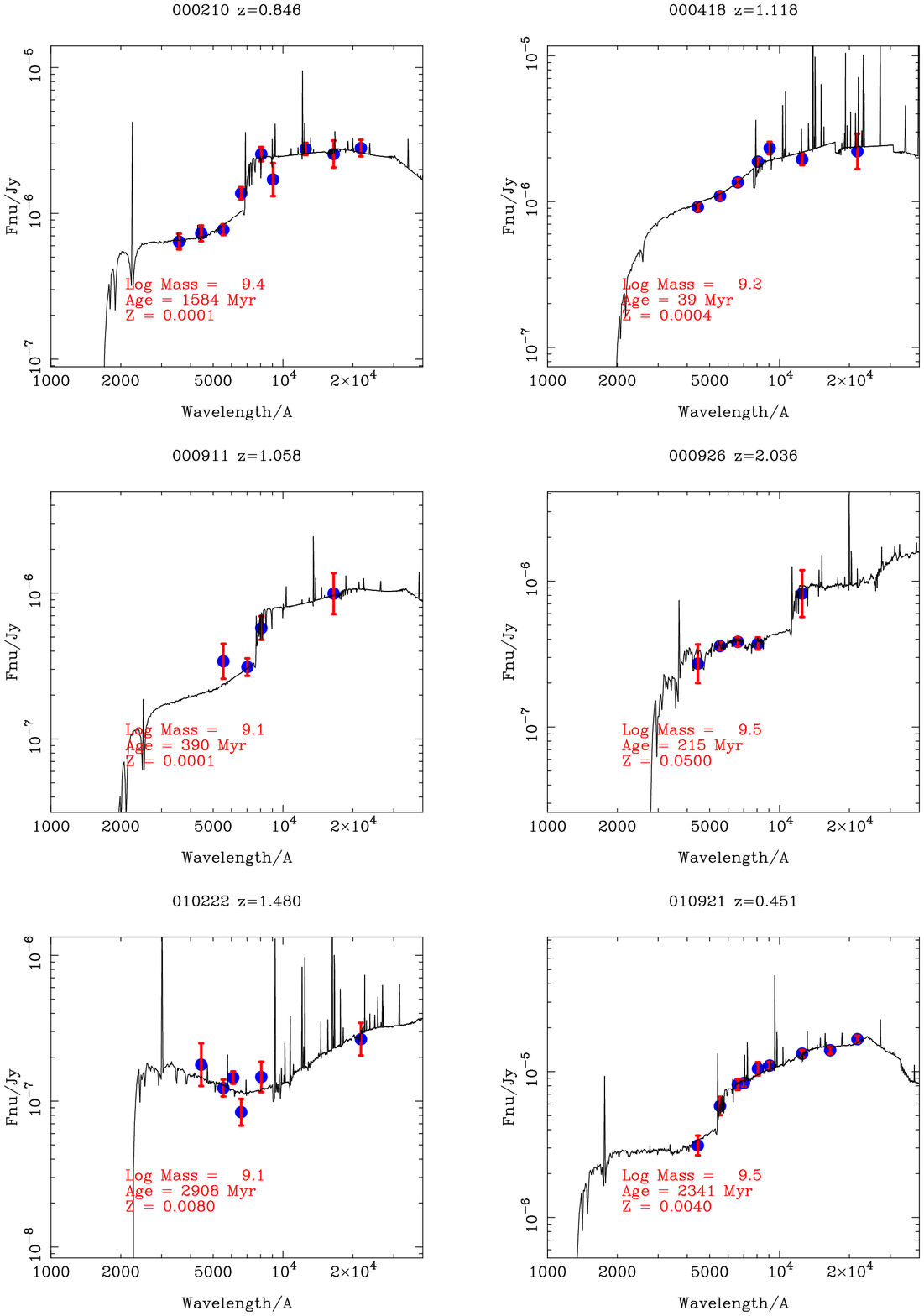}}
\caption {\it Continued.}
\label{fSED3}
\end{figure*}

\begin{figure*}
\setcounter{figure}{3}
%\centerline{\includegraphics[scale = .8]{f4d.pdf}}
\centerline{\includegraphics[scale = .8]{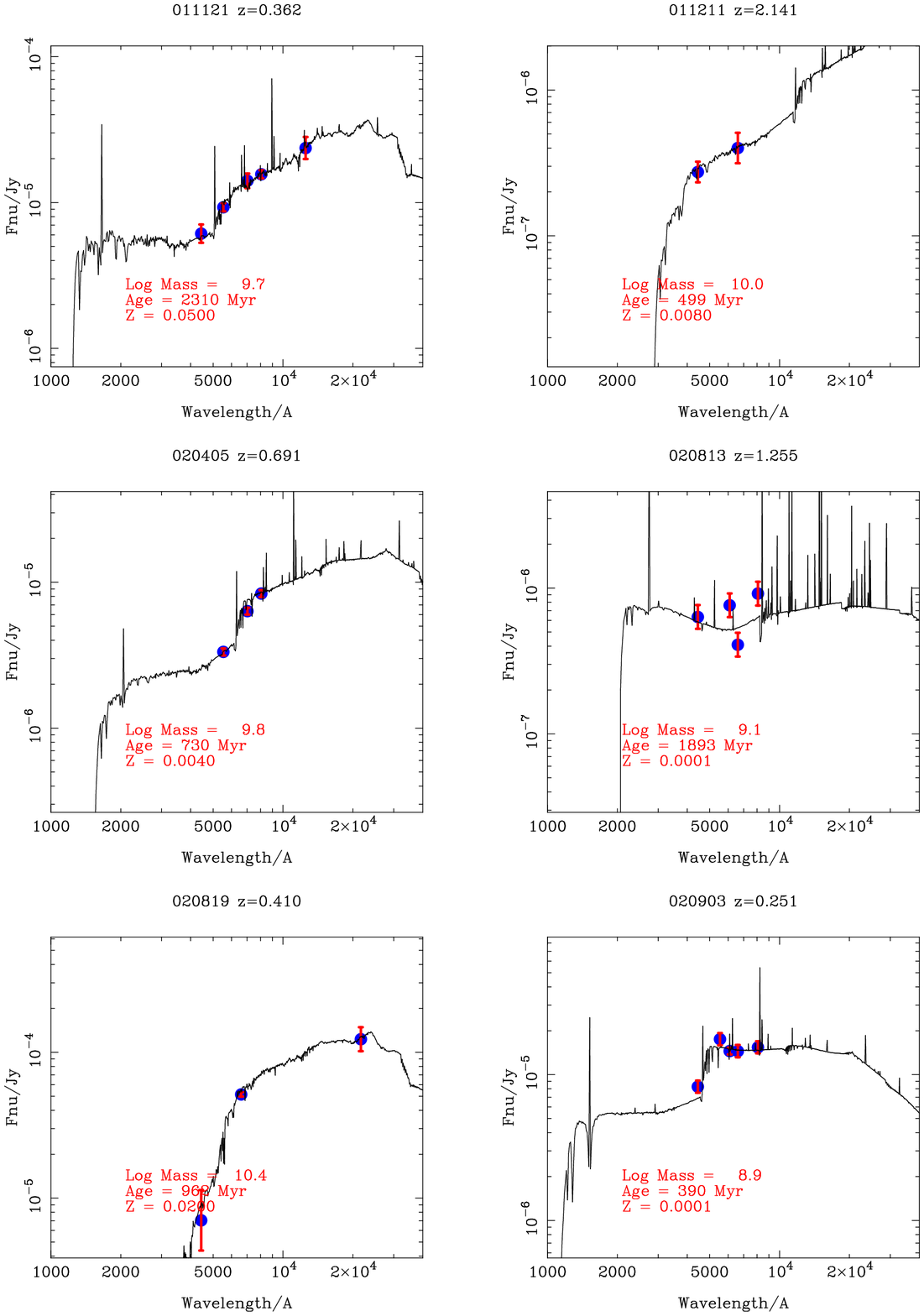}}
\caption {\it Continued.}
\label{fSED4}
\end{figure*}

\begin{figure*}
\setcounter{figure}{3}
%\centerline{\includegraphics[scale = .8]{f4e.pdf}}
\centerline{\includegraphics[scale = .8]{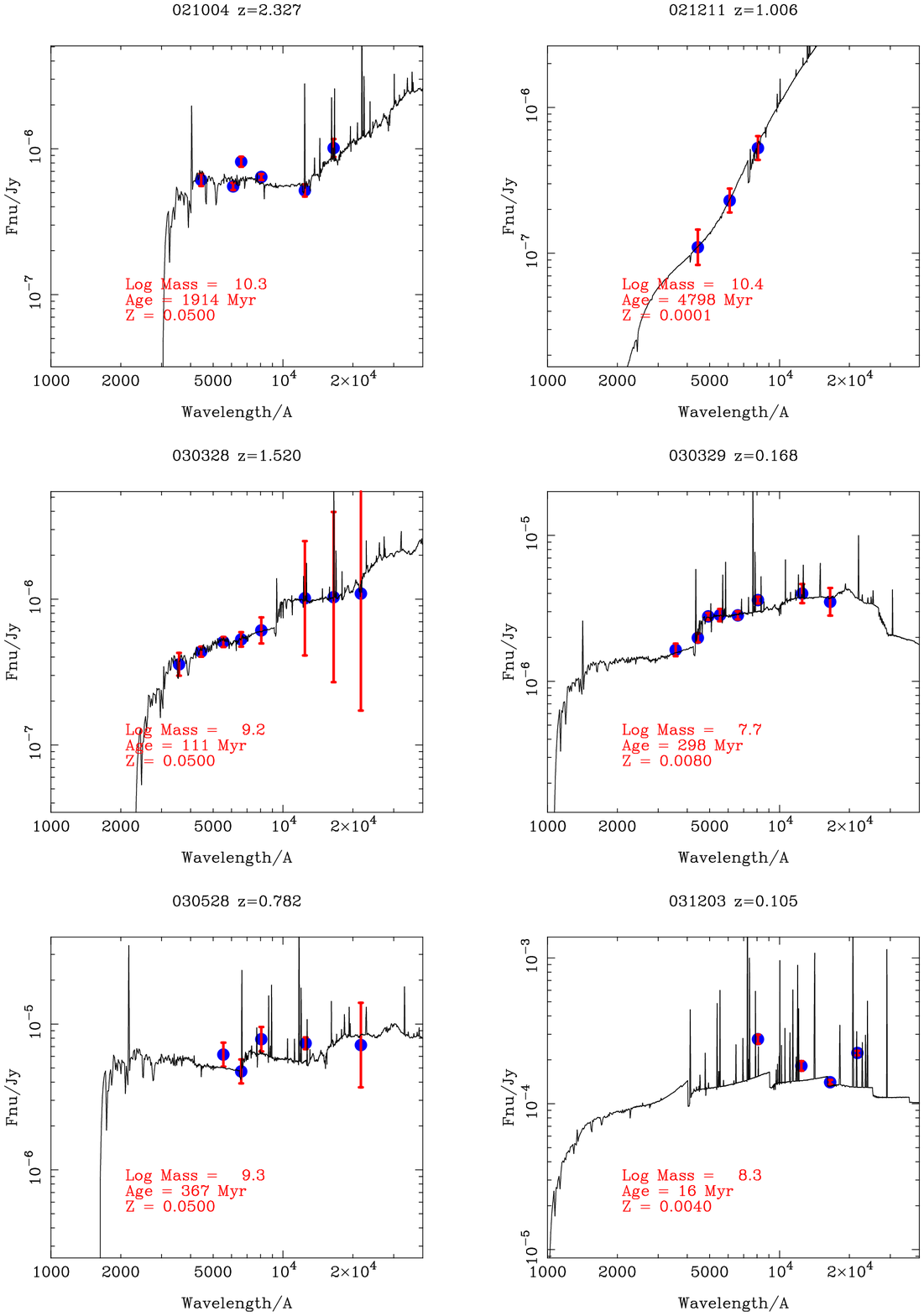}}
\caption {\it Continued.}
\label{fSED5}
\end{figure*}

\begin{figure*}
\setcounter{figure}{3}
%\centerline{\includegraphics[scale = .8]{f4f.pdf}}
\centerline{\includegraphics[scale = .8]{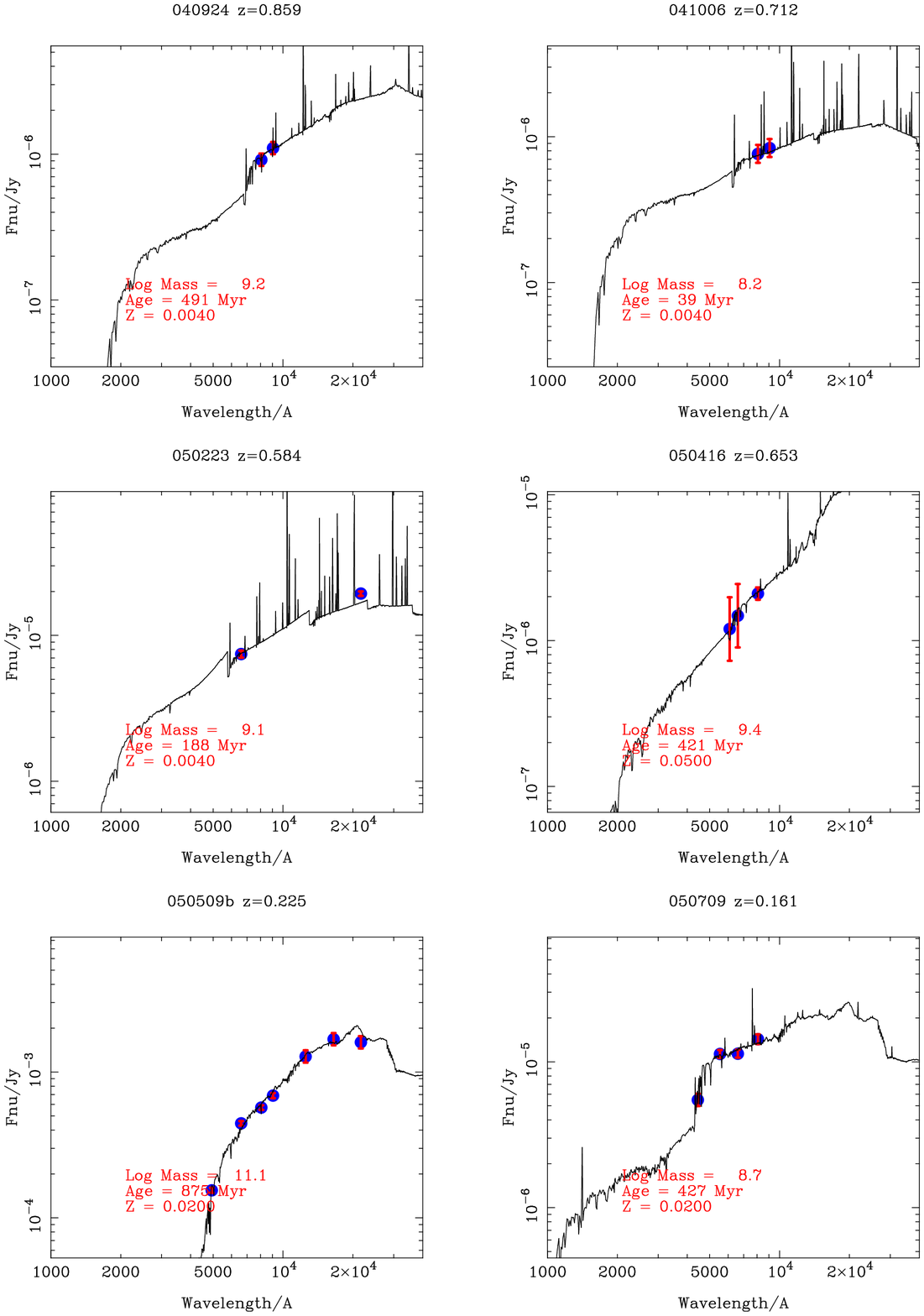}}
\caption {\it Continued.}
\label{fSED6}
\end{figure*}

\begin{figure*}
\setcounter{figure}{3}
%\centerline{\includegraphics[scale = .8]{f4g.pdf}}
\centerline{\includegraphics[scale = .8]{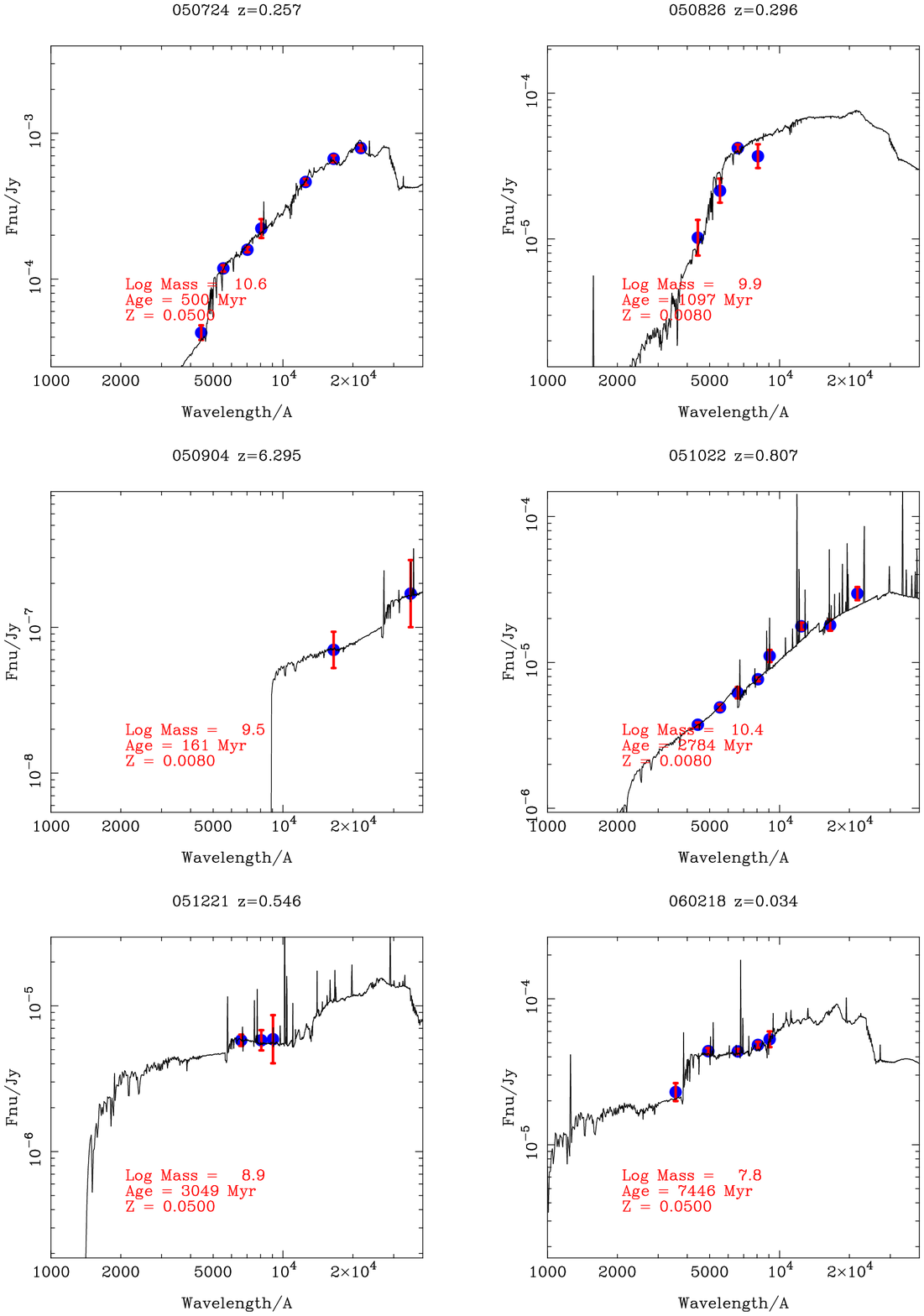}}
\caption {\it Continued.}
\label{fSED7}
\end{figure*}

\begin{figure*}
\setcounter{figure}{3}
%\centerline{\includegraphics[scale = .8]{f4h.pdf}}
\centerline{\includegraphics[scale = .8]{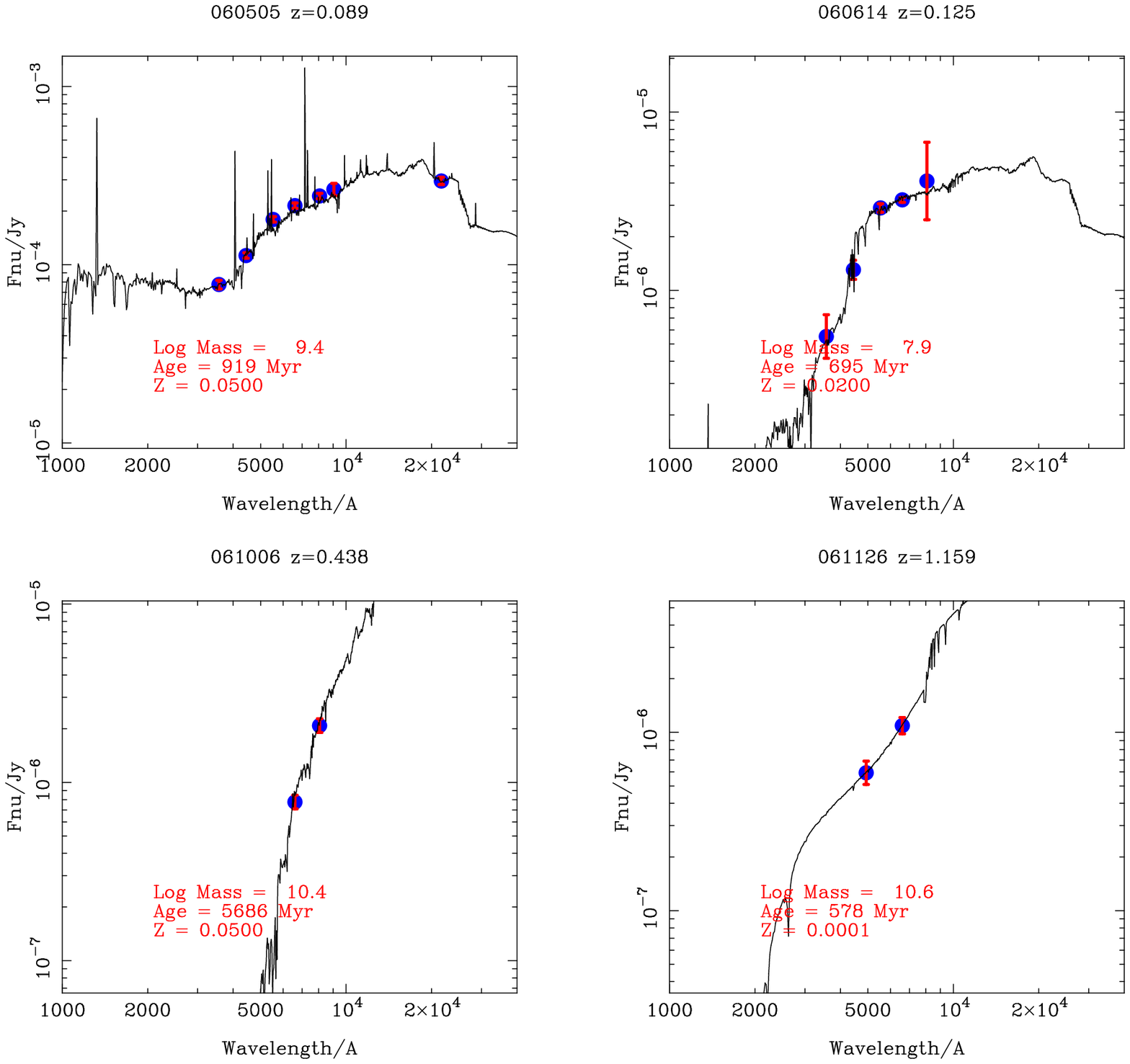}}
\caption {\it Continued.}
\label{fSED8}
\end{figure*}

\section{The Stellar Population}\label{SP}

The stellar properties of GRB hosts in the sample have been studied using the observed multiband photometry. Following the procedure of Glazebrook et al. (2004), we model the stellar population using P\'EGASE.2 (Fioc \& Rocca-Volmerange 1997; 1999) and P\'EGASE-HR (Le Borgne et al.\ 2004).  In order to avoid distorted mass-to-light ratios from young burst components superimposed on older stars, we represent the stellar population  by two components, with SFR $\propto \exp ^{-t/\tau}$, where $t$ is time and $\tau$ is the $e$-folding time. The first component is one of 10 alternate SEDs, each characterized by a different discrete value of $\tau$ (from starburst to constant SFR), covering a range of star formation histories (SFHs). The primary stellar component is coupled with a secondary component representing a star-bursting episode, with $\tau=0.1$ Gyr, and a total stellar mass in the range 1/10,000 up to two times the mass of the primary component. The age of the stellar population is also constrained by the age of the universe at the observed galaxy redshift. The attenuation of the stellar emission, due to dust, is a free parameter and can vary in the range $0<A_V<2$ (the Calzetti law is used). The same for the metallicity, in the range $0.0004 \leq Z \leq 0.02$. The modeling also includes an emission-line nebulosity component which we include as it can have small effects on the photometry near bright lines. We use the  standard P\'EGASE prescription for this and note that this is only used in the SED modeling and not in the interpretation of real galaxy emission-line spectra. 

The IMF used to estimate stellar masses is that which is derived by Baldry \& Glazebrook (2003). This more realistic IMF gives a total stellar mass which is generally 1.8 and 1.3 times lower than that derived by using Salpeter (1955) and Kroupa (2001) IMF, respectively. The full distribution function of allowed masses was calculated by Monte Carlo resampling 
the photometric errors. The Monte Carlo error approach is to: (1) apply a scaled Poisson error distribution of the fluxes and flux-errors (this is superior to using a Gaussian distribution as fluxes come from photon counting detectors); (2) create a new SED by flux and Poisson realization (flux-error) and re-fit the masses; (3) iterate $N$ times to make a distribution. The resulting mass distribution then samples the range of SFHs consistent with the error distribution of the photometry. The final masses and errors are the mean and standard deviation of the Monte Carlo mass distribution. 

Our code is more general than most in the literature because we include a secondary burst component. This is advantageous as it allows for possible mass biases associated with young bursts superimposed on old stellar populations to be explored (by turning off the burst term). In general we find that masses are robust to turning on/off bursts, and varying metallicities and dust, and are good to the $0.2-0.3$ $dex$ level. With this code then the mass of GRB hosts with poor photometry (e.g., few data points or photometry not sampling redder rest frame wavelengths) is poorly constrained, and this is reflected correctly as large error bars in Table~\ref{tsed}.
 
The requirement that one or more photometric detections above the 4000\AA\ Balmer break is known, and that at least two photometric bands are used, provides a reasonably robust total stellar mass determination. At $\lambda > $4000\AA, the SEDs are far less affected by younger stellar populations and uncertain dust modeling.

Figure~\ref{fSED1} shows the best fit to the observed SED.  The stellar mass-to-light ratio in the $K$ band $M_\ast/L_K$ and the stellar mass for each GRB host are listed in Table~\ref{tsed}, together with the absolute $K$ and $B$ magnitudes. The age of the last episode of star formation, the metallicity, the SFR and dust extinction in the stellar component are also calculated. However, these parameters are more uncertain due to the well-known degeneracy, according to which the colors of metal-rich stellar populations are indistinguishable from those of old or dusty stellar populations (Worthey 1994). 

Our method is more general than previous work which make assumptions which speed up the calculation (for example by assuming a single simple SFH or a single metallicity). In contrast with the method adopted by Christensen et al.\ (2004), our metallicity is given by the best fit, and not assumed to be solar. Moreover, Christensen et al.\ (2004) assume a single episode of star formation and do not derive the total stellar mass.

\begin{figure*}
%\centerline{\includegraphics[scale = .85]{f5.pdf}}
\centerline{\includegraphics[scale = .85]{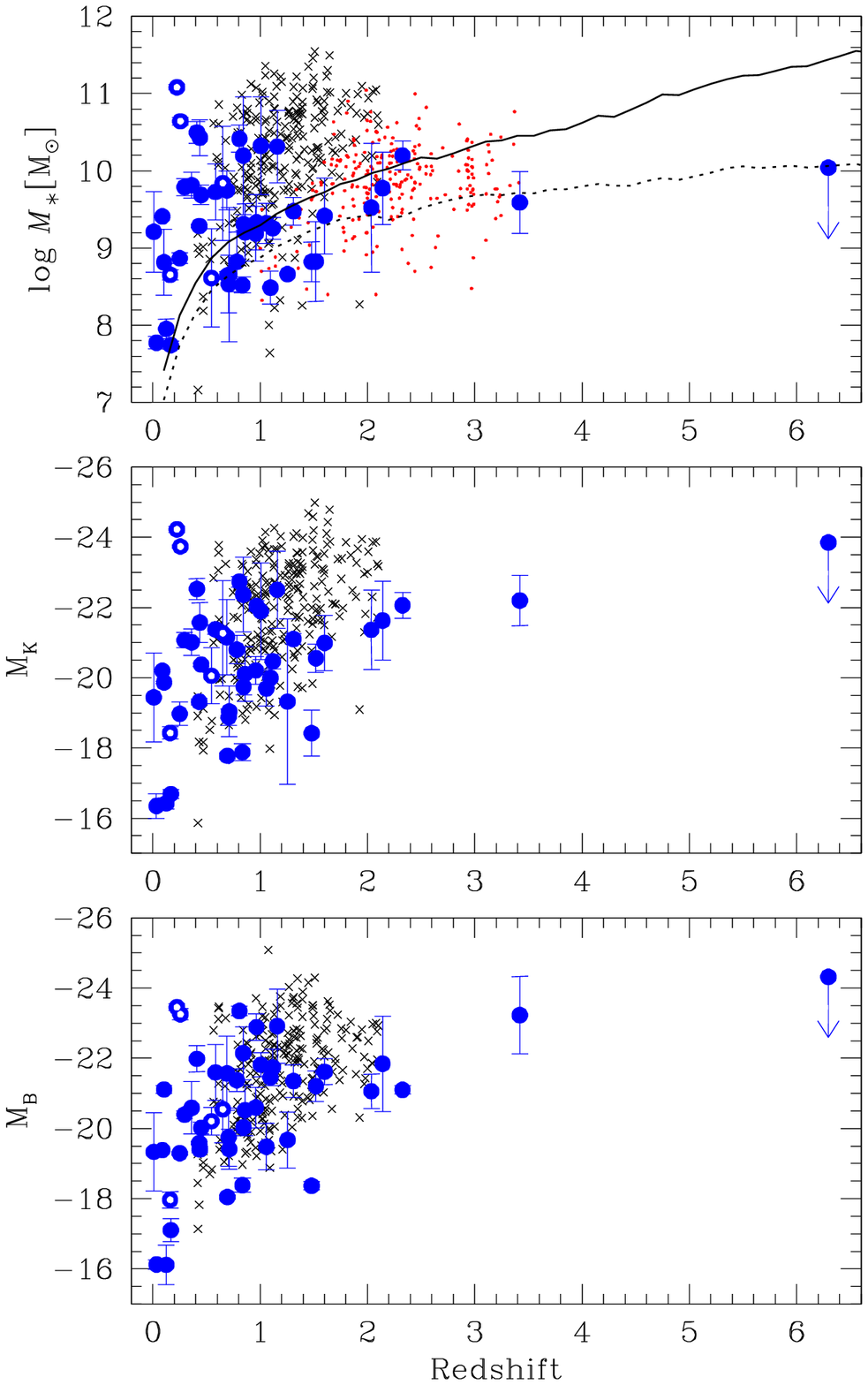}}
\caption {From top to bottom: stellar masses, AB $K$ absolute magnitudes, and AB $B$ absolute magnitudes, respectively. GRB hosts  are presented by filled circles. The filled circles with white dots are short-GRB hosts. In the top panel, only GRB hosts with stellar mass uncertainty $\Delta \log M_\ast<1$ are shown. Crosses represent GDDS
galaxies. Dots in the stellar mass plot represent LBGs (Reddy et al.\ 2006). The solid and dashed lines in the top plot represent the stellar mass as a function of redshift of a galaxy with $m_K=24.3$, and old stellar population or constant SFR, respectively.}
\label{all}
\end{figure*}

\begin{figure}
%\centerline{\includegraphics[scale = .45]{f6.pdf}}
\centerline{\includegraphics[scale = .45]{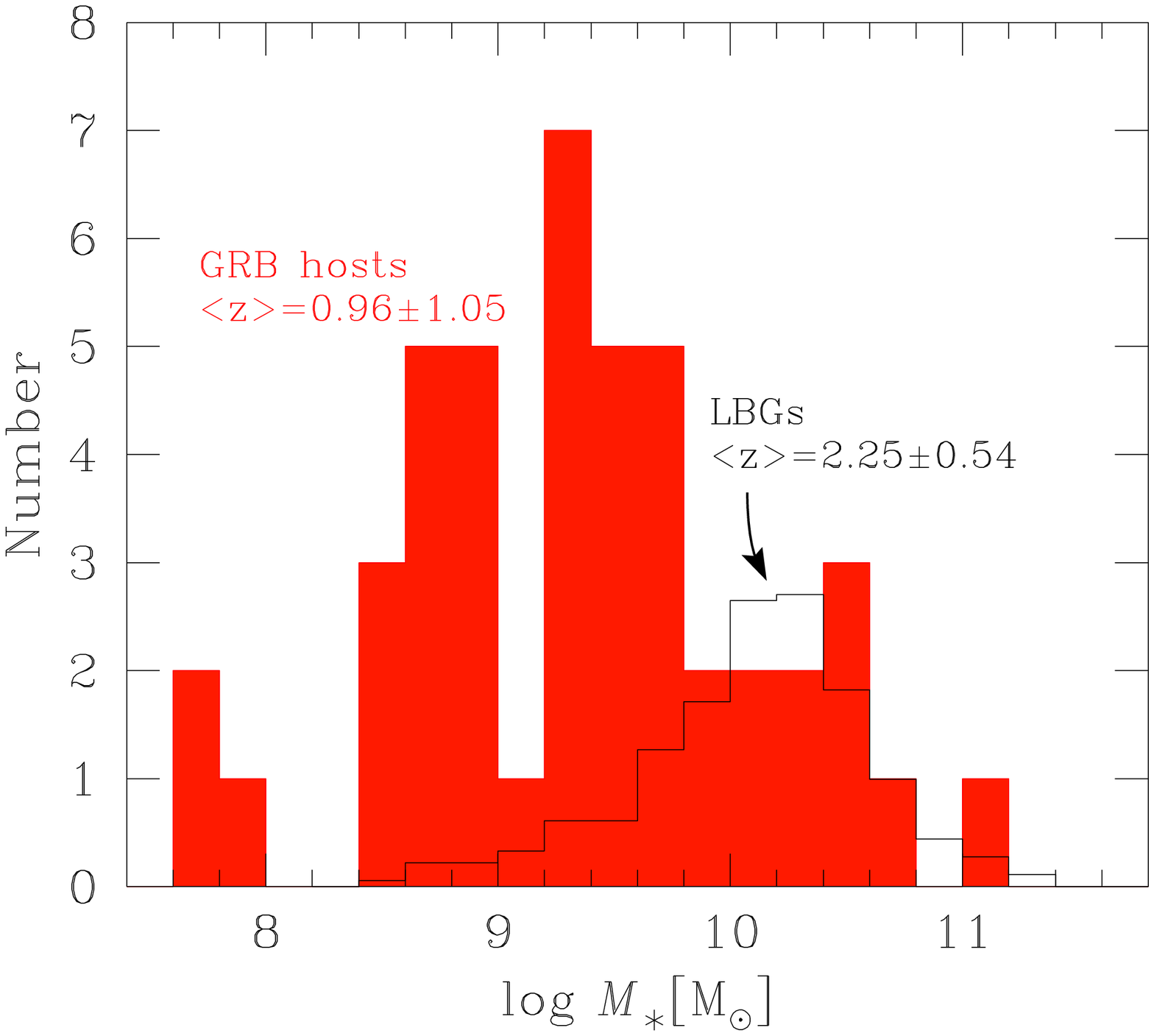}}
\caption {Stellar mass histogram of GRB hosts (filled histogram), mean redshift, and dispersion $z=0.96\pm1.05$. For comparison the empty histogram is derived  from LBGs  (redshift interval $1.3<z<3$; Reddy et al. 2006) normalized to the GRB-host histogram for $M_\ast>10^{10}$ M$_\odot$.}
\label{mass_hist}
\end{figure}

\begin{figure}
%\centerline{\includegraphics[scale = .42]{f7.pdf}}
\centerline{\includegraphics[scale = .42]{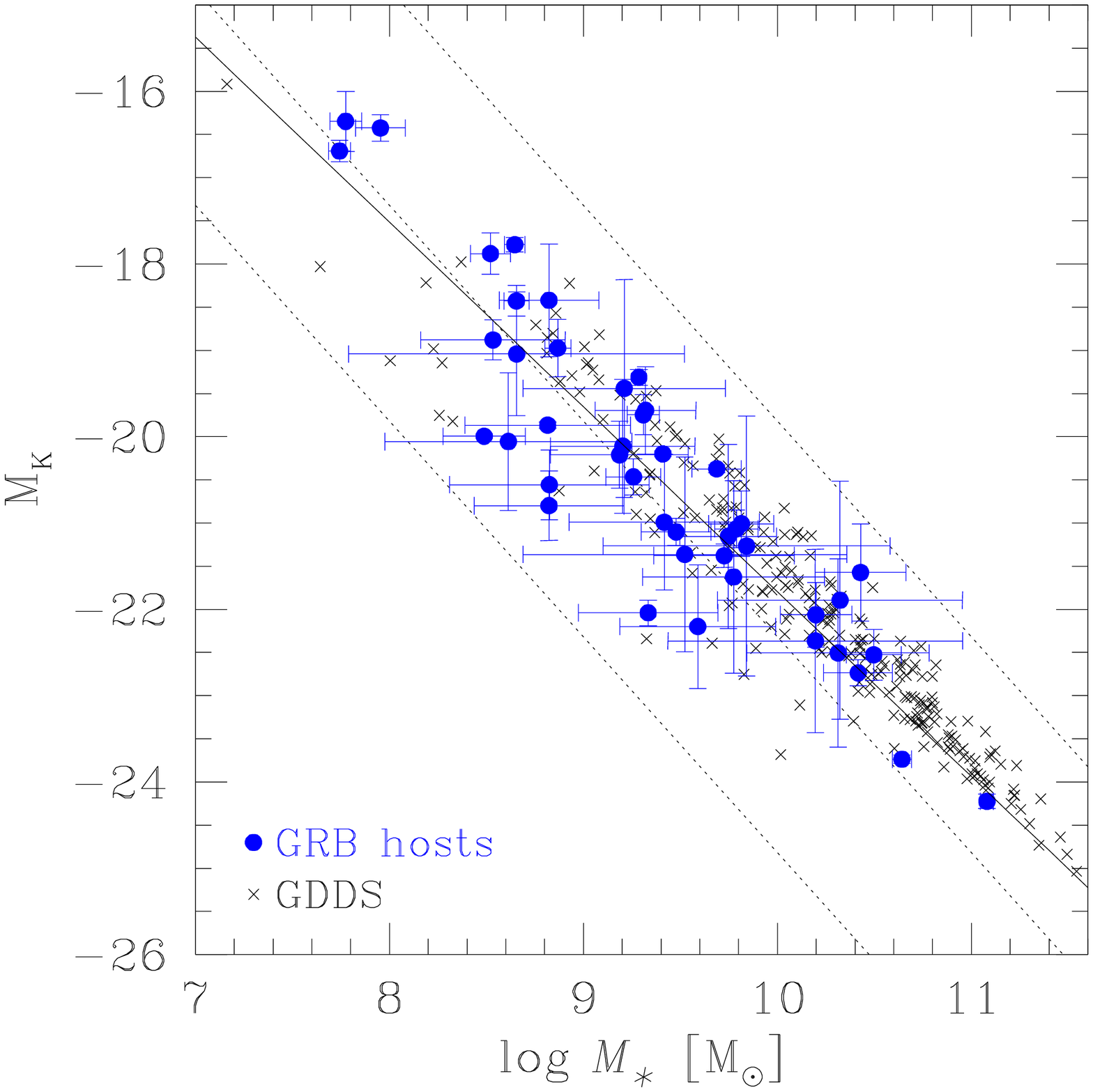}}
\caption {AB $K$ absolute magnitude as a function of stellar mass
for GRB hosts (filled circles) and GDDS galaxies (crosses). Only GRB hosts with stellar mass uncertainty $\Delta \log M_\ast<1$ are shown. The straight line is the linear correlation of the two parameters in the GRB-host sample, defined by Eq.~\ref{emk_m}. The dotted lines indicate, from left to right, constant stellar mass-to-light ratios of $M_\ast/L_K =$ 0.01, 0.1 and 1 $(M/L_K)_\odot$, respectively.}
\label{mk}
\end{figure}

\subsection{Stellar masses}

For the stellar mass fitting, we have utilized a very robust code which has been applied and tested in the literature extensively (e.g., Glazebrook et al.\ 2004; Baldry, Glazebrook, Driver 2008). 

Stellar masses and AB absolute $K$ and $B$ magnitudes are shown in Figure~\ref{all} and Table~\ref{tsed}. Magnitudes are corrected for dust attenuation in the host (a Calzetti law is assumed).  In Figure~\ref{all} we show the comparison with field galaxies, from the GDDS (Glazebrook et al.\ 2004) and LBGs (Reddy et al.\ 2006). The GRB-host median stellar mass is $10^{9.3}$ M$_\odot$, and is much lower than in the GDDS, by about one order of magnitude. For  short-GRB hosts, the median mass is higher, $\sim10^{10.1}$ M$_\odot$. More objects are necessary to conclude that the typical mass of short-GRB hosts is larger than that of long-GRB hosts. The curves in the upper plot of Figure~\ref{all} represent the stellar mass as a function of redshift of a galaxy with AB $K$ observed magnitude $m_K=24.3$, in the case of a old simple stellar population or a young stellar population with constant SFR.

In general, the sample is too small and inhomogeneous to study the mass function of the GRB-host population. Nevertheless, we inspected the fraction of GRB hosts per stellar mass bin. In Figure~\ref{mass_hist}, we compare this with that of the high-$z$ LBG sample (Reddy et al.\ 2006), after normalizing it to the GRB-host sample for galaxy stellar mass $M_\ast>10^{10}$ M$_\odot$, where the LBG sample is less affected by incompleteness. 
From the GRB-host stellar mass histogram (which drops for $M_\ast<10^{9.2}$ M$_\odot$) it is clear that galaxies similar to the typical GRB host are under-represented in high-$z$ spectroscopic galaxy surveys.

As already described before, the stellar mass of galaxies is better represented by the observed fluxes redward of the Balmer break at 4000 \AA. The rest-frame $K$-band luminosity is generally well correlated with the galaxy stellar mass. This correlation is also shown by GRB hosts (Figure~\ref{mk}), and can be expressed by

\begin{equation}\label{emk_m}
 \log M_\ast =  -0.467 \times M_K  -  0.179 ,
\end{equation}

\noindent 
where $M_K$ is the dust-corrected $K$-band absolute magnitude. This relation can be used to estimate the stellar mass of GRB hosts, provided that the SED is known redward of the Balmer break. The dispersion of the sample in Figure~\ref{mk} around Eq.~\ref{emk_m}  is about a factor of 2, and gives the minimum error on the stellar mass that one obtains by using this procedure. Such an error does not take into account systematic uncertainties on the modeling of the stellar mass and errors on the measured photometry.

We note that on average GRB hosts have higher $K$-band luminosities than field galaxies (from GDDS) with the same stellar mass (Figure~\ref{mk}). This means that $M_\ast/L_K$ is lower, which is expected from younger galaxies (Glazebrook et al.\ 2004), suggesting an intrinsic difference between the two galaxy populations.

We compared our results with the stellar masses derived for seven hosts by Chary et al.\ (2002) and another seven hosts by Castro  Cer{\'o}n et al.\ (2006). We find a weak correlation in the former, and no correlation at all in the latter, although the sample is too small to draw any conclusion.

\section{The gas component}\label{gas}

Optical emission lines\footnote{The Ly$\alpha$ line has not been considered because it is rarely detected (Kulkarni et al.\ 1998), and, more importantly, because of its strong attenuation due to resonant scattering  by the hydrogen atoms.} detected in a subsample of  \nemi\ GRB hosts (redshift range $0.009<z<1.31$) are used to estimate SFRs, metallicities, and dust extinction in the gas component. These are all (but one) long GRB hosts. The exception is the host of the short GRB~051221 (Soderberg et al.\ 2006b). Stellar absorption and slit-aperture flux loss (Table~\ref{tbal}) are taken into account to estimate the total emission flux. The dust extinction in the hosts is obtained as described below, and is used when measuring SFR and metallicity. The final adopted SFRs for GRB hosts are listed in the eighth column of Table~\ref{tsfr}. It is the value measured from H$\alpha$ when available, or [OII] otherwise. If no emission line is detected, UV luminosities at 2800 \AA\ are used. Details are in \S~\ref{SFRemi} and \S~\ref{SFRuv}.

\subsection{Dust extinction}\label{dust}

The dust extinction in the gas component is estimated using the expected Balmer decrement (H$\alpha$-to-H$\beta$ ratio) in absence of dust, for case $B$ recombination at a gas temperature of $10^4$K (Osterbrock 1989). We adopt the Milky Way (MW) extinction law, equivalent to Large Magellanic Cloud (LMC) and Small Magellanic Cloud (SMC) extinctions in the optical range. If the real GRB-host extinction law is flatter than in the MW, then our dust-extinction correction is underestimated. The observed Balmer decrement is
available for 10 GRB hosts (Table~\ref{temi}). Extinctions in the visual band $A_V$ for these hosts are derived after correcting for stellar absorption and are reported in Table~\ref{tbal}. The mean value is $A_V=0.53$ (a dispersion of 0.54).

For the other GRB hosts with no Balmer decrement measurement, we used the mean value $A_V=0.53$. For six of these hosts, H$\gamma$ and H$\beta$ are detected simultaneously, thus the line ratio can in principle be used to estimate $A_V$. However H$\gamma$ is much more uncertain and errors are too large, thus we ignore this method.

We note that measuring dust reddening in the gas component using emission  lines does not correspond to measuring the dust attenuation of the stellar component (Calzetti 2001). From emission lines one can estimate the dust extinction valid for point sources. In our case, the HII regions are considered point sources. Galaxies are not uniformly obscured, because the distribution of dust is
clumpy. We can only apply an approximate `mean' extinction correction, or dust attenuation correction. The dust attenuation applied to the stellar component is generally the Calzetti law. 

The visual extinction derived from emission lines of the HII regions is generally larger than the visual attenuation derived for the stellar component. A ratio of about 2 is obtained for two galaxies in the local universe by Cid Fernandes et al.\ (2005). Similar results were obtained by Calzetti et al.\ (2000) using eight star-forming galaxies. The relation between attenuation and extinction can
depend on the galaxy mass and/or SFR.

Although we found that in our sample the gas extinction derived from Balmer lines is higher than the 
stellar dust attenuation derived from the SED best fit, we found no relation between the two parameters. We note, however, that the stellar dust attenuation is very uncertain, because it is degenerate with metallicity and age of the stellar population.

\begin{figure}
%\centerline{\includegraphics[scale = .42]{f8.pdf}}
\centerline{\includegraphics[scale = .42]{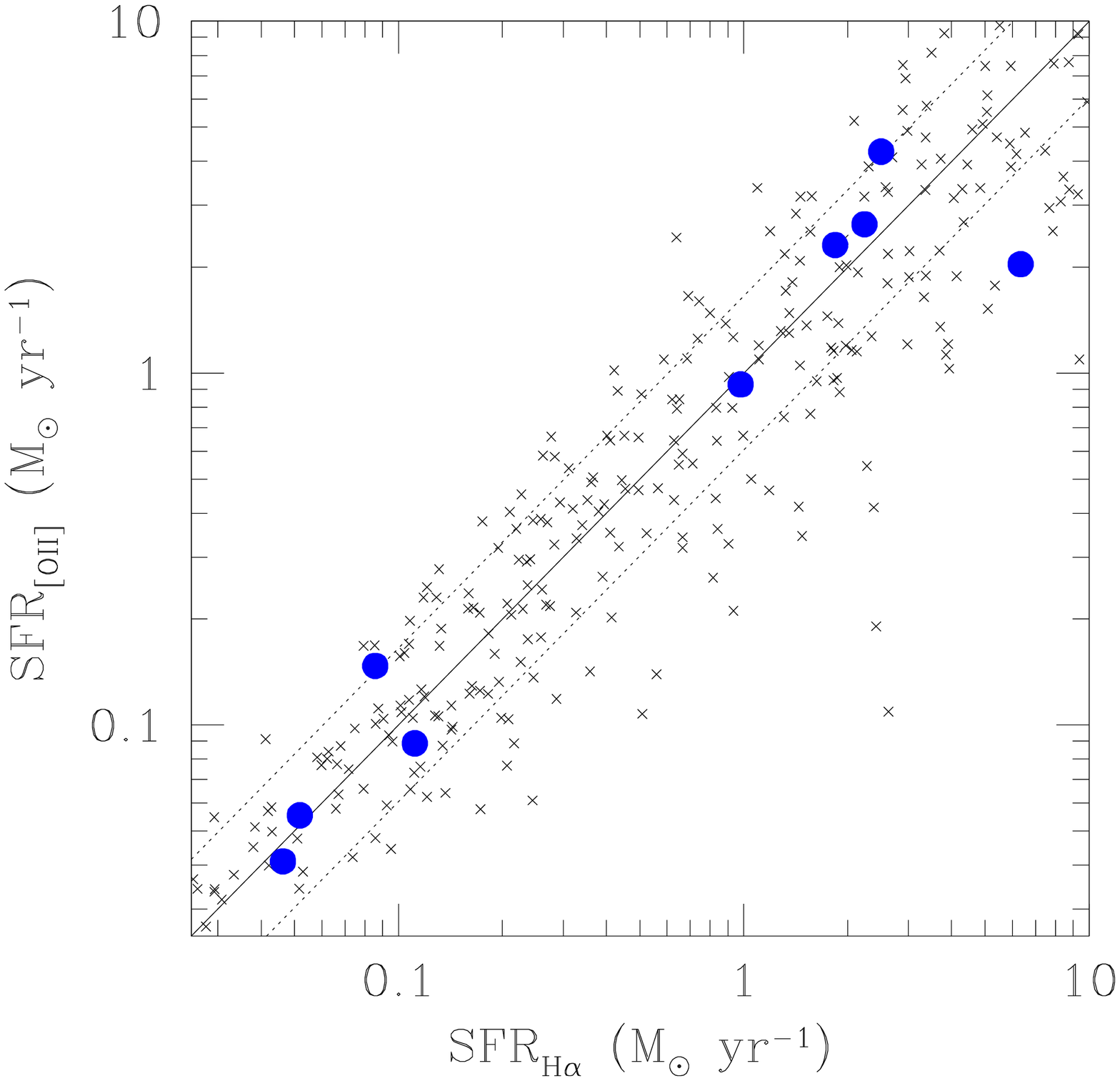}}
\caption {Filled circles are GRB-host SFRs estimated using Eqs.~\ref{sfrha} and \ref{sfroii} for H$\alpha$ and [OII] fluxes, respectively. The solid and dotted lines represent the one-to-one relation and $\pm$rms in the sample, respectively.
The [OII] and H$\alpha$ luminosities are corrected for dust extinction and stellar absorption. The SFR relations do not depend on the slit-aperture flux loss, so the correction is not applied. Uncertainties due to flux errors, when available, are generally small, below 50\%. Crosses represent local galaxies from Moustakas et al.\ (2006), and also include objects from the Nearby Field Galaxy Survey (Jansen et al.\ 2000), all corrected for dust extinction.}
\label{sfr_oii}
\end{figure}

\begin{figure}
%\centerline{\includegraphics[scale = .42]{f9.pdf}}
\centerline{\includegraphics[scale = .42]{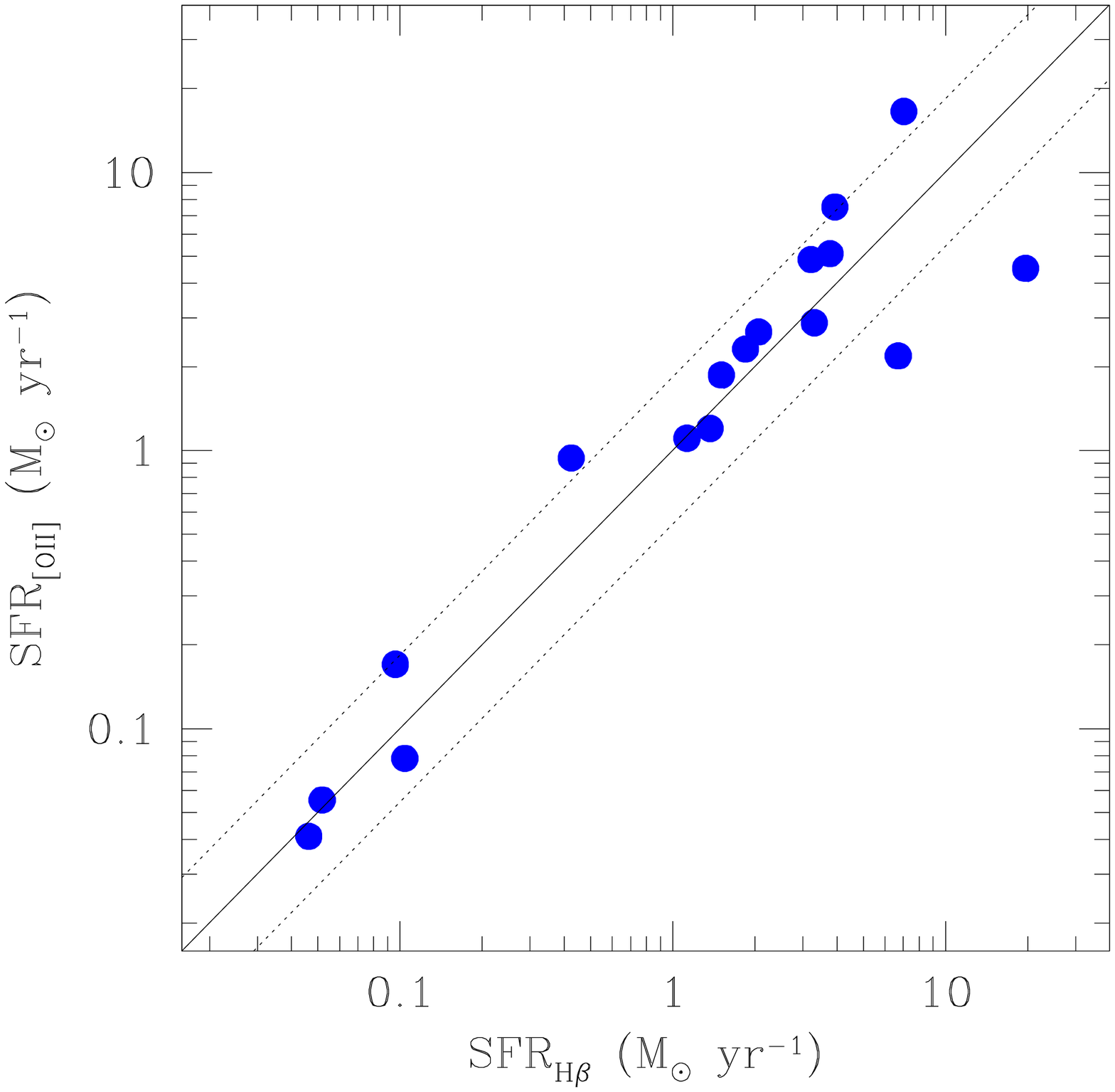}}
\caption {GRB-host SFRs estimated from [OII] and H$\beta$, using Eqs.~\ref{sfroii} and \ref{sfrhb}, respectively. The solid and dotted lines represent the one-to-one relation and $\pm$rms in the sample, respectively. The [OII] and H$\beta$ luminosities are corrected for dust extinction in the host and stellar absorption. The SFR relations do not depend on the slit-aperture flux loss correction, so this is not applied. Uncertainties due to flux measurement errors, when available, are generally small, below 50\%.}
\label{sfr_hb}
\end{figure}

\begin{figure}
%\centerline{\includegraphics[scale = .35]{f10a.pdf}}
%\centerline{\includegraphics[scale = .35]{f10b.pdf}}
%\centerline{\includegraphics[scale = .35]{f10c.pdf}}
\centerline{\includegraphics[scale = .3]{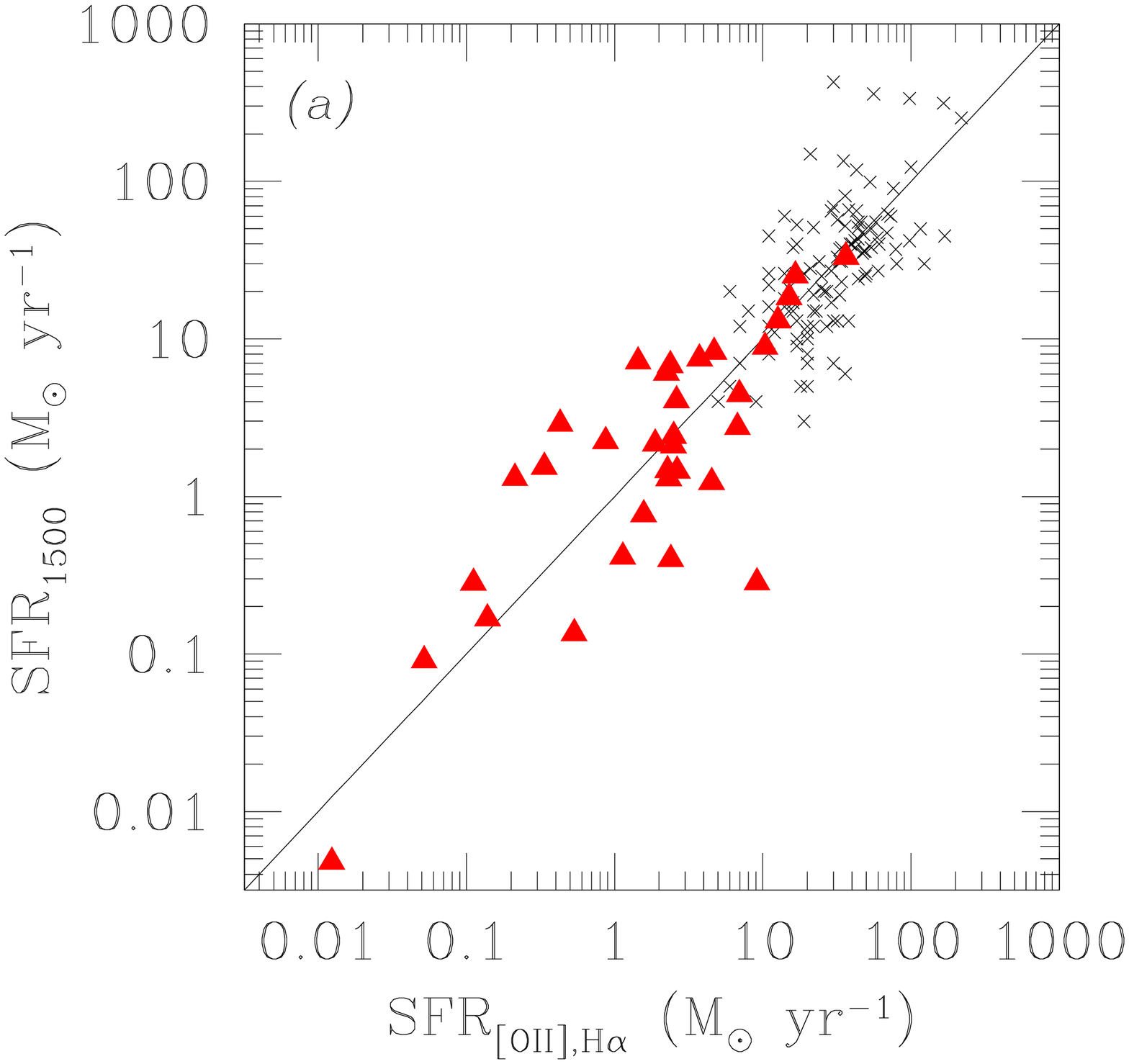}}
\centerline{\includegraphics[scale = .3]{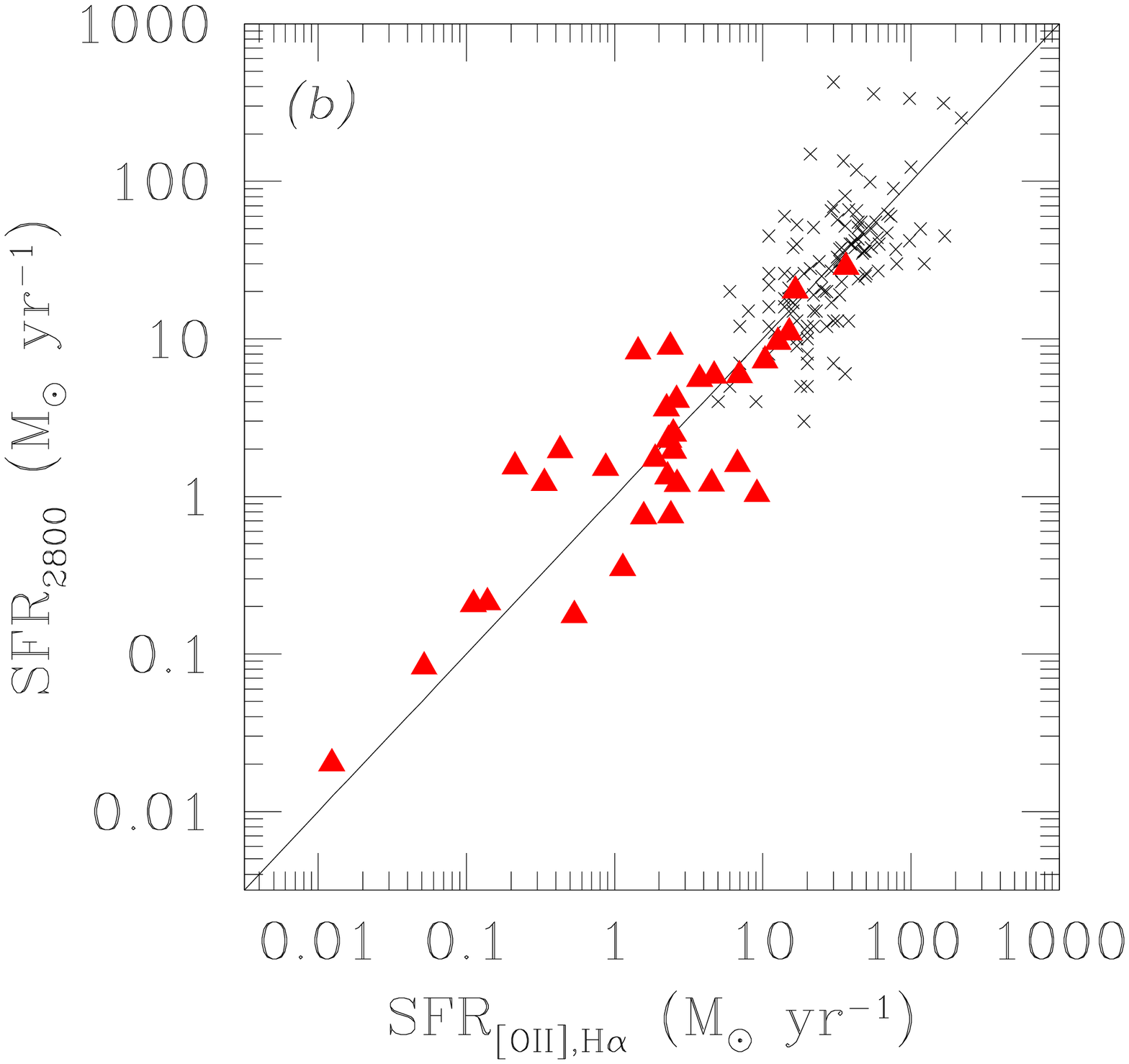}}
\centerline{\includegraphics[scale = .3]{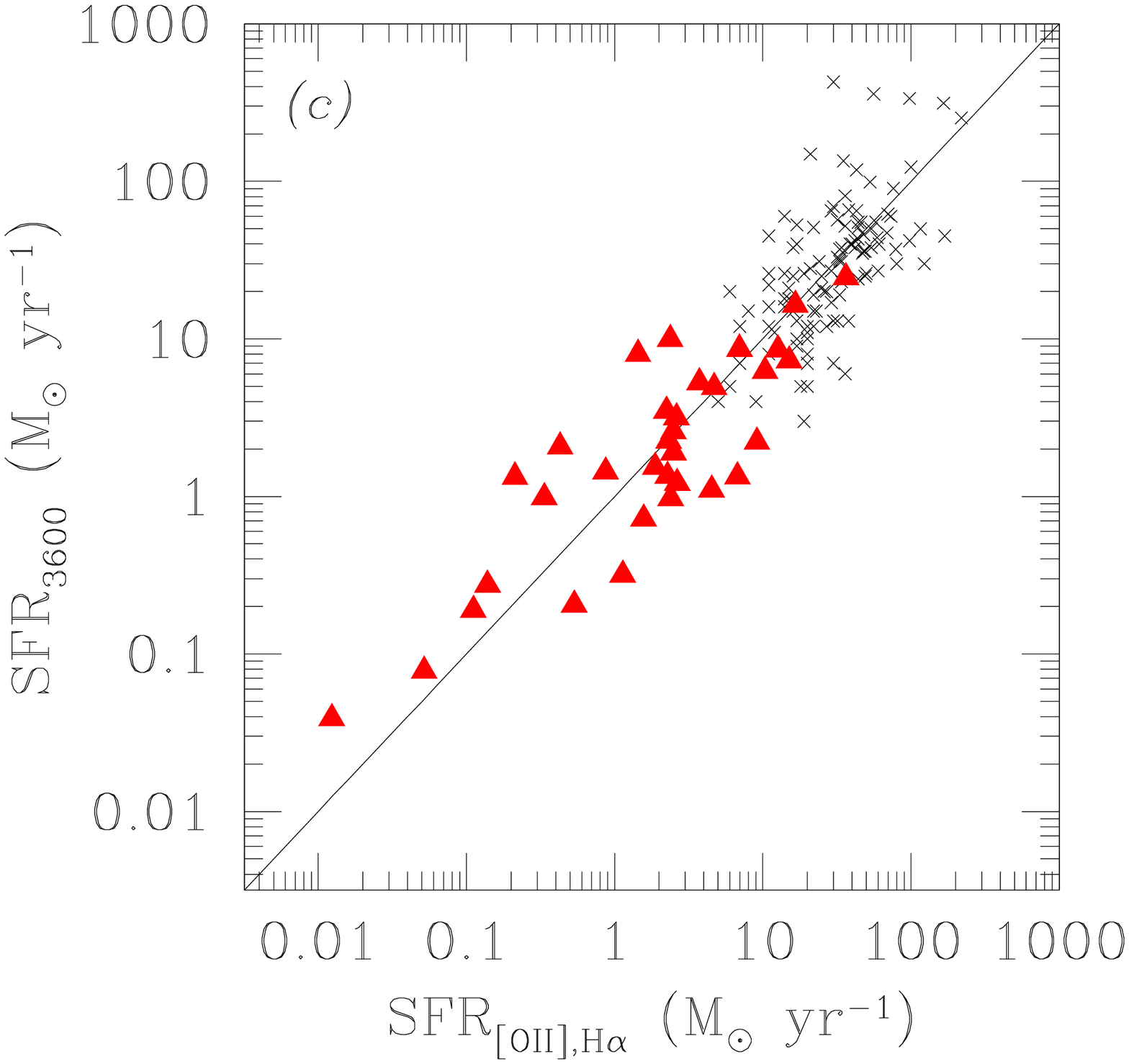}}
\caption {SFRs of GRB hosts (filled triangles) as derived from dust-corrected UV luminosities  ($y$-axis) or [OII] and H$\alpha$ luminosities ($x$-axis). The mean UV luminosities at $(a)$ 1500 \AA, $(b)$ 2800 \AA, and $(c)$ 3600 \AA\ are used, respectively. The UV dust attenuation is derived from the SED best fit of the stellar component. The aperture flux loss correction is also applied to emission lines. The rms dispersions around the straight lines are, from left to right, 0.49, 0.41 and 0.40 dex. Crosses represent SFRs for 114 LBGs derived by Erb et al.\ (2006) using UV fluxes at 1500 \AA\ and H$\alpha$ luminosities. The dispersion in this sample is 0.34 dex. }
\label{sfr_uv}
\end{figure}

\subsection{Star Formation Rates from emission lines}\label{SFRemi}

Traditionally, the best SFR indicator available for the optical-UV is the H$\alpha$ luminosity, corrected for dust extinction and stellar absorption. Otherwise, the more uncertain [OII] or H$\beta$ can be used instead.  The conversions derived by Kennicutt (1998) are the most commonly used. Moustakas, Kennicutt \& Tremonti (2006) also provided conversions, valid for local star-forming galaxies.

We used the dust-corrected [OII], H$\beta$, and H$\alpha$ luminosities. A 
mean  visual extinction $<A_V>=0.53$  corresponds to an SFR correction for [OII], H$\beta$, and H$\alpha$ fluxes of a factor of 2.1, 1.8, and 1.5, respectively (for MW extinction law), is applied when the Balmer decrement is not measured. The Balmer stellar absorption correction is on average 4\% and 10\% of the observed flux, for H$\alpha$ and H$\beta$, respectively (Table~\ref{tbal}).
The slit-aperture flux loss (up to a factor of 5, Table~\ref{temi}) is also taken into account.

To derive SFR from H$\alpha$, we used the prescription given by Kennicutt (1998). We only  correct for the different IMF adopted, which gives an SFR a factor of 1.8 lower, going from Salpeter to the more realistic IMF proposed by Baldry \& Glazebrook (2003). This gives

\begin{equation}\label{sfrha}
SFR_{\rm H\alpha} = 4.39\times10^{-42} \frac{L(\rm H\alpha)_{corr}}{erg\ s^{-1}}
M_\odot\ yr^{-1},
\end{equation}

\noindent
where $L(\rm H\alpha)_{corr}$ is the H$\alpha$ luminosity corrected for stellar absorption and dust extinction. 

Equivalently, the SFR from H$\beta$ can be expressed by

\begin{equation}\label{sfrhb}
SFR_{\rm H\beta} = 12.6\times10^{-42} \frac{L(\rm H\beta)_{corr}}{erg\ s^{-1}}
M_\odot\ yr^{-1},
\end{equation}

\noindent
where $L(\rm H\beta)_{corr}$ is the dust and stellar absorption corrected  H$\beta$ luminosity, and the multiplicative constant 12.6 is the one in Eq.~\ref{sfrha} multiplied  by 2.86 (the recombination factor  assuming $T=10^4$ K; Osterbrock 1989).

The advantage of using H$\beta$ instead of [OII] to estimate SFR is that, to first
order, it does not depend on metallicity. However, the stellar Balmer
absorption is generally uncertain, and important for old stellar populations. On the other hand [OII] is generally stronger and easier to detect than H$\beta$, but it is more affected by an at times undetermined dust extinction. Moreover, the [O II] flux depends on metallicity and ionization level (Kewley et al. 2004). 

Moustakas et al.\ (2006) found that the mean dust-corrected [OII]-to-H$\alpha$ flux ratio in local star-forming galaxies is about 1, with negligible dependence from the galaxy luminosity.  SFRs 
based on the observed [OII] luminosity are 0.4 dex uncertain. The metallicity\footnote{The metallicity is given by the oxygen abundance, expressed in solar units with $12+\log (\rm O/H) = 8.66$, from Asplund et al.\ (2005).} dependence is weak ($\log (\rm [OII]/H\alpha)=-0.1$ to 0.2)  in the metallicity  range $12+\log (\rm O/H)=8.2-8.7$, and becomes linear (in log-log scale) for $12+\log (\rm O/H)<8.2$, down to $12+\log (\rm O/H)=7.5$, where $\log ([OII]/H\alpha)\sim-1.0$. For galaxies with relatively high metallicity, the [OII] SFR is rather robust. 

In nine GRB hosts the dust-corrected ratio is relatively low, in the range $-0.6<\log ([OII]/H\alpha)<0.0$, which can be explained by the generally  low metallicities (see \S\ref{metallicities}).  Metals are an important source of radiative cooling, which gives the relation between metallicity and [OII]/H$\alpha$ (Kewley et al.\ 2001). We note the different mean redshift for the GRB hosts and Moustakas et al.\ (2006) sample ($z=0.23$ and $z\sim0$, respectively). However, this is likely too small to sensibly affect the results. 

We derived an [OII]--SFR relation valid for GRB hosts, by considering the subsample with simultaneous [OII] and H$\alpha$ detection. The [OII]--SFR best conversion is empirically derived by determining the best concordance between the SFRs estimated from H$\alpha$ and [OII] (Figure~\ref{sfr_oii}). This is given by

\begin{equation}\label{sfroii}
SFR_{\rm [OII]} = 5.54\times10^{-42} \frac{L(\rm [OII])_{corr}}{erg\ s^{-1}}
M_\odot\ yr^{-1},
\end{equation}

\noindent
where $L(\rm [OII]_{corr}$ is the dust-corrected [OII] luminosity. The dispersion in Figure~\ref{sfr_oii} is 0.22 dex (a factor of 1.7). Eq.~\ref{sfroii} is proposed as a valuable tool to estimate SFRs  for GRB hosts when H$\alpha$ or H$\beta$ are not detected. Eq.~\ref{sfroii} gives an SFR which is a factor of 1.39 lower than the same relation proposed by Kennicutt (1998), after converting to the same IMF.

To check the validity of our [OII] SFR conversion, we studied the sample of 18 GRB hosts with simultaneous measurement of  [OII] SFR and H$\beta$ SFR (Figure~\ref{sfr_hb}). The correlation is very good, and is basically indistinguishable from the best-fit linear correlation. The dispersion in this sample is 0.26 dex (a factor of 1.8).

When H$\alpha$ is not measured, we estimated SFR from [OII] using Eq.~\ref{sfroii} (Table~\ref{temi}). SFR from H$\beta$ is also estimated for a subsample of GRB hosts (Table~\ref{temi}). The slit-aperture correction is not necessary when deriving SFR relations, but it is required when estimating the total SFR. This correction is uncertain, because it depends on the galaxy size and how the star-forming HII regions are distributed in the galaxy. 
All SFRs derived from emission lines (Table~\ref{tsfr}) are corrected for aperture-slit loss (given in Table~\ref{temi}).

\subsection{Star formation rates from the UV luminosity}\label{SFRuv}

For 13 GRB hosts, there is no information on emission-line luminosities from HII regions. This is because either the redshift of the GRB is too high, so significant emission lines are shifted to the NIR (where observations are much harder), or lines are intrinsically too weak, or simply the host was never spectroscopically observed. The UV luminosity can be used instead to estimate SFR, as this is a tracer of almost instantaneous conversion of gas into stars. However, we should say that residual UV emission can still be detected even in the absence of star formation, for instance from blue horizontal branch stars.

UV luminosities were used several times in the past to study the SFR history of the universe (Madau, Pozzetti, \& Dickinson 1998; Glazebrook et al.\ 1999; Meurer et al.\ 1999; Erb et al.\ 2006; Moustakas et al.\ 2006). The conversion often used is the one provided by Kennicutt (1998) at 1500 \AA. Moustakas et al.\ (2006) use the $U$-band luminosities at 3600 \AA, as this is less affected by dust.

We derived relations between UV luminosities and SFR more suitable for GRB hosts, using the same empirical approach as for [OII]. We compare SFRs derived from emission lines, with UV luminosities obtained from the best fit of  the observed SED (Figure~\ref{fSED1}). We obtain

\begin{equation}\label{esfr_uv1}
SFR_{1500}  = 1.62\times10^{-40} \frac{L_{1500,corr}}{erg\ s^{-1}\ {\rm \AA}^{-1}}\
M_\odot\ yr^{-1} 
\end{equation}

\begin{equation}\label{esfr_uv2}
SFR_{2800}  =4.33\times10^{-40} \frac{L_{2800,corr}}{erg\ s^{-1}\ {\rm \AA}^{-1}}\
M_\odot\ yr^{-1} 
\end{equation}

\begin{equation}\label{esfr_uv3}
SFR_{3600}  = 5.47\times10^{-40} \frac{L_{3600,corr}}{erg\ s^{-1}\ {\rm \AA}^{-1}}\
M_\odot\ yr^{-1} 
\end{equation}

\noindent
for dust-corrected rest-frame 1500 \AA, 2800 \AA\ and 3600 \AA\ luminosities, respectively. Results for \nemi\ GRB hosts are shown in Figure~\ref{sfr_uv} and given in Table~\ref{tsfr} (for $SFR_{2800}$ only). The dispersions of the points around the three relations are  0.49, 0.41 and 0.40 dex, respectively. This is relatively small, given the large interval spanned by the SFRs (3.5 orders of magnitude) and the uncertainties of the UV--SFR relations (Glazebrook et al.\ 1999). The dispersion found by Erb et al.\ (2006), who derived SFRs using H$\alpha$ and UV at 1500 \AA\ and the Kennicutt (1998) conversion, is 0.34 dex. Moustakas et al.\ (2006) used the UV luminosities  at 3600 \AA, on the subsample of galaxies with $B$ absolute magnitude $M_B<-18.3$. When converted to take into account the different IMF and the monochromatic luminosity, the constant in our Eq.~\ref{esfr_uv3} is a factor of 2.0 larger than theirs. This factor is 1.5 if dwarf galaxies only are considered ($M_B>-18.3$). Meurer et al.\ (1999) estimated an SFR conversion in local starbursts using UV luminosities at 1600 \AA. After correcting for the same IMF, Eq.~\ref{esfr_uv1} gives SFRs which are 2.7 times higher than theirs. These differences can be explained by variations in SFH, dust attenuation, and redshift among the samples (effectively, different sample selections). Eqs.~\ref{esfr_uv1}-\ref{esfr_uv3}, proposed for GRB hosts, can be used to estimate SFR when no emission lines are detected, likely for low SFRs or $z> 1.6$. 

The good agreement between SFRs from emission lines and SFR$_{\rm 3600}$ is surprising as one expects these longer continuum wavelengths to be increasingly affected by older stellar populations (Madau, Pozzetti, Dickinson 1998). The fact that the dispersion is still small indicates again that GRB galaxies are dominated by young stellar populations. We note that 3600 \AA\ is much less affected by dust than 1500 \AA, but likely marginally affected by UV photons from older stars. The best measure of SFR, on balance, might be that derived at 2800 \AA. 

\begin{figure}
%\centerline{\includegraphics[scale = .45]{f11.pdf}}
\centerline{\includegraphics[scale = .45]{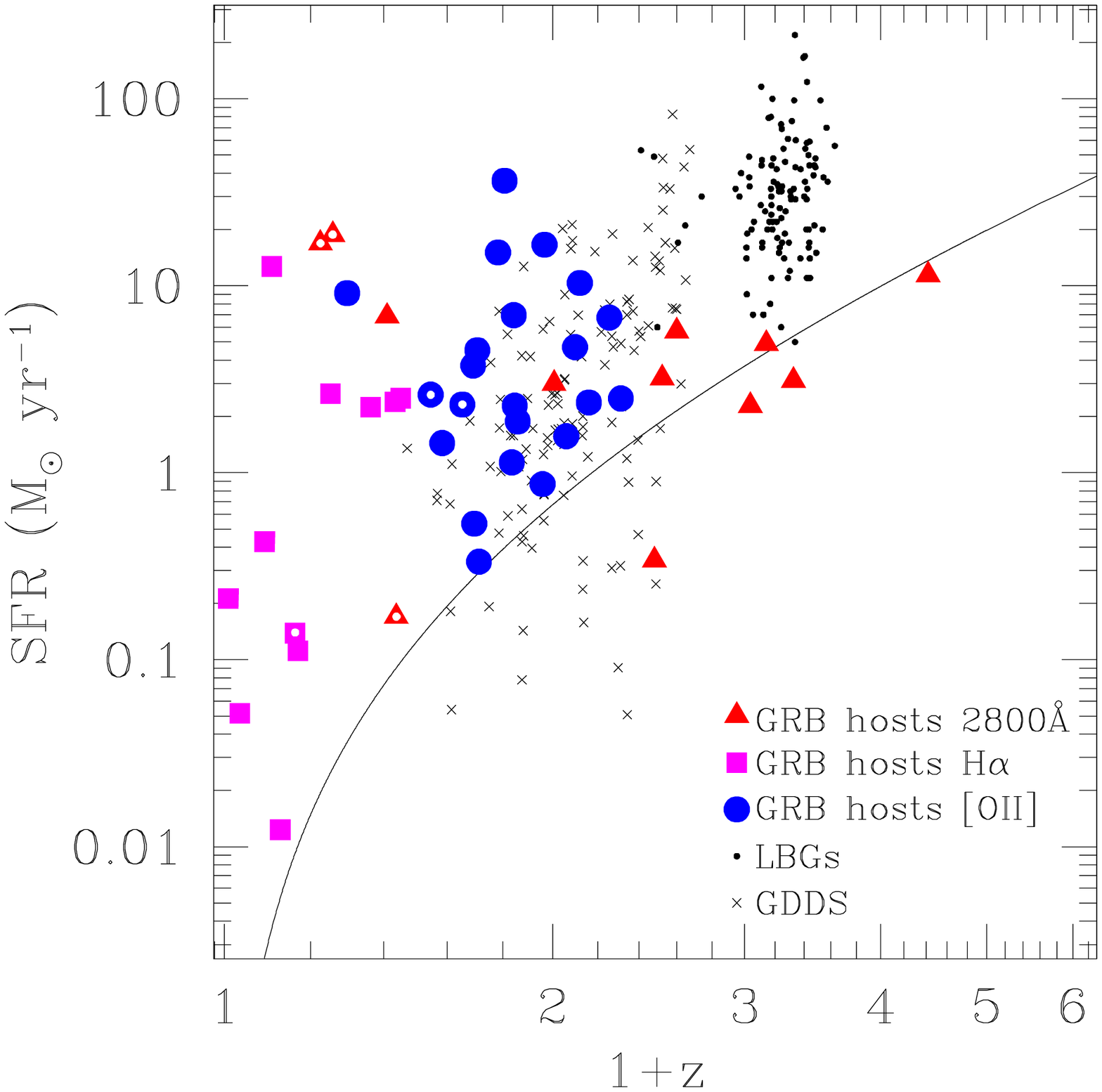}}
\caption {SFRs as a function of redshift for a complete sample of \nhosts\ GRB hosts. The filled squares and circles are derived from H$\alpha$ at $z<0.5$ and [OII] at $0.3<z<1.3$, respectively.  Values are corrected for dust extinction and slit-aperture flux loss. The H$\alpha$ is also corrected for stellar absorption. Filled triangles are SFRs derived from UV luminosity at 2800 \AA.  Dispersions in the SFR relations for emission lines and UV luminosities (Eqs.~\ref{sfroii}-\ref{esfr_uv3}) indicate an uncertainty in these SFRs larger than 2.
Symbols with a white dot mark short-GRB hosts. Crosses  and dots are GDDS galaxies (Juneau et al.\ 2005; Savaglio et al.\ 2005) and  LBGs from Erb et al.\ (2006), respectively. The line represents an H$\alpha$ or [OII] emission flux of $1.3\times10^{-17}$ erg s$^{-1}$ or $0.7\times10^{-17}$ erg s$^{-1}$, respectively, assuming a dust extinction in the visual band $A_V=0.53$, and a slit-aperture correction of  1.5.}
\label{sfr_uv_z}
\end{figure}

In Figure~\ref{sfr_uv_z}, we show SFR$_{2800}$ and SFR derived from emission lines, as a function of redshift. UV luminosity are very useful to estimate SFRs at $z>1.6$, and in general are more sensitive to low SFRs. In the same figure also SFRs from GDDS galaxies and LBGs (Erb et al.\ 2006).

\begin{figure}
%\centerline{\includegraphics[scale = .45]{f12.pdf}}
\centerline{\includegraphics[scale = .45]{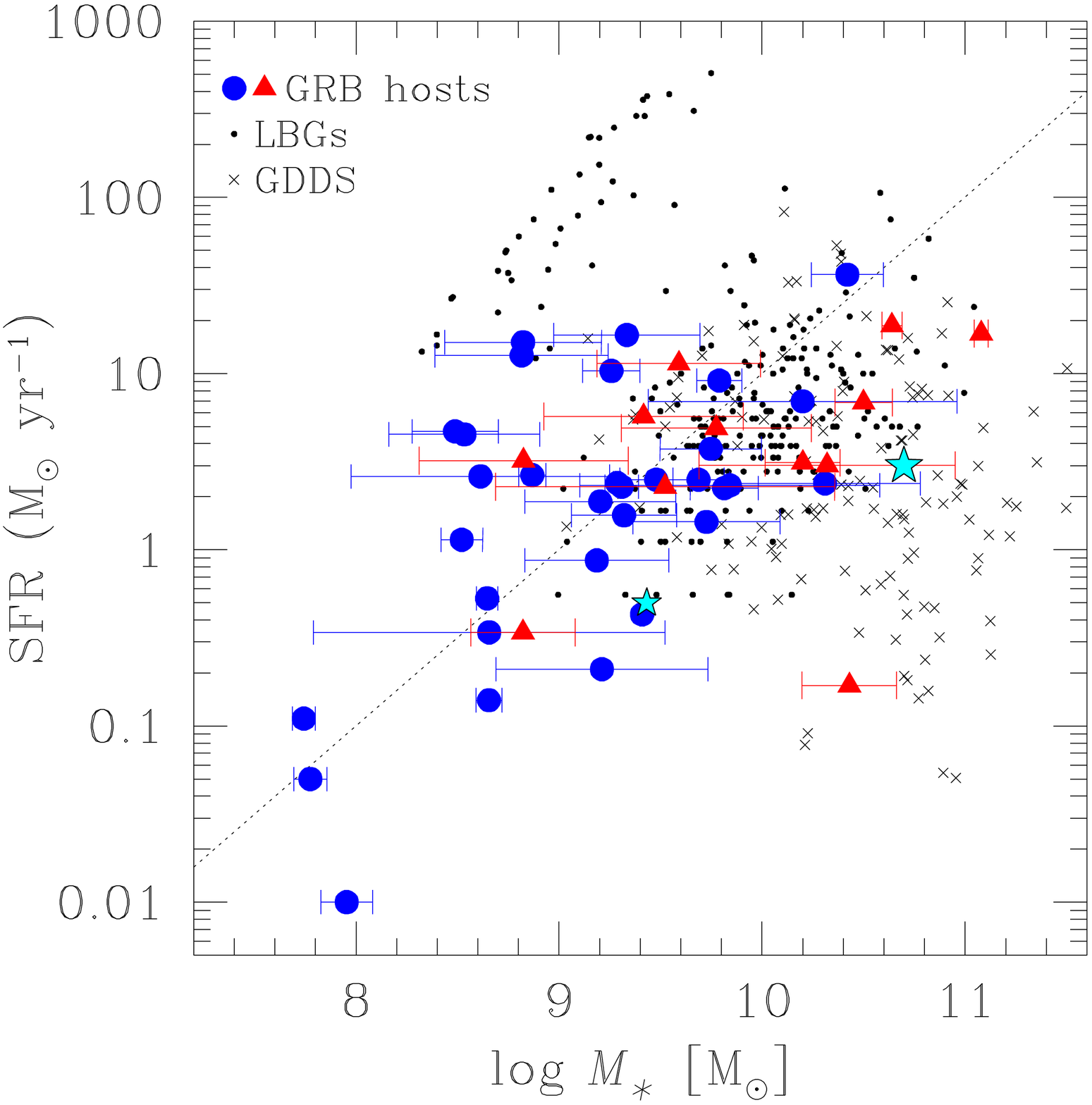}}
\caption {Star formation rate as a function of stellar
mass. The filled circles are GRB hosts with SFRs from H$\alpha$ and [OII] ($0.01<z<1.3$). The filled triangles are SFRs from UV 2800\AA\ luminosities ($0.2<z<6.3$). Only GRB hosts with stellar mass uncertainty $\Delta \log M_\ast<1$ are shown. Crosses are star-forming GDDS galaxies at $0.5<z<1.7$  (Juneau et al.\ 2005; Savaglio et al.\ 2005).  Dots are LBGs at $1.3\lsim z\lsim 3$ (Reddy et al.\ 2006). The large and small stars represent the MW and LMC, respectively. The dotted line marks a constant specific SFR of 1 Gyr$^{-1}$.}
\label{sfr_m}
\end{figure}

An interesting finding is a possible correlation between the SFR and the stellar mass (Figure~\ref{sfr_m}).  This has not been seen before in high-redshift or local galaxy samples. If confirmed, it might be peculiar to the GRB-host sample, and suggest an approximately constant specific SFR (SSFR). Whether or not it has physical significance needs to be determined from more extensive data.

\begin{figure*}
%\centerline{\includegraphics[scale = .6]{f13.pdf}}
\centerline{\includegraphics[scale = .6]{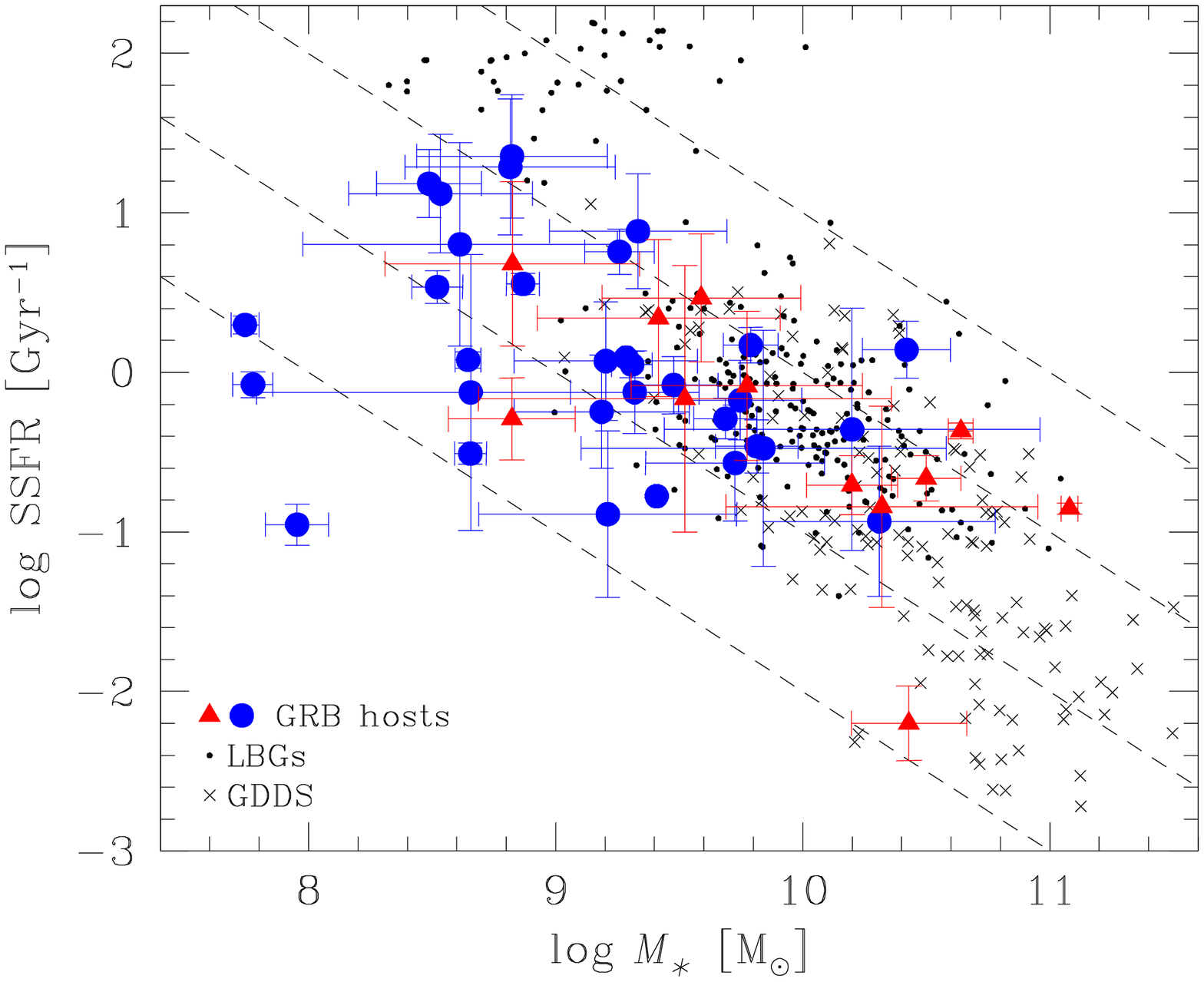}}
\caption {SSFR as a function of stellar
mass. The filled circles and triangles are GRB hosts with SFRs measured from emission lines ($0<z<1.3$) and UV luminosities  ($0.2<z<6.3$), respectively. Only GRB hosts with stellar mass uncertainties $\Delta \log M_\ast<1$ are shown. Crosses are star-forming GDDS galaxies at $0.5<z<1.7$  (Juneau et al.\ 2005; Savaglio et al.\ 2005).  Dots are LBGs at $1.3\lsim z\lsim 3$ (Reddy et al.\ 2006). The dashed lines, from left to right, mark constant SFRs of 0.1, 1, 10, and 100 M$_\odot$ yr$^{-1}$.}
\label{ssfrm}
\end{figure*}

\begin{figure}
%\centerline{\includegraphics[scale = .45]{f14.pdf}}
\centerline{\includegraphics[scale = .45]{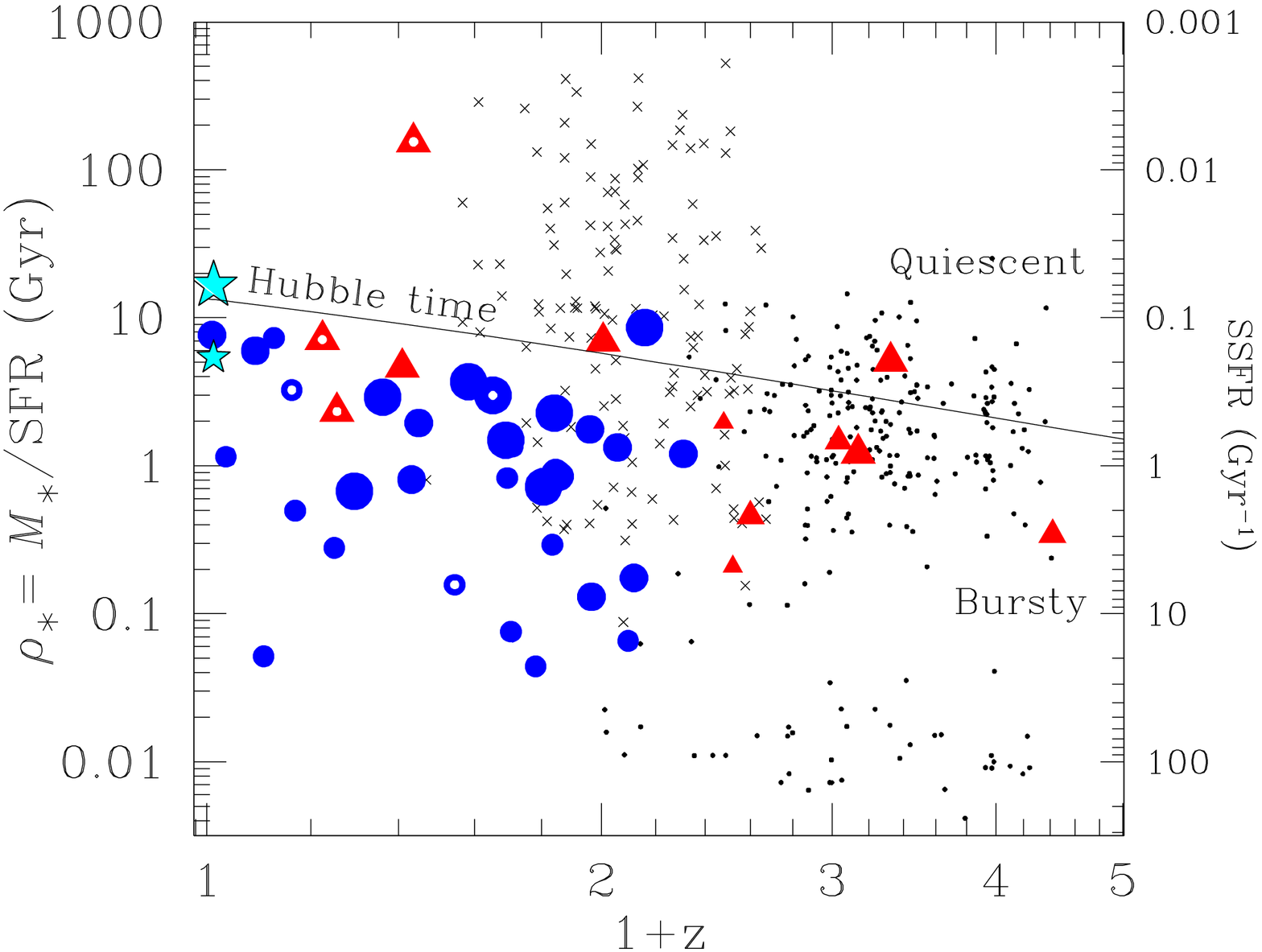}}
\caption {Growth timescale $\rho_\ast=M_\ast/SFR$ (left y-axis) or specific star
formation rate SSFR (right y-axis) as a function of redshift.
Filled circles and triangles are GRB hosts with SFRs measured from emission lines and UV luminosities, respectively. Only hosts with stellar mass uncertainties $\Delta \log M_\ast<1$ are shown. Small, medium and large symbols are hosts with
 $M_\ast \leq 10^{9.0}$ M$_\odot$, $10^{9.0}$ M$_\odot < M_\ast \leq 10^{9.7}$
M$_\odot$, and $M_\ast> 10^{9.7}$ M$_\odot$, respectively. Hosts with small white dots are associated with short GRBs. The curve is the Hubble time as a function of redshift, and indicates the transition from bursty to quiescent mode for galaxies. Crosses are GDDS galaxies at $0.5<z<1.7$ (Juneau et al.\ 2005; Savaglio et al.\ 2005). Dots are LBGs at $1.3\lsim z\lsim 3$, for which SSFRs are derived by assuming an exponential decline for star formation  (Reddy et al.\ 2006). The big and small stars at zero redshift represent the growth timescale for the Milky Way and the Large Magellanic Cloud, respectively.}
\label{ssfr}
\end{figure}

\subsection{Specific Star Formation Rates}

The SFR by itself does not tell how active the galaxy is. Another parameter to consider is the SSFR, that is the SFR normalized to the total stellar mass of the galaxy . The SSFR has been studied in galaxies  at low redshift (P\'erez-Gonz\'alez et al.\ 2003; Brinchmann et al.\ 2004) and high redshift (Juneau et al.\ 2005; Bauer et al.\ 2005; Noeske et al.\ 2007). 

In Figure~\ref{ssfrm} we show the SSFR as a function of stellar mass, investigated for the first time for a large sample of GRB hosts. These are compared with SSFRs of star-forming galaxies of the GDDS in the redshift interval $0.4<z<1.7$  (Juneau et al.\ 2005; Savaglio et al.\ 2005) and LBGs at $1.3\lsim z\lsim 3$ (Reddy et al.\ 2006). Field galaxies of the AEGIS at $0.70<z<0.85$ cover a similar region  in the SSFR--$M_\ast$ plot as the GDDS (Noeske et al.\ 2007). The SSFRs in GRB hosts tend to show properties different from field galaxies in the GDDS, with lower masses and higher SSFRs. The median SSFR for the GRB hosts is 0.8 Gyr$^{-1}$, similar to LBGs, but the mean stellar mass is six times lower, $10^{9.3}$ M$_\odot$ for the former and $10^{10.1}$ M$_\odot$ for the latter. 

The inverse of the SSFR is called the growth timescale $\rho_\ast$. It defines the time required by the galaxy to form its observed stellar
mass, assuming that the measured SFR has been constant over the entire galaxy history. In Figure~\ref{ssfr} we show $\rho_\ast$ (left $y$-axis) or SSFR (right $y$-axis) as a function of redshift. The symbol size is bigger for larger
stellar masses. For all but one GRB host with detected emission lines, $\rho_\ast$ is smaller than the age of the universe (Hubble time) at the observed redshift. This is true for four more GRB hosts, with SFR estimated from UV luminosities. For the total sample of 46 hosts, the median value is $\rho_\ast=1.3$ Gyr, and for 2/3 $\rho_\ast<2$ Gyr. It is also
apparent that the growth time-scale is longer for more-massive galaxies. This is different than what has been observed in other  galaxy samples, which have a large fraction of quiescent galaxies ($\rho_\ast$ larger than the Hubble time). When $\rho_\ast$ is $<1$ Gyr, true for
about 2/5 of the GRB-host sample, the galaxy is in a  ``bursty mode". For instance, a galaxy with a stellar mass of a few times $10^9$ M$_\odot$ (the stellar mass of the LMC) and SFR = 5 M$_\odot$ yr$^{-1}$ (10 times larger than LMC) is a starburst, with $\rho_\ast\sim 500$ Myr. 

\section{Metallicities}\label{metallicities}

Metallicity in HII regions is typically measured using detected emission lines.  At redshift $z>0.2$, the spatial resolution of the data is generally too low to derive metallicity in small regions, so what is measured from integrated fluxes is an optical luminosity-weighted mean value in the galaxy. This is the case for most GRB hosts analyzed here, as 85\% of the spectroscopy sample is at $z>0.2$. For low-$z$ hosts, like that of GRB~060505 at $z=0.089$ studied in great detail by Th{\"o}ne et al.\ (2008), we use integrated fluxes over the entire galaxy to treat it as the rest of the sample.

Metallicities can be derived from different emission-line sets, according to the spectral coverage, redshift, and galaxy properties. The numerous methods, developed in the last few years, have complicated the effort of having a tool for a realistic estimate. Each method is affected by systematic errors that are not easy to determine (for a review, see Kewley \& Ellison 2008). 

A direct estimate is possible through  a measurement of the electron temperature $T_e$ (Izotov et al.\ 2006), possible when several lines of the same element with very different ionization levels are detected (Bresolin 2007). For instance, these could be the generally weak auroral line [OIII]$\lambda4363$ and the nebular lines [OIII]$\lambda\lambda4959,5007$. For its definition, the $T_e$ method is sensitive to low-metallicity systems. As this is based on the generally weak [OIII]$\lambda4363$ line, it is very hard to measure in small and distant galaxies.

Other calibrators are easier to measure, but give more systematically uncertain results. The O3N2 calibrator, which uses a combination of [OIII]$\lambda5007$, [NII]$\lambda6583$, H$\beta$, and H$\alpha$ lines (Alloin et al. 1979), needs NIR spectroscopy when exceeding redshift $z=0.5$. 

For higher redshift and relatively low spectral sensitivity, the very popular $R_{23}$ calibration (Pagel et al.\ 1979) and its refinement, the $P$-method (Pilyugin 2000), both requiring H$\beta$, [OII], and [OIII] fluxes, can be used up to redshift $z\simeq1$. The $R_{23}$ calibration is notoriously problematic. First, it gives two solutions (lower and upper branch), which are not easy to disentangle. Moreover,
the upper branch solution, when applied to integrated fluxes, is affected by systematic errors which could be as large as a factor of 3 (Kewley \& Dopita 2002; Kennicutt et al.\ 2003;  Stasi{\'n}ska 2005; Bresolin 2006). A value of $\log ({\rm NII]/[OII]}) <-1.2$ can point to the lower branch solution (Kewley \& Dopita 2002).  Despite the difficulties, the $R_{23}$ calibrator is an important resource because it is easier to measure than other methods.

Our choice for the final metallicity is done as follows: the $T_e$ metallicity is preferred, measurable for four GRB hosts with detected [OIII]$\lambda4363$. When  this is not detected, we give priority to the O3N2 metallicity,  available for five GRB hosts, and use the prescription given by Pettini \& Pagel (2004). 

For all other GRB hosts, we used the $R_{23}$ calibrator, available for a total of 18 hosts. Recently, Kewley \& Ellison (2008; hereafter KE08) derived conversion relations between different methods. Following their prescriptions, we used the $R_{23}$ calibrators given by  Kewley \& Dopita (2002) and Kobulnicky \& Kewley (2004), for the lower and upper branch solutions, respectively. We converted $R_{23}$ metallicities into O3N2 metallicities using the relations proposed by KE08. These are tested for metallicities in the ranges $8.05 < \log 12+\log (\rm O/H) < 8.3$ and $8.2 < \log 12+\log (\rm O/H) < 8.9$, for the  lower and upper branch solution, respectively. For metallicities outside these ranges, we use the recently tested $R_{23}$--metallicity relation proposed by Nagao et al.\ (2006, hereafter N06).

The hosts for which we could measure metallicity are all at $z<1$. For $z>1$, no NIR spectra of GRB hosts are good enough to measure metallicity. In the following subsections, we list measurements for different methods. 
%Results are summarized in Table~\ref{tr23}.

\begin{figure}
%\centerline{\includegraphics[scale = .42]{f15.pdf}}
\centerline{\includegraphics[scale = .42]{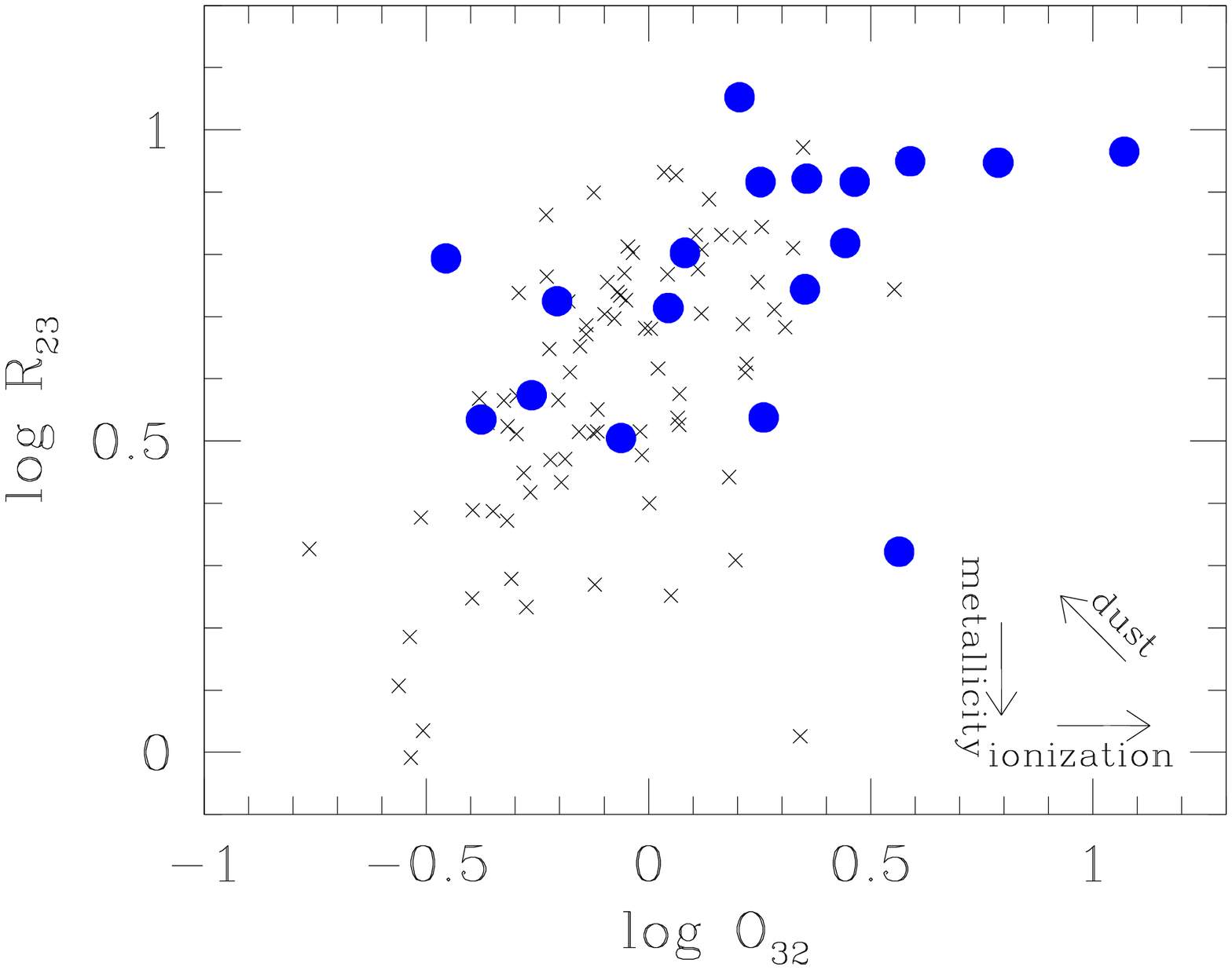}}
\caption {$R_{23}$ vs.\ $O_{32}$, not corrected for dust extinction. Filled circles are GRB hosts. Crosses are $0.4<z<1$ field galaxies from GDDS (Savaglio et al.\ 2005) and CFRS (Lilly et al.\ 2003). The large fraction of GRB hosts with $R_{23} >2$ with respect to other star-forming galaxies indicates lower metallicity and/or higher dust extinction. The arrows indicate the effects of higher metallicity, dust extinction and ionization.}
\label{r23}
\end{figure}

\subsection{The $T_e$ metallicities}

The temperature sensitive [OIII]$\lambda$4363/[OIII]$\lambda5007$ line
ratio has been detected in four hosts (Table~\ref{tte}). These are
the hosts of GRB~980425, GRB~020903, GRB~031203, and GRB~060218 (Hammer
et al.\ 2006; Prochaska et al.\ 2004; Wiersema et al.\ 2007). Though according to Wiersema et al.\ (2007) 
the [OIII]$\lambda$4363 line in GRB~060218 is 5$\sigma$ significant, the line in the spectrum is somehow doubtful, 
because located in a noisy part of the spectrum. We will consider
this detection with caution. We recalculated the oxygen abundance using the prescriptions
provided by Izotov et al.\ (2006), which are not very sensitive to the electron density, provided that this is smaller than $10^3$ cm$^{-3}$. Metallicities derived with this method are low (Table~\ref{tte}).

Temperatures in the HII regions are slightly larger than $10^4$ K (Table~\ref{tte}), with the exception of the host of GRB~060218 (Wiersema et al.\ 2007), for which the temperature is twice larger. However, as already said, the detection of O$\lambda4363$ in this host is weak.

\subsection{O3N2 metallicities}

The O3N2 method, originally proposed by Alloin et al.\ (1979), was updated by Pettini \& Pagel (2004). In the new formulation, the metallicity is given by

\begin{equation}
12+\log (\rm O/H) = 8.73-0.32\times O3N2,
\end{equation}

\noindent
where $\rm O3N2 = \log \{([OIII]5007 / H\beta)/([NII]6583 / H\alpha)\}$. This method is useful only for $-1<{\rm O3N2}<1.9$, and much more uncertain for ${\rm O3N2}\gsim2$.

This metallicity can be measured in five GRB hosts, and in two additional objects an upper limit is set
(Table~\ref{to3n2}). Derived values are low, with a mean metallicity of 1/3 solar.

\subsection{$R_{23}$ metallicities}

The $R_{23}$ calibrator uses the following combination of [OII], [OIII] and H$\beta$ lines:

\begin{equation}\label{r23}
\log R_{23} = \log \left(\frac{\rm [OII]\lambda3727+[OIII]\lambda\lambda4959,5007}{\rm H\beta}\right) ,
\end{equation}

\noindent
and the $O_{32}$ parameter: 

\begin{equation}\label{o32}
\log O_{32} = \log \left(\frac{\rm [OIII]\lambda\lambda4959,5007}{\rm [OII]\lambda3727}\right) ,
\end{equation}

\noindent
which takes into account different ionization levels of the gas. In Figure~\ref{r23} we show $R_{23}$ and $O_{32}$ for GRB hosts, and comparison with galaxies of the GDDS and Canada France Redshift Survey (CFRS; Lilly et al.\ 2003) at $0.4<z<1$, before dust-extinction correction (the dust extinction in GDDS and CFRS galaxies is not measured). GRB hosts tend to have higher $R_{23}$ values, which could be due to low metallicity or/and lower dust extinction.

To measure $R_{23}$ metallicities, we followed the prescriptions given by KE08. Namely, we used Kewley \& Dopita (2002) and Kobulnicky \& Kewley (2004) for the lower and upper branch solutions, respectively. Then we used the relations given by KE08 to convert to O3N2 metallicities. These relations are valid for $8.05 < 12+\log (\rm O/H) < 8.3$ and $8.2 < 12+\log (\rm O/H) < 8.9$ only (for the input metallicity). See \S~\ref{nagao} for more $R_{23}$ metallicities.

The $R_{23}$ and $O_{32}$ values used are corrected for dust extinction, and are reported in Table~\ref{tr23}, together with the derived metallicities. 

\subsection{More $R_{23}$ metallicities}\label{nagao}

For metallicities outside the intervals $8.05 < 12+\log (\rm O/H) < 8.3$ and $8.2 < 12+\log (\rm O/H) < 8.9$ we used the $R_{23}$--metallicity relation provided by N06. This is described by a best-fit polynomial, whose coefficients are in Table 6 of N06. Six GRB hosts have $R_{23}$ in the turnover region, for which $12+\log (\rm O/H) \sim 8.1$ (Table~\ref{tr23}). For the other GRB hosts, in the case of the lower branch part of the relation, we estimated the correction from N06 to KE08, by comparing results for those hosts with metallicity calculated with both calibrations. The conversion relation from N06 to KE08 is given by

\begin{equation}\label{enagao}
12+\log (\rm O/H)_{KE08} =  0.8885\times[12+\log (\rm O/H)_{N06}] +1.177 ,
\end{equation}
 
\noindent 
valid in $\log 12+\log (\rm O/H)=7.2 - 8.1$ for the N06 metallicity. 
Results for the lower branch metallicities, after applying Eq.~\ref{enagao}, are given in the sixth column of Table~\ref{tr23}.

The upper branch solution of N06 is not used, as this does not add anything to our analysis. 

\subsection{The electron density}

The [OII] doublet at $\lambda\lambda=3726,3729$ \AA\ is sensitive to the electron density (Osterbrock \& Ferland 2006). A low density was measured in the host of  GRB~060218 (Wiersema et al.\ 2007). Thanks to the use of the high-resolution spectroscopy, [OII]$\lambda\lambda2726,3729$ is resolved in 6 GRB hosts. Assuming that the typical temperature in the star-forming regions is around $10^4$ K (Table~\ref{tte}), we derive the electron density using the definition of Osterbrock \& Ferland (2006). Results are in the range  $n_e \sim5 \times 10^2$ cm$^{-3}$ to $13\times 10^2$ cm$^{-3}$ (Table~\ref{tne}).

\begin{figure*}
%\centerline{\includegraphics[scale = .65]{f16.pdf}}
\centerline{\includegraphics[scale = .65]{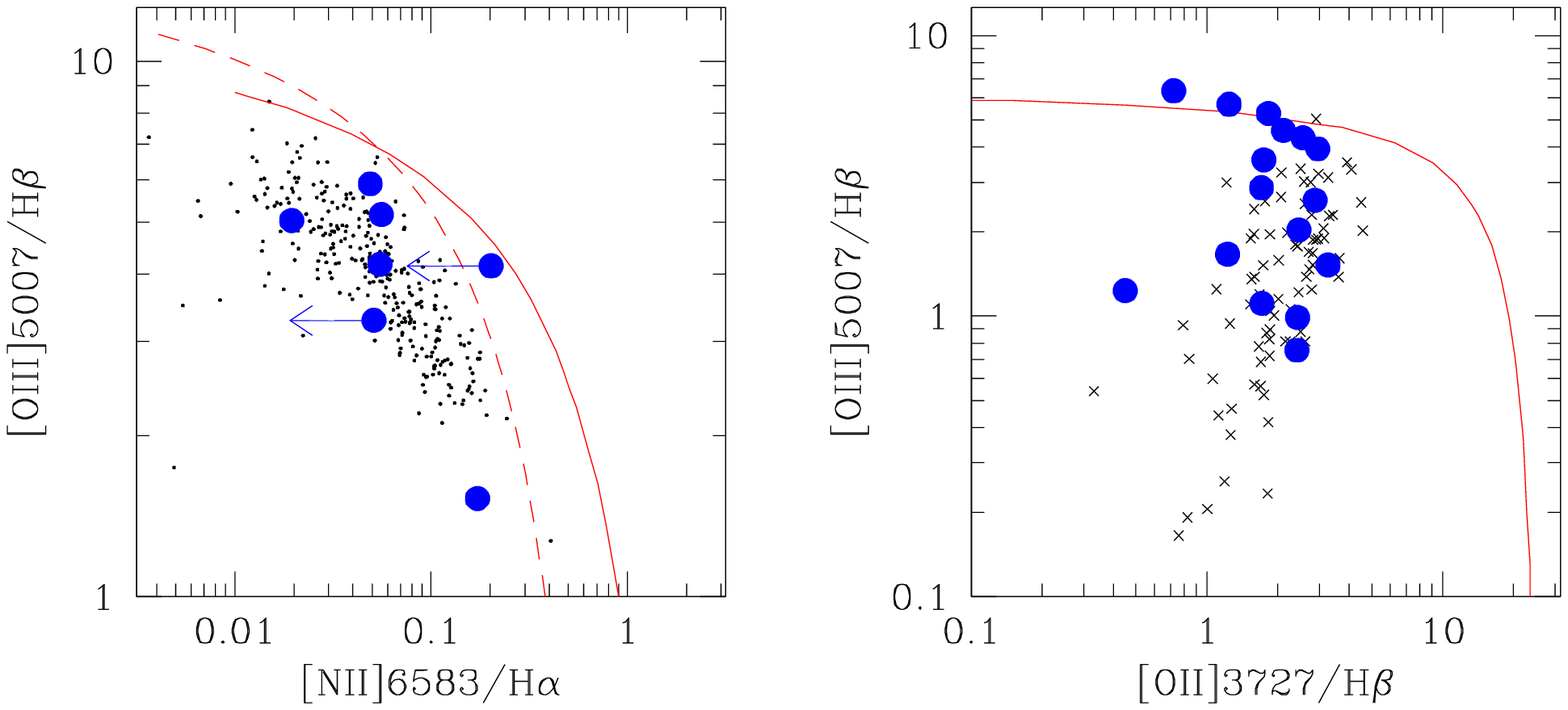}}
\caption {AGN diagnostic diagrams: [OIII]/H$\beta$ vs.\ NII]/H$\alpha$ (left plot) and vs.\ [OII]$\lambda3737$/H$\beta$ (right plot). Filled circles are GRB hosts. Dots on the left and crosses on the right are metal poor galaxies in the local universe
(Izotov et al.\ 2006), and $0.4<z<1$ star forming galaxies (Savaglio et al.\ 2005), respectively. 
Solid and dashed lines in the left plot mark the AGN selection criteria by Kewley et al.\ (2001) and  Kauffmann et al.\ (2003), respectively. Solid line in the right plot is the AGN selection criterion proposed by Lamareille et al.\
(2004).}
\label{agn}
\end{figure*}

\subsection{AGN contamination}

Figure~\ref{agn} shows the location of GRB hosts, in the diagrams that use line ratios to distinguish galaxies with bright HII regions from AGN dominated galaxies. For comparison, we also show GDDS and CFRS star-forming galaxies at high redshift, and local metal-poor galaxies (Izotov et al.\ 2006).

The [OIII]/H$\beta$ versus [NII]/H$\alpha$ relations proposed by Kewley et al.\ (2001) and Kauffmann et al.\ (2003) are relatively robust diagnostics and generally used in galaxy surveys. The [OIII]/H$\beta$ versus [OII]/H$\beta$ relation is more uncertain (Lamareille et al.\ 2004). However, this is easier to measure at $z>0.5$, when NIR spectra are not available. 

From Figure~\ref{agn}, we conclude that the AGN contamination in GRB hosts is likely not significant. The [OIII]/H$\beta$ versus [NII]/H$\alpha$ plot indicates that GRB hosts behave like local metal-poor galaxies, i.e.\ with high values of [OIII]/H$\beta$. In fact, the majority of galaxies in the local universe from the SDSS have [OIII]/H$\beta<1$ (Stasi{\'n}ska et al.\ 2006).

\begin{figure*}
%\centerline{\includegraphics[scale = .42]{f17a.pdf}\includegraphics[scale = .42]{f17b.pdf}}
\centerline{\includegraphics[scale = .42]{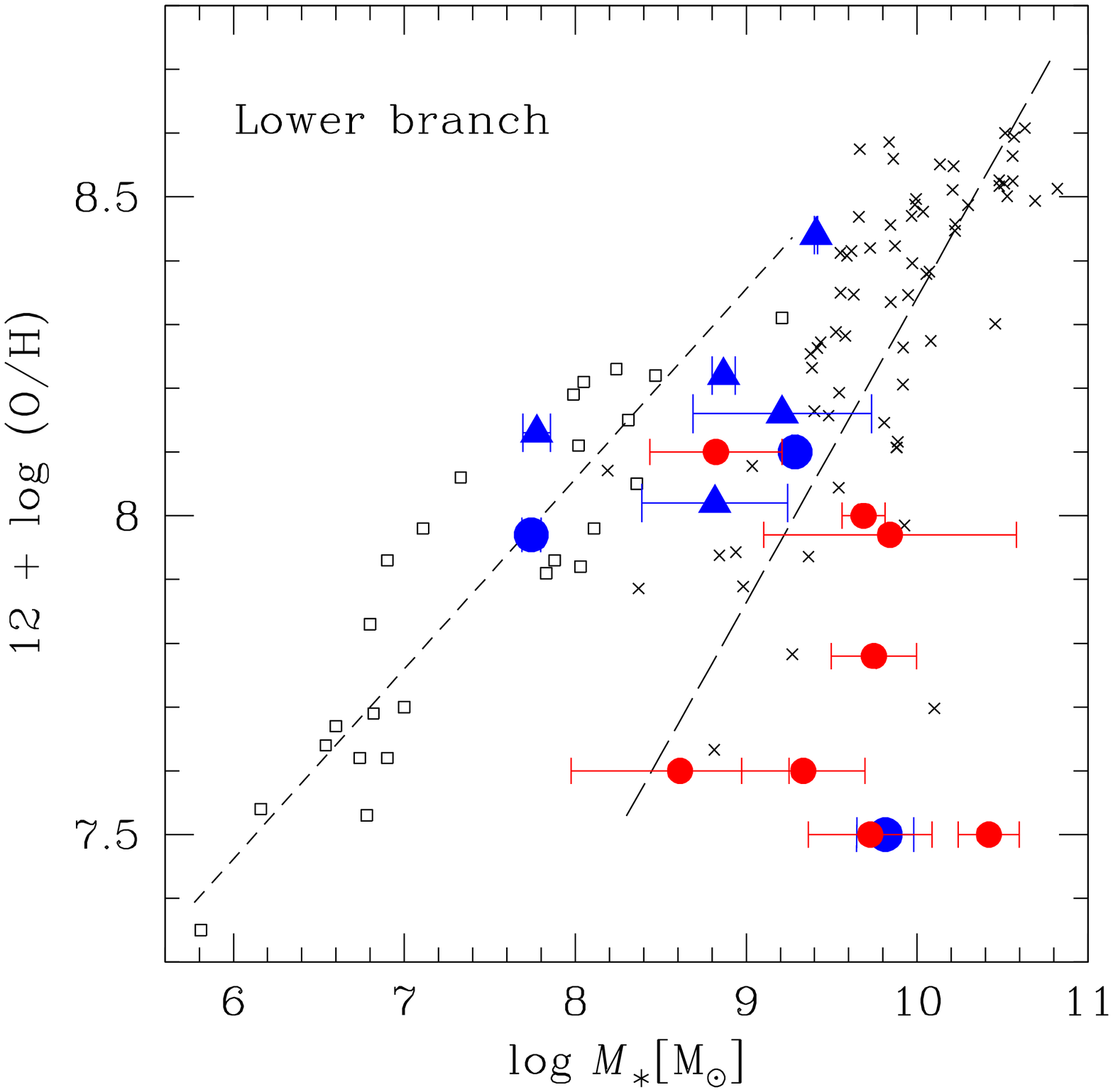}\includegraphics[scale = .42]{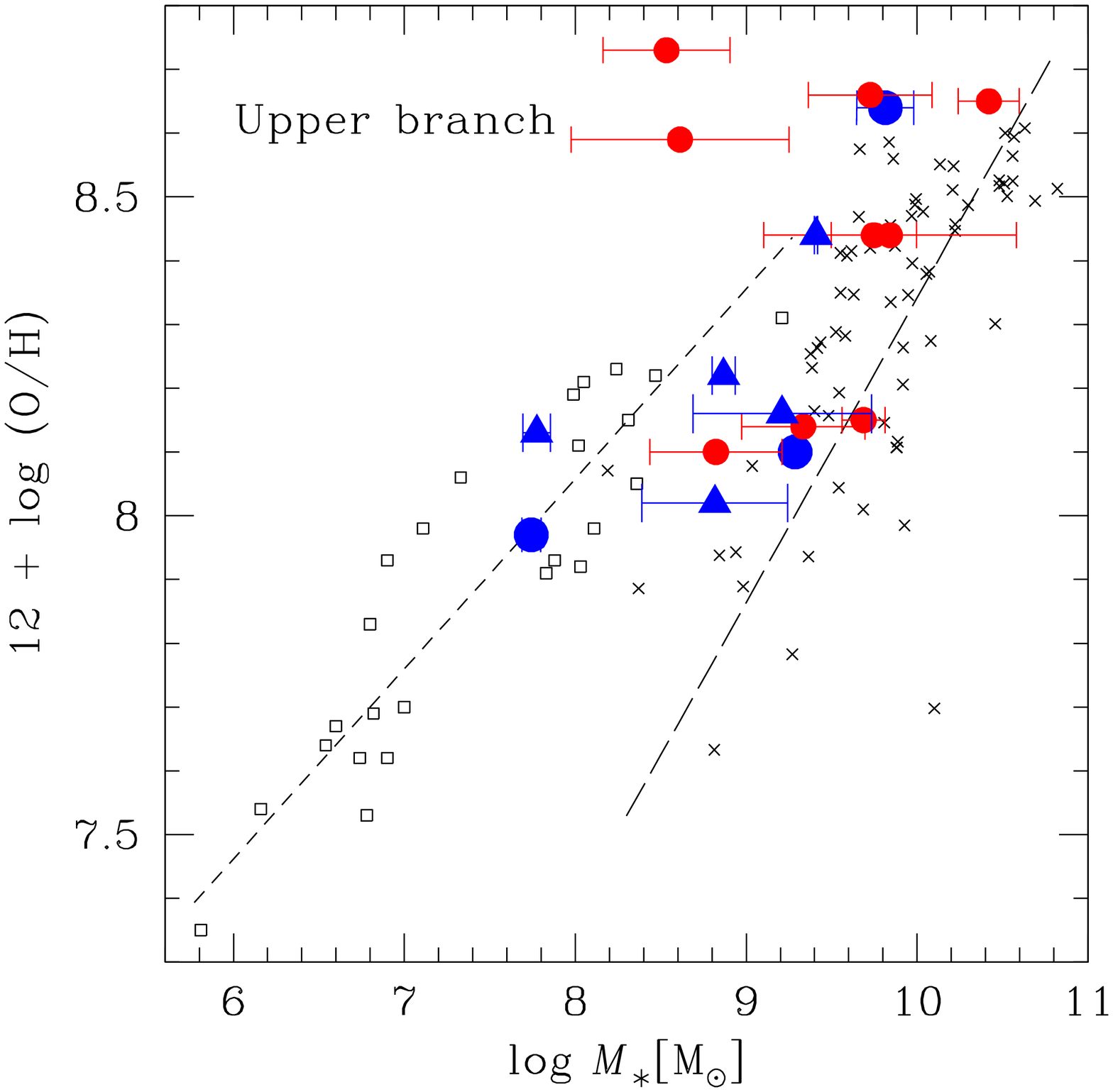}}
\caption {Mass-metallicity distribution for different galaxy samples. Filled circles in the left and right side plots are GRB host metallicities determined with the $R_{23}$ calibration, when choosing the lower and upper branch solutions, respectively. Filled triangles are GRB host metallicities measured with the $T_e$ and O3N2 method. Small filled circles are GRB hosts at $z\geq0.45$. Large circles and triangles are GRB hosts at $z<0.45$.  Open squares are dwarf star forming galaxies at low redshift, whose metallicities were determined with the $T_e$ method  (Lee et al.\ 2006). Crosses are star forming galaxies at $z\sim0.7$ from the GDDS (Savaglio et al.\ 2005). Here the $R_{23}$ metallicities were converted to O3N2 metallicities assuming a constant shift of $-0.5$ dex (see Figure 2 of Kewley \& Ellison 2008). Short-dashed and long-dashed lines are the linear correlations for the local and the $z\sim0.7$ galaxy samples, respectively.}
\label{MZ}
\end{figure*}

\section{Discussion}\label{discussion}

The main goal of this work is to characterize  the galaxy population hosting GRBs. In the past, several authors conducted studies over a small sample of GRB hosts and concluded that this population is not representative of the bulk of all galaxies as a function of cosmic time (Fruchter et al.\ 1999; Le Floc'h et al.\ 2003; Christensen et al.\ 2004; Berger et al.\ 2003, Tanvir et al.\ 2004; Fruchter et al.\ 2006; Le Floc'h et al.\ 2006). In fact, they are on average  bluer, younger, and fainter than the field galaxy population. However, we consider that faint, blue, and young galaxies are by far the most common galaxies that existed in the past. Galaxies can only get bigger over time and not smaller, through merging  and star formation processes. Nevertheless, they are the most elusive to find in the distant universe. Through a GRB event, our capabilities of digging deeper in terms of galaxy mass and distance is considerably increased.

Fruchter et al.\ (2006) showed that long GRBs happen much more in the brightest regions of their host galaxies than type II SNe, suggesting that GRBs are associated with different galaxies. This result is not confirmed by Kelly et al.\ (2008), who considered SN~Ic hosts and found that SN~Ic occur in an environment similar to that of GRBs. Indeed,  some nearby GRBs are associated with SN~Ic, a subclass of core-collpase supernovae  (Galama et al.\ 1998;  Stanek et al.\ 2003;  Hjorth et al.\ 2003; Modjaz et al.\ 2006). 

The connection between GRB events and the nature of their hosts is advocated by Modjaz et al.\ (2008), who found that five GRB hosts are significantly more metal poor than sample of SN~Ic hosts in a similar luminosity range in the $B$ band. 
However, SN discoveries are observationally biased toward luminous (and hence on average more metal rich) galaxies.  The result of Modjaz et al.\ (2008) is affected by the small number statistics. Prieto et al.\ (2008) found in a statistically significant sample a much larger metallicity spread, and a significant mass--metallicity relation which would include GRB hosts.
Moreover, one should consider that the $B$-band absolute luminosity in galaxies is weakly correlated with mass.  The $K$-band luminosity would instead better represent the galaxy stellar mass (Figure~\ref{mk}), and this would better support the special nature of GRB hosts. Wolf \& Podsiadlowski (2007) quantitatively showed that any GRB rate dependence on galaxy mass (or ultimately metallicity) is at best small. 

To better understand the impact of GRB hosts on our knowledge of galaxy formation and evolution, we have analyzed the properties of the largest possible sample, \nhosts\ GRB hosts, using multiband photometry and optical spectroscopy. Relevant parameters are summarized in Table~\ref{tsummary}. The median redshift, $K$-band absolute magnitude, stellar mass, dust extinction, metallicity, SFR, and SSFR are reported at the bottom of the table.

The ultimate goal is to understand whether  GRB hosts are special galaxies, related to the occurrence of a GRB event, or just normal galaxies, but small and metal poor because these are the most common galaxies in the universe. Due to the large redshift interval spanned by the sample, $0<z<6.3$, different populations likely coexist. The one at low redshift is different from that at high redshift because of the intrinsic evolution of galaxies with time, and because of the redshift dependence of the selection effects.

From the colors of the sample, we confirm that GRB hosts tend to be blue galaxies. The best-fit SED can constrain the fraction of stars involved in the burst phase, which in our sample is generally less than 10\% of the total mass. A reliable estimate of the $K$-band absolute magnitudes is provided. Due to observational limits, at high redshift only bright galaxies are detected. GRB hosts below $z=0.4$ have $-24<M_K<-16$. Most galaxies (83\%) have stellar masses in the range $10^{8.5}$ M$_\odot$ to $10^{10.3}$ M$_\odot$. The stellar mass completeness limit of the typical present-day high-redshift galaxy survey is higher than $10^{10.3}$ M$_\odot$. For $M_\ast<10^{9.2}$ M$_\odot$, the GRB-host sample is highly incomplete. For the six short-GRB hosts the median stellar mass is $M_\ast \simeq 10^{10.1}$ M$_\odot$, indicating that the association of short GRBs with elliptical more-massive galaxies needs a larger sample to be proven. We provide a relation between stellar mass and absolute magnitude $K$ (Eq.~\ref{emk_m}), which is based on the finding of a nearly constant stellar mass-to-light ratio in the $K$ band,  of the order of $M_\ast/L_K=0.1$ $(M/L_K)_\odot$.

The SFRs derived from emission lines, after dust-extinction, aperture-slit loss, and stellar Balmer absorption correction, are the best approximation to the true total SFR.  When emission lines are not measured, we propose an SFR relation which uses the UV luminosity at 2800 \AA. SFRs in the total sample span three and a half  orders of magnitudes (SFR $= 0.01-36$ M$_\odot$ yr$^{-1}$). Two out of six short-GRB hosts have high SFRs, more than 10 M$_\odot$ yr$^{-1}$. The SFR normalized by the stellar mass SSFR, or its inverse, the growth timescale $\rho_\ast$ indicate that GRB hosts are active and young systems, which is expected given the generally low mass of the hosts, the high redshifts, and the fact that redshifts are often found trough the detection of emission lines.

\begin{figure*}
%\centerline{\includegraphics[scale = .45]{f18a.pdf}\includegraphics[scale = .45]{f18b.pdf}}
\centerline{\includegraphics[scale = .45]{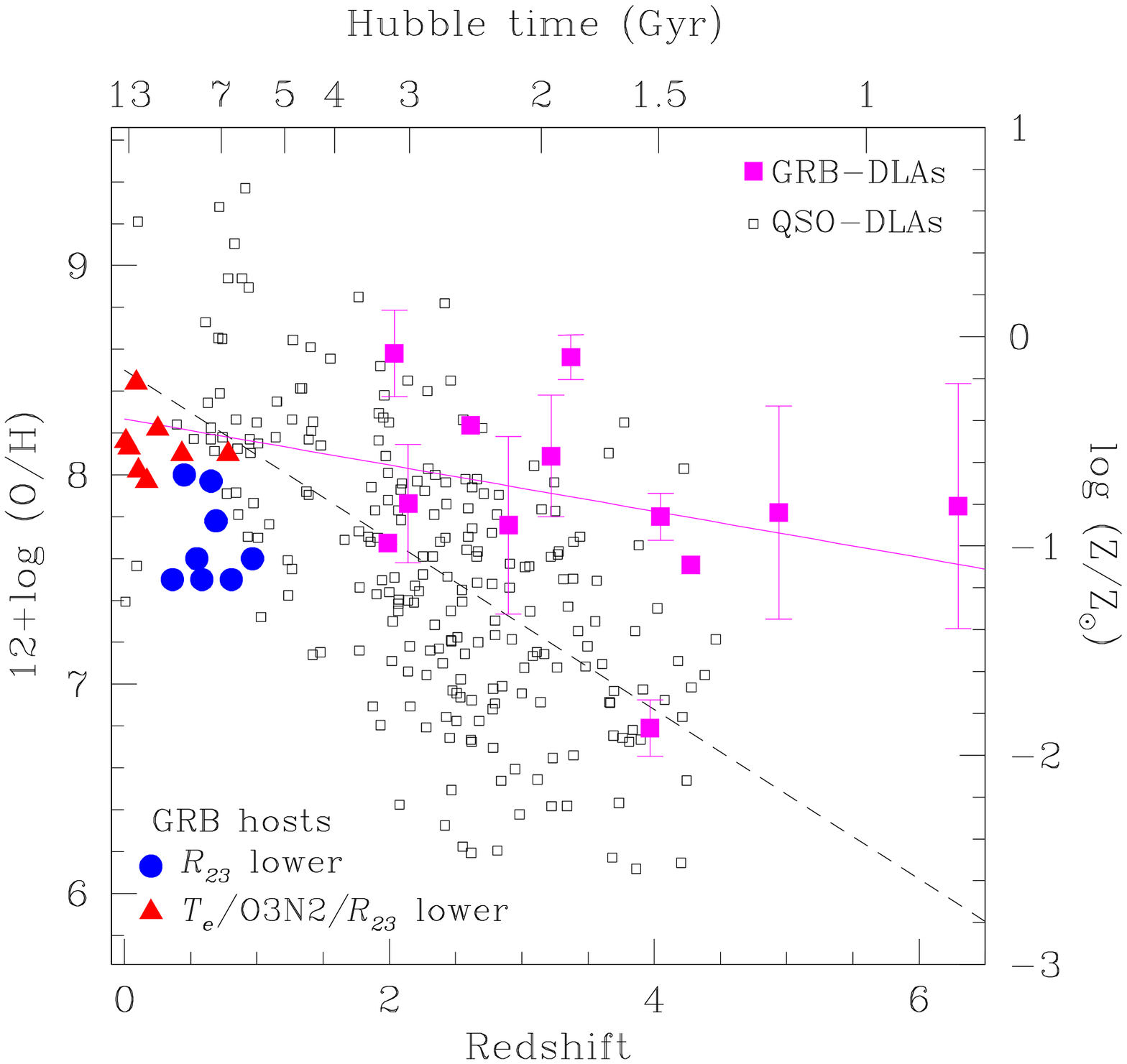}\includegraphics[scale = .45]{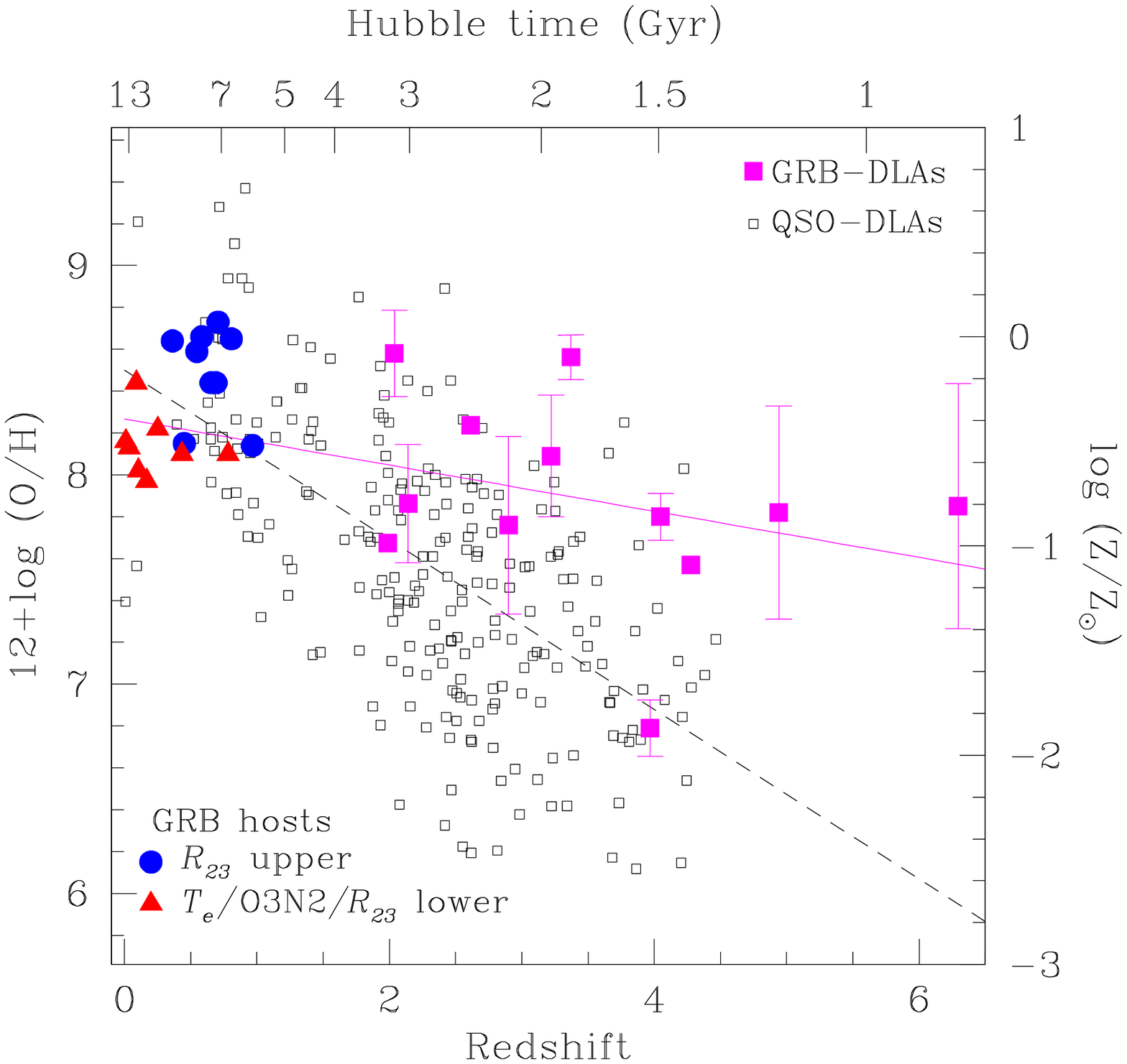}}
\caption {Metallicity as a function of redshift (lower x-axis) or Hubble time (upper x-axis). Filled circles in the left and right side plots are GRB host metallicities determined with the $R_{23}$ calibration, when choosing the lower and upper branch solutions, respectively. Filled triangles are GRB hosts with the $T_e$ and O3N2 metallicities. Filled stars are GRB-DLA metallicities, derived from the absorption lines detected in the afterglow spectra. Open squares are DLA metallicities measured in QSO spectra. Solid and dashed lines are the linear correlation for GRB-DLAs and QSO-DLAs, respectively (see Savaglio 2006).}
\label{zZ}
\end{figure*}

Berger et al.\ (2003) and Le Floc'h et al.\ ( 2006) have claimed much higher SFRs, based on radio and mid-IR detections. The host of GRB~980613 looks involved into a merger of several subcomponents (Hjorth et al. 2002; Djorgovski et al. 2003). Our SFR estimate refers to component A that hosted the GRB. The high SFR derived by Le Floc'h et al.\ (2006) refers to component D, which is more than 30 kpc away from component A. 
Regarding GRB~980703 and GRB~000418, Le Floc'h et al. (2006) commented that AGN contamination would explain the high radio emission and no Spitzer detection. However, the existence of two GRB hosts with AGN on a relatively small sample observed in the radio is hard to explain (Micha{\l}owski et al.\ 2008). If the radio is associated with SF, then the nondetection in the mid-IR would imply a strong absorption by the silicates around 10 $\mu$m (E. Le Floc'h, 2008, private communication). The blue optical colors for some of the GRB hosts can be explained if large regions of these galaxies are totally obscured by dust. Optical light might be dominated by relatively less obscured stars outside star-forming regions, leading to blue colors (Micha{\l}owski et al.\ 2008). 

We further investigate the relation between GRB hosts and other galaxies, by analyzing the  mass-metallicity ($MZ$) relation. This is the correlation between the stellar mass of galaxies and their metallicity (Lequeux et al.\ 1979; Tremonti et al.\ 2004; Savaglio et al.\ 2005). In the $MZ$ plot of  Figure~\ref{MZ}, we compare  GRB hosts with local dwarf galaxies (Lee et al.\ 2006), and GDDS and CFRS galaxies at $0.4<z<1$ (Savaglio et al.\ 2005). The local sample metallicities are estimated using the $T_e$ method. Since the $T_e$ or O3N2 metallicities are not measured for GDDS galaxies and GRB hosts, proper conversions are applied, as explained in \S~\ref{metallicities}. In the left and right panels of Figure~\ref{MZ}, we show the lower and the upper branch solution for metallicity in a subsample of 9 GRB hosts. The picture is not very clear, and we do not detect a $MZ$ relation in the GRB hosts of any kind, especially because for 9 hosts the metallicity is poorly constrained. However, we can say that there is no indication that GRB hosts have metallicity lower than expected, given their stellar mass (Stanek et al.\ 2006). GRB hosts are not a special class of galaxies. Given the redshift and the stellar mass, the lower branch metallicity is favored for the 9 GRB hosts with double solution (right hand side of Figure~\ref{MZ}), 

As regards the redshift evolution, we compare GRB host metallicities, with the metallicities measured in 11 damped Lyman-$\alpha$ systems through afterglow spectroscopy (GRB-DLAs; Savaglio 2006) at redshift $z>2$ (Figure~\ref{zZ}). Again, the situation is not very clear. GRB-DLAs indicate that metallicities of the order of half solar are expected at $z<1$. This is actually observed in the subsample of 6 GRB hosts whose metallicity (determined with the $T_e$ or O3N2 calibration) is better constrained. However, for the rest of the sample, the lower branch solution suggests no redshift evolution in $0<z<6.3$, an interval of more than 13 Gyr. Any conclusion needs more and deeper observations, to break the degeneracy in those objects already observed, or to study new objects. For instance, the O3N2 calibrator can be detected using NIR spectroscopy for the brightest objects at $1\lsim z \lsim1.6$.

The typical GRB host is a small star-forming galaxy with likely sub-solar metallicity, but a non-negligible dust extinction (Table~\ref{tsummary}). It is in some regards similar to a young LMC, observed at redshift $z\sim0.7$, when it was more active than now, with an SFR which is 5 times higher than today. We still do not know whether hosts associated with short GRBs are considerably different from those associated with long GRBs.

\section{Summary}\label{summary}

We have presented a complete study of the largest sample of galaxies hosting GRBs, \nhosts\ objects, distributed along the redshift interval $0<z<6.3$. GRB hosts can be used as important  probes of the cosmic history of galaxy formation and evolution. 
Most GRBs are associated with the death of young massive stars, which are more common in star-forming galaxies. Therefore, GRBs are an effective tool to detect star-forming galaxies. As shown by recent studies (Glazebrook et al.\ 2004; Juneau et al.\ 2005; Borch et al.\ 2006), the star-formation density in the $z<1$ universe is carried out by small, faint, low-mass star-forming  galaxies, similar to the typical GRB host.  Moreover, in the $z\sim5$ universe,  GRB hosts observed with Spitzer are $\sim3$ times fainter than the typical spectroscopically confirmed galaxy in the Great Observatories Origins Deep Survey (GOODS; Yan et al.\ 2006), suggesting that not all star-forming galaxies at these redshifts are detected in deep surveys (Chary, Berger \& Cowie (2007;  Y{\"u}ksel et al.\ 2008).
We consider however, that our view of GRB hosts is still partial, as we mainly detect those at $z<1.5$. Hosts at higher redshift are harder to observe. It is possible that high-$z$ GRB hosts are more massive than those at low redshift, because star formation could be carried by more-massive galaxies in the remote universe. Future deeper multiband observations of high-$z$ hosts are  mandatory to help solve this issue.

In summary, our conclusions are (Table~\ref{tsummary}):

\begin{itemize}

\item GRB hosts are generally small star-forming galaxies. The mean stellar mass is similar to the stellar mass of the LMC, $M_\ast \sim 10^{9.3}$ M$_\odot$. About 83\% of the sample has stellar mass in the interval $10^{8.5}-10^{10.3}$ M$_\odot$). The median SFR $=2.5$ M$_\odot$ yr$^{-1}$ is 5 times higher than in the LMC;
 
\item To estimate SFR, we derived new relations, suitable for GRB hosts. These give the total SFR when H$\alpha$ is not detected, but [OII]  or UV are detected. Our SFRs span an interval of more than three orders of magnitudes, from 0.01 M$_\odot$ yr$^{-1}$ to 36 M$_\odot$ yr$^{-1}$;

\item The dust extinction in the visual band is on average $A_V=0.53$. The Balmer stellar absorption is generally small, but not negligible. Dust extinction, Balmer absorption and slit-aperture flux loss are considered when measuring SFR;

\item The median star-formation rate per unit stellar mass SSFR is $\sim0.8$ Gyr$^{-1}$,  with a small scatter, such that SFR $\propto M_\ast$, a somewhat surprising result. The median SSFR is about 5 times higher than in the LMC.  A large fraction of GRB hosts are the equivalent of local starbursts;

\item Metallicities derived from emission lines in the host galaxies at  $z<1$ are relatively low, likely in the range 1/10 solar to  solar;
 
\item Metallicities measured from UV absorption lines in the cold medium of GRB hosts at $z>2$ (GRB-DLAs) are in a similar range. Combining this with the results for $z<1$ GRB hosts, we see no significant evolution of metallicity in GRB hosts in the interval $0<z<6$;

\item The subsample of 6 short-GRB hosts have stellar masses $10^{8.7}$ M$_\odot <M_\ast < 10^{11.0}$ M$_\odot$ and SSFRs in the range $0.006-6$ Gyr$^{-1}$. The suggestion that short-GRB hosts are large quiescent galaxies requires a larger sample to be confirmed;

\item There is no clear indication that GRB host galaxies belong to a special population. Their properties are those expected for normal star-forming galaxies, from the local to the most distant universe.

\end{itemize}

Our investigation will continue, and results will be made public in the GHostS public database. In the final sample, we plan to include all GRB hosts detected with Swift, for a total of a few hundred objects. This will give a more complete picture on the nature of GRB hosts and their relation with all other galaxies.

\acknowledgments 
We thank Avishay Gal-Yam, Emeric Le Floc'h, Lisa Kewley, Maryam Modjaz, John Moustakas, Daniele Pierini, Gra$\dot{\rm z}$yna Stasi{\'n}ska and Christian Wolf for stimulating discussions. We are indebted to the anonymous referees for the many valuable comments. We also acknowledge the inspiring collaboration with Tam\'as Budav\'ari, the programmer behind the GHostS database. Gerhardt Meurer is acknowledged for constant scientific support.

{}

\newpage

\begin{deluxetable}{lcccccc}
\tablecaption{GRB host sample\label{tsample}}
\tablecolumns{7}
\tablewidth{0pc}
\tabletypesize{\footnotesize}
\tablehead{
  \colhead{GRB}  & 
  \colhead{$z$} & 
  \colhead{Type \tablenotemark{a}} & 
  \colhead{Ref} &  
  \colhead{Morphology\tablenotemark{b}} & 
  \colhead{Morphology\tablenotemark{c}} &
    \colhead{E$_G(B-V)$\tablenotemark{d}}
}
\startdata
GRB~970228       & 0.695 & long & [1,2,3]        & Late & Disk & 0.234 \\
GRB~970508       & 0.835 & long & [3,4,5]        & Early & Spheroid & 0.026 \\
GRB~970828       & 0.960 & long & [5,6,7]        & . . . & Asymmetric, Merger & 0.036 \\
GRB~971214       & 3.420 & long & [2,8,5]        & Late & Asymmetric & 0.016 \\
GRB~980425       & 0.009 & long & [9,10,5,11]    & . . . & . . . & 0.065 \\
GRB~980613       & 1.097 & long & [2,12]         & Merger & Asymmetric, Merger & 0.087 \\
GRB~980703       & 0.966 & long & [3,13,5]       & Late & Spheroid & 0.061 \\
GRB~990123       & 1.600 & long & [3,5]  & Late & Disk, Merger & 0.016 \\
GRB~990506       & 1.310 & long & [14,5]         & Early & Spheroid & 0.068 \\
GRB~990705       & 0.842 & long & [9,15,16,5]    & Late & Disk & 0.120 \\
GRB~990712       & 0.433 & long & [3,17]         & Late & Disk, Merger & 0.033 \\
GRB~991208       & 0.706 & long & [3,18]         & Early & Spheroid, Merger? & 0.016 \\
GRB~000210       & 0.846 & long & [3,19]         & . . . & . . . & 0.019 \\
GRB~000418       & 1.118 & long & [3,14]         & Early & Spheroid & 0.032 \\
GRB~000911       & 1.058 & long & [20,21]        & . . . & . . . & 0.107 \\
GRB~000926       & 2.036 & long & [3]    & Late & Merger & 0.023 \\
GRB~010222       & 1.480 & long & [22,23,24,5,25]        & Merger & Spheroid & 0.023 \\
GRB~010921       & 0.451 & long & [3,25,26]      & Late & Disk & 0.148 \\
GRB~011121       & 0.362 & long & [27,28]        & Late & Disk & 0.370 \\
GRB~011211       & 2.141 & long & [29]   & Late & Merger & 0.045 \\
GRB~020405       & 0.691 & long & [30,25]        & Merger & Asymmetric, Merger & 0.055 \\
GRB~020813       & 1.255 & long & [31,7,25]      & Early & . . . & 0.111 \\
GRB~020819B    & 0.410 & long & [32]   & . . . & . . . & 0.070 \\
XRF~020903       & 0.251 & long & [33,34,25]     & Merger & Merger & 0.033 \\
GRB~021004       & 2.330 & long & [35,36,37]     & Late & Spheroid & 0.060 \\
GRB~021211       & 1.006 & long & [25]   & Late & Spheroid & 0.028 \\
GRB~030328       & 1.520 & long & [38]   & . . . & . . . & 0.047 \\
GRB~030329       & 0.168 & long & [39,40]        & Late & . . . & 0.025 \\
XRF~030528       & 0.782 & long & [41,42]        & . . . & . . . & 0.620 \\
GRB~031203       & 0.105 & long & [43,44]        & . . . & . . . & 1.040 \\
GRB~040924       & 0.859 & long & [25,45]        & . . . & . . . & 0.058 \\
GRB~041006       & 0.712 & long & [25,46]        & . . . & Asymmetric & 0.070 \\
GRB~050223       & 0.584 & long & [47]   & . . . & . . . & 0.090 \\
XRF~050416       & 0.653 & short & [48]   & . . . & . . . & 0.059 \\ 
GRB~050509B      & 0.225 & short & [49]   & . . . & . . . & 0.019 \\
GRB~050709       & 0.161 & short & [50,51]        & . . . & . . . & 0.012 \\
GRB~050724       & 0.257 & short & [52,53]        & . . . & . . . & 0.613 \\
GRB~050826       & 0.296 & long & [54,55]        & . . . & . . . & 0.590 \\
GRB~050904       & 6.295 & long & [56,57,58]     & . . . & . . . & 0.060 \\
GRB~051022       & 0.807 & long & [59,60]        & . . . & . . . & 0.040 \\
GRB~051221       & 0.546 & short & [61]   & . . . & . . . & 0.069 \\
XRF~060218       & 0.034 & long & [62,63,64,65]  & . . . & . . . & 0.140 \\
GRB~060505       & 0.089 & long? & [66]   & . . . & . . . & 0.021 \\
GRB~060614       & 0.125 & long & [67,68,69,70]     & . . . & . . . & 0.070 \\
GRB~061006       & 0.438 & short & [71]   & . . . & . . . & 0.320 \\
GRB~061126       & 1.159 & long & [72]   & . . . & . . . & 0.182 \\
\enddata
\tablenotetext{a}{GRB type as defined by its duration. According to the canonical definition, long or short GRBs (for which the duration of 90\% of its $\gamma$-ray energy is emitted in more or less than $\sim2$ seconds, respectively) are generally associated with core-collapse SNe or merger of compact objects, respectively. For several GRBs this definition does not hold. For instance GRB~060505 and GRB~060614 (with a duration of 4 and 102 seconds, respectively) show no evidence of underlying SN (Fynbo et al.\ 2006; Gal-Yam et al.\ 2006) despite the low redshift ($z\simeq 0.1$.)}
\tablenotetext{b}{Morphology of the host, as given by Conselice et al.\ (2005).}
\tablenotetext{c} {Morphology of the host, as given by Wainwright et al.\ (2007).}
\tablenotetext{d}{Galactic color excess, estimated from HI maps by Schlegel et al.\ (1998).}
\tablecomments{References: 
[1] Bloom et al.\ (2001)
[2] Chary et al.\ (2002)
[3] Christensen et al.\ (2004)
[4] Bloom et al.\ (1998)
[5] Le Floc'h et al.\ (2006)
[6] Djorgovski et al.\ (2001)
[7] Le Floc'h et al.\ (2003)
[8] Kulkarni et al.\ (1998)
[9] Bloom et al.\ (2002)
[10] Hammer et al.\ (2006)
[11] Sollerman et al.\ (2005)
[12] Djorgovski et al.\ (2003)
[13] Djorgovski et al.\ (1998)
[14] Bloom et al.\ (2003)
[15] Holland et al.\ (2000)
[16] Le Floc'h et al.\ (2002)
[17] K\"upc\"u Yolda\c{s} et al.\ (2006)
[18] Castro-Tirado et al.\ (2001)
[19] Piro et al.\ (2002)
[20] Masetti et al.\ (2005)
[21] Price et al.\ (2002b)
[22] Frail et al.\ (2002)
[23] Fruchter et al.\ (2001)
[24] Galama et al.\ (2003)
[25] Wainwright et al.\ (2007)
[26] Price et al.\ (2002a)
[27] Garnavich et al.\ (2003)
[28] K\"upc\"u Yolda\c{s} et al.\ (2007)
[29] Fynbo et al.\ (2003)
[30] Price et al.\ (2003)
[31] Barth et al.\ (2003)
[32] Jakobsson et al.\ (2005)
[33] Bersier et al.\ (2006)
[34] Soderberg et al.\ (2004)
[35] de Ugarte Postigo et al.\ (2005)
[36] Mirabal et al.\ (2002)
[37] M{\o}ller et al.\ (2002)
[38] Gorosabel et al.\ (2005a)
[39] Gorosabel et al.\ (2005b)
[40] Th{\"o}ne et al.\ (2007)
[41] Rau et al.\ (2004)
[42] Rau et al.\ (2005)
[43] Cobb et al.\ (2004)
[44] Prochaska et al.\ (2004)
[45] Wiersema et al.\ (2008)
[46] Soderberg et al.\ (2006a)
[47] Pellizza et al.\ (2006)
[48] Soderberg et al.\ (2007)
[49] Castro-Tirado et al.\ (2005)
[50] Covino et al.\ (2006)
[51] Hjorth et al.\ (2005)
[52] Berger et al.\ (2005)
[53] Gorosabel et al.\ (2006)
[54] Mirabal et al.\ (2007)
[55] Ovaldsen et al.\ (2007)
[56] Aoki et al.\ (2006)
[57] Berger et al.\ (2007a)
[58] Totani et al.\ (2006)
[59] Castro-Tirado et al.\ (2007)
[60] Rol et al.\ (2007)
[61] Soderberg et al.\ (2006b)
[62] Cobb et al.\ (2006a)
[63] Pian et al.\ (2006)
[64] Sollerman et al.\ (2006)
[65] Wiersema et al.\ (2007)
[66] Th{\"o}ne et al.\ (2008)
[67] Cobb et al.\ (2006b)
[68] Della Valle et al.\ (2006)
[69] Gal-Yam et al.\ (2006)
[70] Mangano et al.\ (2007)
[71] Berger et al.\ (2007b)
[72] Perley et al.\ (2008)
}
\end{deluxetable}

\begin{deluxetable}{lccccccccc}
%\rotate 
\tablecaption{GRB-host photometry\label{tcol}}
\tablecolumns{10}
\tablewidth{0pc}
\tabletypesize{\footnotesize}
\tablehead{
 \colhead{GRB}  & 
 \colhead{Redshift} & 
 \colhead{$u$'} & 
 \colhead{B} &  
 \colhead{V} & 
 \colhead{R} & 
 \colhead{F814} &
 \colhead{J} & 
 \colhead{H} & 
 \colhead{K}
 }
\startdata
970228 &  0.695 & . . . & $26.28\pm0.31$ & $25.20\pm0.25$ & $24.92\pm0.20$ & $24.45\pm0.20$ & . . . & $24.46\pm0.31$ &$24.37\pm0.20$ \\ 
970508 &  0.835 & . . . & $25.70\pm0.11$ & $25.50\pm0.17$ & $25.21\pm0.09$ & $24.50\pm0.29$ & . . . & . . . & $24.59\pm0.20$ \\ 
970828 &  0.960 & . . . & . . . & . . . & $25.27\pm0.31$ & . . . & . . . & . . . & $23.34\pm0.31$ \\ 
971214 &  3.420 & . . . & . . . & $26.59\pm0.20$ & $25.77\pm0.17$ & . . . & . . . & . . . & $24.28\pm0.20$ \\ 
980425 &  0.009 & . . . & $14.79\pm0.60$ & $14.57\pm0.60$ & $14.28\pm0.05$ & $14.13\pm0.60$ & . . . & . . . & . . . \\ 
980613 &  1.097 & . . . & $24.98\pm0.31$ & $24.20\pm0.20$ & $23.98\pm0.06$ & $23.85\pm0.10$ & . . . & . . . & $23.53\pm0.22$ \\ 
980703 &  0.966 & . . . & $23.09\pm0.11$ & $22.88\pm0.08$ & $22.57\pm0.06$ & $22.27\pm0.25$ & $21.71\pm0.11$ & $21.84\pm0.25$ & $21.48\pm0.12$ \\ 
990123 &  1.600 & $24.49\pm0.16$ & $24.08\pm0.10$ & $24.16\pm0.15$ & $23.90\pm0.10$ & $24.04\pm0.15$ & . . . & . . . & $23.60\pm0.31$ \\ 
990506 &  1.310 & . . . & . . . & . . . & $25.67\pm0.20$ & . . . & . . . & . . . & $23.29\pm0.20$ \\ 
990705 &  0.842 & . . . & . . . & $22.79\pm0.20$ & $22.17\pm0.10$ & . . . & . . . & . . . & . . . \\ 
990712 &  0.433 & $23.91\pm0.09$ & $23.15\pm0.08$ & $22.31\pm0.05$ & $21.90\pm0.05$ & $21.79\pm0.05$ & $21.68\pm0.17$ & $21.60\pm0.19$ & $21.85\pm0.10$ \\ 
991208 &  0.706 & . . . & $25.05\pm0.17$ & $24.51\pm0.16$ & $24.40\pm0.15$ & $23.69\pm0.20$ & . . . & . . . & $23.60\pm0.21$ \\ 
000210 &  0.846 & $24.38\pm0.13$ & $24.24\pm0.13$ & $24.18\pm0.09$ & $23.56\pm0.10$ & $22.89\pm0.12$ & $22.80\pm0.10$ & $22.88\pm0.23$ & $22.78\pm0.14$ \\ 
000418 &  1.118 & . . . & $23.99\pm0.06$ & $23.81\pm0.06$ & $23.57\pm0.05$ & $23.22\pm0.05$ & $23.18\pm0.10$ & . . . & $23.04\pm0.31$ \\ 
000911 &  1.058 & . . . & . . . & $25.07\pm0.31$ & . . . & $24.50\pm0.20$ & . . . & $23.91\pm0.36$ & . . . \\ 
000926 &  2.036 & . . . & $25.31\pm0.34$ & $25.01\pm0.06$ & $24.94\pm0.07$ & $24.97\pm0.10$ & $24.11\pm0.42$ & . . . & . . . \\ 
010222 &  1.480 & . . . & $25.77\pm0.38$ & $26.18\pm0.14$ & $26.59\pm0.23$ & $25.99\pm0.26$ & . . . & . . . & $25.34\pm0.28$ \\ 
010921 &  0.451 & . . . & $22.67\pm0.17$ & $21.99\pm0.15$ & $21.62\pm0.09$ & $21.35\pm0.11$ & $21.09\pm0.05$ & $21.03\pm0.05$ & $20.84\pm0.05$ \\ 
011121 &  0.362 & . . . & $21.93\pm0.15$ & $21.48\pm0.06$ & . . . & $20.92\pm0.06$ & $20.47\pm0.19$ & . . . & . . . \\ 
011211 &  2.141 & . . . & $25.31\pm0.18$ & . . . & $24.89\pm0.26$ & . . . & . . . & . . . & . . . \\ 
020405 &  0.691 & . . . & . . . & $22.59\pm0.05$ & . . . & $21.59\pm0.05$ & . . . & . . . & . . . \\ 
020813 &  1.255 & . . . & $24.39\pm0.20$ & . . . & $24.87\pm0.20$ & $24.00\pm0.20$ & . . . & . . . & . . . \\ 
020819B &  0.410 & . . . & $21.78\pm0.56$ & . . . & $19.62\pm0.02$ & . . . & . . . & . . . & $18.68\pm0.21$ \\ 
020903 &  0.251 & . . . & $21.61\pm0.10$ & $20.79\pm0.10$ & $21.00\pm0.10$ & $20.93\pm0.10$ & . . . & . . . & . . . \\ 
021004 &  2.327 & . . . & $24.43\pm0.10$ & . . . & $24.12\pm0.08$ & $24.38\pm0.04$ & . . . & $23.89\pm0.15$ & . . . \\ 
021211 &  1.006 & . . . & $26.30\pm0.31$ & . . . & . . . & $24.60\pm0.20$ & . . . & . . . & . . . \\ 
030328 &  1.520 & $25.02\pm0.19$ & $24.80\pm0.08$ & $24.63\pm0.08$ & $24.59\pm0.12$ & $24.44\pm0.22$ & . . . & $23.9$ & $23.8$ \\ 
030329 &  0.168 & $23.36\pm0.10$ & $23.16\pm0.07$ & $22.77\pm0.10$ & $22.77\pm0.06$ & $22.51\pm0.05$ & $22.40\pm0.16$ & $22.54\pm0.24$ & . . . \\ 
030528 &  0.782 & . . . & . . . & $21.92\pm0.21$ & $22.21\pm0.21$ & $21.66\pm0.21$ & $21.73\pm0.10$ & . . . & $21.76\pm0.87$ \\ 
031203 &  0.105 & . . . & . . . & . . . & . . . & $17.79\pm0.07$ & . . . & $18.53\pm0.02$ & $18.03\pm0.02$ \\ 
040924 &  0.859 & . . . & . . . & . . . & . . . & $24.00\pm0.10$ & . . . & . . . & . . . \\ 
041006 &  0.712 & . . . & . . . & . . . & . . . & $24.20\pm0.15$ & . . . & . . . & . . . \\ 
050223 &  0.584 & . . . & . . . & . . . & $21.72\pm0.05$ & . . . & . . . & . . . & $20.68\pm0.02$ \\ 
050416 &  0.653 & . . . & . . . & . . . & $23.47\pm0.60$ & $23.09\pm0.10$ & . . . & . . . & . . . \\ 
050509B &  0.225 & $21.29\pm0.13$ & . . . & . . . & $17.28\pm0.02$ & $17.01\pm0.02$ & $16.13\pm0.10$ & $15.84\pm0.10$ & $15.89\pm0.10$ \\ 
050709 &  0.161 & . . . & $22.05\pm0.10$ & $21.27\pm0.06$ & $21.26\pm0.06$ & $21.01\pm0.08$ & . . . & . . . & . . . \\ 
050724 &  0.257 & . . . & $19.82\pm0.12$ & $18.72\pm0.04$ & . . . & $18.03\pm0.16$ & $17.23\pm0.04$ & $16.84\pm0.05$ & $16.65\pm0.04$ \\ 
050826 &  0.296 & . . . & $21.38\pm0.31$ & $20.57\pm0.21$ & $19.84\pm0.05$ & $19.98\pm0.20$ & . . . & . . . & . . . \\ 
050904 &  6.295 & . . . & . . . & . . . & . . . & . . . & . . . & $26.79\pm0.32$ & . . . \\ 
051022 &  0.807 & . . . & $22.47\pm0.02$ & $22.17\pm0.04$ & $21.92\pm0.09$ & $21.69\pm0.01$ & . . . & $20.76\pm0.09$ & $20.22\pm0.11$ \\ 
051221 &  0.546 & . . . & . . . & . . . & $21.99\pm0.09$ & $21.99\pm0.17$ & . . . & . . . & . . . \\ 
060218 &  0.034 & $20.50\pm0.15$ & . . . & . . . & $19.80\pm0.03$ & $19.69\pm0.04$ & . . . & . . . & . . . \\ 
060505 &  0.089 & $19.18\pm0.05$ & $18.77\pm0.02$ & $18.26\pm0.02$ & $18.07\pm0.02$ & $17.94\pm0.02$ & . . . & . . . & $17.73\pm0.04$ \\ 
060614 &  0.125 & $24.55\pm0.31$ & $23.61\pm0.13$ & $22.75\pm0.05$ & $22.63\pm0.01$ & $22.37\pm0.60$ & . . . & . . . & . . . \\ 
061006 &  0.438 & . . . & . . . & . . . & $24.18\pm0.09$ & $23.10\pm0.09$ & . . . & . . . & . . . \\ 
061126 &  1.159 & . . . & . . . & . . . & $23.81\pm0.11$ & . . . & . . . & . . . & . . . \\ 
\enddata
\tablecomments{All magnitudes (corrected for Galactic extinction) are in the AB system. }
\end{deluxetable}

\clearpage
\LongTables
\begin{landscape}

\begin{deluxetable}{lcccccccccccc}
%\rotate
\tablecaption{GRB-host emission line fluxes\label{temi}}
\tablecolumns{13}
\tablewidth{0pc}
\tabletypesize{\tiny}
\tablehead{
  \colhead{GRB} & 
  \colhead{$z$} & 
  \colhead{[OII]$\lambda3727$} & 
  \colhead{[NeIII]$\lambda3869$} & 
  \colhead{H$\gamma$} &
  \colhead{H$\beta$} & 
\colhead{[OIII]$\lambda4959$} & 
\colhead{[OIII]$\lambda5007$} &
\colhead{H$\alpha$} & 
\colhead{[NII]$\lambda6583$} & 
\colhead{[SII]$\lambda6716$} & 
\colhead{[SII]$\lambda6731$} & 
\colhead{Ape.}
}
\startdata
970228	 & 0.695	 & $2.2\pm0.1$	 & $0.2$	 & . . .	 & $<0.34$	 & . . .	 & $1.55\pm0.12$	 & . . .	 & . . .	 & . . .	 & . . .	& 1\\
970508	 & 0.835	 & $2.98\pm0.22$	 & $1.25\pm0.1$	 & . . .	 & . . .	 & . . .	 & . . .	 & . . .	 & . . .	 & . . .	 & . . .	& 1\\
970828	 & 0.958	 & $1.63\pm0.07$	 & $0.5$	 & . . .	 & . . .	 & . . .	 & . . .	 & . . .	 & . . .	 & . . .	 & . . .	& 1\\
980425	 & 0.0085	 & $440$	 & $113$	 & . . .	 & $354$	 & $497$	 & $2012$	 & $1839$	 & $113$	 & $109$	 & $86$	& 4.6\\
980613	 & 1.097	 & $5.25\pm0.15$	 & $0.5$	 & . . .	 & . . .	 & . . .	 & . . .	 & . . .	 & . . .	 & . . .	 & . . .	& 1.2\\
980703	 & 0.966	 & $30.4\pm1.5$	 & . . .	 & $2.80\pm0.14$	 & $6.1\pm0.3$	 & $2.7\pm0.2$	 & . . .	 & . . .	 & . . .	 & . . .	 & . . .	& 1\\
990506	 & 1.31	 & $2.16\pm0.18$	 & . . .	 & . . .	 & . . .	 & . . .	 & . . .	 & . . .	 & . . .	 & . . .	 & . . .	& 1\\
990705	 & 0.8424	 & $17.92$	 & . . .	 & . . .	 & . . .	 & . . .	 & . . .	 & . . .	 & . . .	 & . . .	 & . . .	& 1\\
990712	 & 0.434	 & $35.6\pm1.2$	 & $5.42\pm0.30$	 & $4.49\pm0.32$	 & $13.24\pm0.28$	 & $22.04\pm0.48$	 & $60.36\pm0.46$	 & $45.0\pm1.0$	 & $<10$	 & . . .	 & . . .	& 1.3\\
991208	 & 0.706	 & $17.9\pm2.2$	 & . . .	 & . . .	 & $38.4\pm3.3$	 & $16.1\pm3.2$	 & $49.0\pm3.3$	 & . . .	 & . . .	 & . . .	 & . . .	& 1 \\
000210	 & 0.846	 & $5.8\pm1.5$	 & . . .	 & . . .	 & . . .	 & . . .	 & . . .	 & . . .	 & . . .	 & . . .	 & . . .	& 1\\
000418	 & 1.118	 & $6.97$	 & $0.21$	 & $0.417$	 & . . .	 & . . .	 & . . .	 & . . .	 & . . .	 & . . .	 & . . .	& 1.9\\
000911	 & 1.0585	 & $2.3\pm0.3$	 & . . .	 & . . .	 & . . .	 & . . .	 & . . .	 & . . .	 & . . .	 & . . .	 & . . .	& 1\\
010921	 & 0.451	 & $24.2\pm2.3$	 & $<9.8$	 & . . .	 & $6.9\pm1.1$	 & $4.1\pm1.3$	 & $21.8\pm1.4$	 & $32.9\pm5.9$	 & . . .	 & . . .	 & . . .	& 1\\
011121	 & 0.362	 & $63.63$	 & $<20$	 & . . .	 & $21.59$	 & . . .	 & $19.95$	 & $83.23$	 & . . .	 & . . .	 & . . .	& 1\\
020405	 & 0.691	 & $12.02\pm0.60$	 & $1.31\pm0.51$	 & $1.99\pm0.53$	 & $6.63\pm0.57$	 & $5.11\pm0.39$	 & $20.19\pm0.37$	 & . . .	 & . . .	 & . . .	 & . . .	& 1.3\\
020813	 & 1.255	 & $6.5\pm0.6$	 & . . .	 & . . .	 & . . .	 & . . .	 & . . .	 & . . .	 & . . .	 & . . .	 & . . .	& 1 \\
020903	 & 0.251	 & $39.18$	 & $9.027$	 & $7.14$	 & $20.47$	 & $34.49$	 & $113.5$	 & $75.0$	 & $1.61$	 & $4.09$	 & $2.38$	& 2.7\\
030329	 & 0.168	 & $20.40\pm0.39$	 & $4.04\pm0.28$	 & $4.82\pm0.20$	 & $11.70\pm0.36$	 & $14.76\pm0.38$	 & $42.21\pm0.31$	 & $32.40\pm0.47$	 & $<1.8$	 & . . .	 & $3.27\pm0.32$	& 1\\
030528	 & 0.782	 & $15\pm1$	 & $<1$	 & $<2.5$	 & $4.8\pm0.4$	 & $4.1\pm1$	 & $20\pm1$	 & . . .	 & . . .	 & . . .	 & . . .	& 3.1\\
031203	 & 0.1055	 & $1238\pm103$	 & $731\pm65$	 & $699\pm35$	 & $1677\pm26$	 & $3584\pm33$	 & $10880\pm54$	 & $4935\pm21$	 & $265.9\pm8.4$	 & $149.9\pm5.9$	 & $125.3\pm4.9$	& 2\\
040924	 & 0.858	 & $2.31\pm0.14$	 & $0.47\pm0.17$	 & . . .	 & $0.44\pm0.20$	 & $0.94\pm0.24$	 & . . .	 & . . .	 & . . .	 & . . .	 & . . .	& 2\\
041006	 & 0.712	 & $1.3$	 & . . .	 & . . .	 & . . .	 & . . .	 & . . .	 & . . .	 & . . .	 & . . .	 & . . .	& 1\\
050223	 & 0.584	 & $8.2\pm1.8$	 & . . .	 & . . .	 & $3.85\pm0.94$	 & . . .	 & $5.31\pm0.61$	 & . . .	 & . . .	 & . . .	 & . . .	& 1.2\\
050416	 & 0.6528	 & $11.11\pm0.46$	 & . . .	 & $1.35\pm0.22$	 & $4.07\pm0.55$	 & $2.74\pm0.33$	 & $9.19\pm0.44$	 & . . .	 & . . .	 & . . .	 & . . .	& 1\\
050709	 & 0.1606	 & . . .	 & . . .	 & . . .	 & $11.5$	 & . . .	 & . . .	 & $35.3$	 & . . .	 & . . .	 & . . .	& 1.15\\
050826	 & 0.296	 & $110\pm10$	 & . . .	 & . . .	 & . . .	 & . . .	 & . . .	 & . . .	 & . . .	 & . . .	 & . . .	& 2.6\\
051022	 & 0.8070	 & $104.1\pm3.0$	 & . . .	 & . . .	 & $83.4\pm6.0$ & . . .	 & $141.1\pm5.0$	 & . . .	 & . . .	 & . . .	 & . . .	& 1\\
051221	 & 0.5459	 & $13.8$	 & . . .	 & $1.43$	 & $4.91$	 & $1.75$	 & $5.62$	 & . . .	 & . . .	 & . . .	 & . . .	& 1.4\\
060218	 & 0.03345 & $196.3\pm5.3$	 & $34.1\pm3.5$	 & $48.0\pm3.5$	 & $92.9\pm1.9$	 & $139\pm2$	 & $426\pm2$	 & $315\pm25$	 & $19\pm2$	 & $14\pm2$	 & $14\pm2$	& 1\\
060505	 & 0.0889	 & $56.97\pm0.68$	 & . . .	 & $4.42\pm0.06$	 & $15.78\pm0.05$	 & $9.37\pm0.32$	 & $26.49\pm0.25$	 & $58.22\pm0.12$	 & $11.01\pm0.12$	 & $12.33\pm0.14$	 & $9.22\pm0.14$	& 5\\
060614	 & 0.1254	 & . . .	 & . . .	 & . . .	 & . . .	 & . . .	 & . . .	 & $4.1$	 & . . .	 & . . .	 & . . .	& 1\\
061126	 & 1.1588	 & $2.14\pm0.27$	 & . . .	 & . . .	 & . . .	 & . . .	 & . . .	 & . . .	 & . . .	 & . . .	 & . . .	& 1.3\\
\enddata
\tabletypesize{\footnotesize}
\tablecomments{Emission line fluxes (corrected for Galactic extinction) are in units of $10^{-17}$ erg s$^{-1}$ cm$^{-2}$. The number in the last column is the multiplicative factor used to correct for the slit-aperture flux loss, not used here.
Stellar Balmer absorption and host dust extinction are not considered here (see text and Table~\ref{tbal}). Flux errors are available for a subsample of lines.}
\end{deluxetable}

\clearpage
\end{landscape}

\begin{table}
\caption[t1]{Balmer stellar absorption and dust extinction corrections}
\begin{center} 
\begin{tabular}{lcccc} 
\tableline\tableline&&&&\\[-5pt]
 GRB  & (H$\alpha_i$/H$\alpha)$\tablenotemark{a} & (H$\beta_i$/H$\beta$)\tablenotemark{a} & (H$\gamma_i$/H$\gamma$)\tablenotemark{a} & $A_V$\tablenotemark{b}  \\
[5pt]\tableline&&&&\\[-5pt] 
980425 & 1.00 & 1.00 & . . . & $1.73$ \\
980703 & . . . & 1.08 & 1.26 & . . . \\ 
990712 & 1.02 & 1.05 & 1.19 & $0.39\pm0.09$ \\
991208 & . . . & 1.04 & . . . & . . . \\
000418 & . . . & . . . & 1.22 & . . . \\
010921 & 1.06 & 1.22 & . . . & $1.06\pm0.62$ \\
011121 & 1.03 & 1.22 & . . . & $0.38$ \\
020405 & . . . & 1.06 & 1.29 & . . . \\
020903 & 1.00 & 1.05 & . . . & $0.59$ \\
030329 & 1.00 & 1.00 & 1.00 & $<0.1$ \\ 
030528 & . . . & 1.06 & . . . & . . . \\
031203 & 1.00 & 1.02 & 1.06 & $0.03\pm0.05$ \\
040924 & . . . & 1.21 & . . . & . . . \\
050223 & . . . & 1.24 & . . . & . . . \\
050416 & . . . & 1.11 & 1.41 & . . . \\
050709 & 1.10 & 1.30 & . . . & $\sim 0$ \\
051022 & . . . & 1.02 & . . . & . . . \\
051221 & . . . & 1.16 & 1.56 & . . . \\
060218 & 1.00 & 1.00 & 1.00 & $0.49\pm0.24$ \\
060505 & 1.06 & 1.10 & 1.50 & $0.63\pm0.01$ \\
060614 & 1.13 & . . . & . . . & . . . \\
[2pt]\tableline
\end{tabular}
\end{center}
\tablenotetext{a}{Ratio of Balmer lines after and before stellar Balmer-absorption correction.}
\tablenotetext{b}{Visual extinction in the gas component of the GRB host.}
\label{tbal}
\end{table}

\begin{deluxetable}{lcccccc}
\tablecaption{GRB host parameters\label{tsed}}
\tablecolumns{7}
\tablewidth{0pc}
\tabletypesize{\small}
\tablehead{
  \colhead{GRB}  & 
  \colhead{$z$} & 
  \colhead{$M_K$} & 
  \colhead{$M_B$} & 
  \colhead{$\log M_\star$} &
  \colhead{$M_\ast/L_K$} &
  \colhead{Age} \\
 \colhead{} & 
  \colhead{} & 
 \colhead{} & 
 \colhead{} & 
  \colhead{[M$_\odot$]} &
 \colhead{$(M/L_K)_\odot$} & 
\colhead{(Myr)}
 }
\startdata
970228   & 0.695 & $ -17.78\pm0.09 $     & $ -18.05\pm0.06 $     & $ 8.65\pm0.05 $       & $ 0.294\pm0.049 $     & 1022 \\ 
970508   & 0.835 & $ -17.88\pm0.24 $     & $ -18.38\pm0.20 $     & $ 8.52\pm0.10 $       & $ 0.203\pm0.053 $     & 499 \\ 
970828   & 0.960 & $ -20.21\pm0.39 $     & $ -20.59\pm0.58 $     & $ 9.19\pm0.36 $       & $ 0.139\pm0.101 $     & 833 \\ 
971214   & 3.420 & $ -22.20\pm0.72 $     & $ -23.23\pm1.10 $     & $ 9.59\pm0.40 $       & $ 0.056\pm0.052 $     & 166 \\ 
980425   & 0.009 & $ -19.44\pm1.26 $     & $ -19.33\pm1.11 $     & $ 9.21\pm0.52 $       & $ 0.319\pm0.234 $     & 6928 \\ 
980613   & 1.097 & $ -19.99\pm0.03 $     & $ -21.46\pm0.04 $     & $ 8.49\pm0.21 $       & $ 0.029\pm0.018 $     & 21 \\ 
980703   & 0.966 & $ -22.04\pm0.15 $     & $ -22.89\pm0.37 $     & $ 9.33\pm0.36 $       & $ 0.040\pm0.043 $     & 951 \\ 
990123   & 1.600 & $ -20.99\pm0.79 $     & $ -21.62\pm0.37 $     & $ 9.42\pm0.49 $       & $ 0.094\pm0.032 $     & 1803 \\ 
990506   & 1.310 & $ -21.10\pm0.15 $     & $ -21.35\pm0.46 $     & $ 9.48\pm0.18 $       & $ 0.099\pm0.041 $     & 713 \\ 
990705   & 0.842 & $ -22.37\pm1.06 $     & $ -22.14\pm0.76 $     & $ 10.20\pm0.76 $      & $ 0.191\pm0.138 $     & 1351 \\ 
990712   & 0.433 & $ -19.31\pm0.09 $     & $ -19.59\pm0.04 $     & $ 9.29\pm0.02 $       & $ 0.309\pm0.024 $     & 1052 \\ 
991208   & 0.706 & $ -18.88\pm0.23 $     & $ -19.75\pm0.83 $     & $ 8.53\pm0.37 $       & $ 0.122\pm0.115 $     & 809 \\ 
000210   & 0.846 & $ -19.75\pm0.23 $     & $ -20.04\pm0.15 $     & $ 9.31\pm0.08 $       & $ 0.226\pm0.062 $     & 1186 \\ 
000418   & 1.118 & $ -20.47\pm0.21 $     & $ -21.74\pm0.53 $     & $ 9.26\pm0.14 $       & $ 0.112\pm0.070 $     & 235 \\ 
000911   & 1.058 & $ -19.70\pm0.51 $     & $ -19.48\pm0.66 $     & $ 9.32\pm0.26 $       & $ 0.243\pm0.079 $     & 1799 \\ 
000926   & 2.036 & $ -21.36\pm1.13 $     & $ -21.06\pm0.49 $     & $ 9.52\pm0.84 $       & $ 0.117\pm0.090 $     & 720 \\ 
010222   & 1.480 & $ -18.42\pm0.65 $     & $ -18.37\pm0.12 $     & $ 8.82\pm0.26 $       & $ 0.247\pm0.053 $     & 2062 \\ 
010921   & 0.451 & $ -20.37\pm0.05 $     & $ -20.02\pm0.17 $     & $ 9.69\pm0.13 $       & $ 0.304\pm0.099 $     & 4287 \\ 
011121   & 0.362 & $ -21.01\pm0.38 $     & $ -20.59\pm0.75 $     & $ 9.81\pm0.17 $       & $ 0.244\pm0.128 $     & 3889 \\ 
011211   & 2.141 & $ -21.62\pm1.12 $     & $ -21.84\pm1.36 $     & $ 9.77\pm0.47 $       & $ 0.125\pm0.063 $     & 603 \\ 
020405   & 0.691 & $ -21.16\pm1.07 $     & $ -21.57\pm1.05 $     & $ 9.75\pm0.25 $       & $ 0.187\pm0.092 $     & 874 \\ 
020813   & 1.255 & $ -19.32\pm2.36 $     & $ -19.68\pm0.80 $     & $ 8.66\pm1.41 $       & $ 0.133\pm0.141 $     & 1710 \\ 
020819B   & 0.410 & $ -22.53\pm0.30 $     & $ -21.98\pm0.37 $     & $ 10.50\pm0.14 $      & $ 0.292\pm0.188 $     & 1875 \\ 
020903   & 0.251 & $ -18.97\pm0.33 $     & $ -19.30\pm0.05 $     & $ 8.87\pm0.07 $       & $ 0.172\pm0.058 $     & 386 \\ 
021004   & 2.327 & $ -22.06\pm0.38 $     & $ -21.10\pm0.11 $     & $ 10.20\pm0.18 $      & $ 0.201\pm0.018 $     & 1894 \\ 
021211   & 1.006 & $ -21.89\pm1.38 $     & $ -21.82\pm0.34 $     & $ 10.32\pm0.63 $      & $ 0.339\pm0.138 $     & 2979 \\ 
030328   & 1.520 & $ -20.56\pm0.40 $     & $ -21.20\pm0.43 $     & $ 8.83\pm0.52 $       & $ 0.061\pm0.086 $     & 137 \\ 
030329   & 0.168 & $ -16.69\pm0.12 $     & $ -17.11\pm0.33 $     & $ 7.74\pm0.06 $       & $ 0.100\pm0.014 $     & 1281 \\ 
030528   & 0.782 & $ -20.80\pm0.40 $     & $ -21.37\pm0.33 $     & $ 8.82\pm0.39 $       & $ 0.034\pm0.024 $     & 154 \\ 
031203   & 0.105 & $ -19.87\pm0.04 $     & $ -21.11\pm0.11 $     & $ 8.82\pm0.43 $       & $ 0.091\pm0.085 $     & 3120 \\ 
040924   & 0.859 & $ -20.11\pm0.78 $     & $ -20.52\pm0.72 $     & $ 9.20\pm0.37 $       & $ 0.142\pm0.104 $     & 789 \\ 
041006   & 0.712 & $ -19.04\pm0.72 $     & $ -19.41\pm0.56 $     & $ 8.66\pm0.87 $       & $ 0.199\pm0.262 $     & 1361 \\ 
050223   & 0.584 & $ -21.38\pm0.14 $     & $ -21.60\pm0.80 $     & $ 9.73\pm0.36 $       & $ 0.162\pm0.091 $     & 1211 \\ 
050416   & 0.653 & $ -21.27\pm1.50 $     & $ -20.54\pm0.96 $     & $ 9.84\pm0.74 $       & $ 0.236\pm0.140 $     & 2961 \\ 
050509B  & 0.225 & $ -24.22\pm0.08 $     & $ -23.45\pm0.09 $     & $ 11.08\pm0.03 $      & $ 0.209\pm0.003 $     & 865 \\ 
050709   & 0.161 & $ -18.43\pm0.18 $     & $ -17.97\pm0.23 $     & $ 8.66\pm0.07 $       & $ 0.166\pm0.038 $     & 975 \\ 
050724   & 0.257 & $ -23.74\pm0.05 $     & $ -23.25\pm0.16 $     & $ 10.64\pm0.05 $      & $ 0.120\pm0.017 $     & 1161 \\ 
050826   & 0.296 & $ -21.07\pm0.22 $     & $ -20.40\pm0.11 $     & $ 9.79\pm0.11 $       & $ 0.196\pm0.022 $     & 951 \\ 
050904\tablenotemark{a}   & 6.295 & $ >-23.8 $  & $ >-24.3 $  & $ <10.0 $ & $ . . . $     & . . . \\ 
051022   & 0.807 & $ -22.73\pm0.15 $     & $ -23.34\pm0.14 $     & $ 10.42\pm0.18 $      & $ 0.188\pm0.079 $     & 2974 \\ 
051221   & 0.546 & $ -20.06\pm0.80 $     & $ -20.21\pm0.39 $     & $ 8.61\pm0.64 $       & $ 0.056\pm0.076 $     & 792 \\ 
060218   & 0.034 & $ -16.35\pm0.35 $     & $ -16.13\pm0.13 $     & $ 7.78\pm0.08 $       & $ 0.150\pm0.039 $     & 5297 \\ 
060505   & 0.089 & $ -20.20\pm0.03 $     & $ -19.39\pm0.03 $     & $ 9.41\pm0.01 $       & $ 0.181\pm0.010 $     & 910 \\ 
060614   & 0.125 & $ -16.42\pm0.15 $     & $ -16.12\pm0.56 $     & $ 7.95\pm0.13 $       & $ 0.210\pm0.052 $     & 770 \\ 
061006   & 0.438 & $ -21.57\pm0.56 $     & $ -19.40\pm0.10 $     & $ 10.43\pm0.23 $      & $ 0.536\pm0.053 $     & 7009 \\ 
061126   & 1.159 & $ -22.50\pm1.09 $     & $ -22.92\pm1.05 $     & $ 10.31\pm0.47 $      & $ 0.184\pm0.068 $     & 760 \\ 
\enddata
\tablecomments{Absolute magnitudes $M_K$ and $M_B$ are estimated in the AB system, and are corrected for dust attenuation in the host. Errors on stellar mass are  $< 0.1$ $dex$ are unreliable, as they reflect outlying points in parameter space where the model grid is sparse compared to the photometric errors. For these objects, a more realistic error is 0.1 dex}.
\tablenotetext{a}{Conservative limits, based on non detections with HST and Spitzer (Berger et al.\ 2007b),  are derived assuming a young stellar population.}
\end{deluxetable}

\begin{deluxetable}{lcccccccc}
\tablecaption{Star Formation Rates\label{tsfr}}
\tablecolumns{9}
\tablewidth{0pc}
\tabletypesize{\small}
\tablehead{
  \colhead{GRB}  & 
  \colhead{$z$} & 
  \colhead{$\log M_\ast$} & 
  \colhead{SFR$_{\rm H\alpha}$\tablenotemark{a}} & 
  \colhead{SFR$_{\rm [OII]}$\tablenotemark{b}} & 
    \colhead{SFR$_{\rm H\beta}$\tablenotemark{c}} & 
  \colhead{SFR$_{2800}$\tablenotemark{d}} & 
  \colhead{SFR\tablenotemark{e}} & 
  \colhead{$\log$ SSFR} \\
  \colhead{}& 
  \colhead{}& 
  \colhead{[M$_\odot$]} & 
  \colhead{} & 
  \colhead{} & 
  \colhead{} &
  \colhead{} & 
  \colhead{(Adopted)}  & 
  \colhead{[Gyr$^{-1}$]}
  }
  \startdata
970228 & 0.695 	 & $ 8.65\pm0.05$ 	 & . . . 	 & 0.53 	 & . . .	 & 0.18 	 & 0.53 	 & 0.082 \\ 
970508 & 0.835 	 & $ 8.52\pm0.10$ 	 & . . . 	 & 1.14 	 & . . . 	 & 0.35 	 & 1.14 	 & 0.534 \\ 
970828 & 0.960 	 & $ 9.19\pm0.36$ 	 & . . .   	 & 0.87 	 & . . . 	 & 1.51 	 & 0.87 	 & -0.246 \\ 
971214 & 3.420 	 & $ 9.59\pm0.40$ 	 & . . . 	 & . . . 	 & . . . 	 & 11.40 	 & 11.40 	 & 0.467 \\ 
980425 & 0.009 	 & $ 9.21\pm0.52$ 	 & 0.21 	 & 0.19 	 & . . . 	 & 1.54 	 & 0.21 	 & -0.883 \\ 
980613 & 1.097 	 & $ 8.49\pm0.21$ 	 & . . . 	 & 4.70 	 & . . . 	 & 5.85 	 & 4.70 	 & 1.184 \\ 
980703 & 0.966 	 & $ 9.33\pm0.36$ 	 & . . . 	 & 16.57 	 & 7.03 	 & 20.22 	 & 16.57 	 & 0.885 \\ 
990123 & 1.600 	 & $ 9.42\pm0.49$ 	 & . . . 	 & . . . 	 & . . . 	 & 5.72 	 & 5.72 	 & 0.340 \\ 
990506 & 1.310 	 & $ 9.48\pm0.18$ 	 & . . . 	 & 2.50 	 & . . . 	 & 2.51 	 & 2.50 	 & -0.081 \\ 
990705 & 0.842 	 & $ 10.20\pm0.76$ 	 & . . . 	 & 6.96 	 & . . . 	 & 5.86 	 & 6.96 	 & -0.357 \\ 
990712 & 0.433 	 & $ 9.29\pm0.02$ 	 & 2.39 	 & 3.01 	 & . . . 	 & 0.76 	 & 2.39 	 & 0.093 \\ 
991208 & 0.706 	 & $ 8.53\pm0.37$ 	 & . . . 	 & 4.52 	 & 19.63 	 & 1.20 	 & 4.52 	 & 1.121 \\ 
000210 & 0.846 	 & $ 9.31\pm0.08$ 	 & . . . 	 & 2.28 	 & . . . 	 & 1.34 	 & 2.28 	 & 0.049 \\ 
000418 & 1.118 	 & $ 9.26\pm0.14$ 	 & . . . 	 & 10.35 	 & . . . 	 & 7.29 	 & 10.35 	 & 0.757 \\ 
000911 & 1.058 	 & $ 9.32\pm0.26$ 	 & . . . 	 & 1.57 	 & . . . 	 & 0.75 	 & 1.57 	 & -0.124 \\ 
000926 & 2.036 	 & $ 9.52\pm0.84$ 	 & . . . 	 & . . . 	 & . . . 	 & 2.28 	 & 2.28 	 & -0.165 \\ 
010222 & 1.480 	 & $ 8.82\pm0.26$ 	 & . . . 	 & . . . 	 & . . . 	 & 0.34 	 & 0.34 	 & -0.290 \\ 
010921 & 0.451 	 & $ 9.69\pm0.13$ 	 & 2.50 	 & 4.26 	 & . . . 	 & 1.94 	 & 2.50 	 & -0.289 \\ 
011121 & 0.362 	 & $ 9.81\pm0.17$ 	 & 2.24 	 & 2.65 	 & . . . 	 & 3.61 	 & 2.24 	 & -0.464 \\ 
011211 & 2.141 	 & $ 9.77\pm0.47$ 	 & . . . 	 & . . . 	 & . . . 	 & 4.90 	 & 4.90 	 & -0.084 \\ 
020405 & 0.691 	 & $ 9.75\pm0.25$ 	 & . . . 	 & 3.74 	 & 4.29 	 & 5.55 	 & 3.74 	 & -0.174 \\ 
020813 & 1.255 	 & $ 8.66\pm1.41$ 	 & . . . 	 & 6.76 	 & . . . 	 & 1.60 	 & 6.76 	 & 1.167 \\ 
020819B & 0.410 	 & $ 10.50\pm0.14$ 	 & . . . 	 & . . . 	 & . . . 	 & 6.86 	 & 6.86 	 & -0.664 \\ 
020903 & 0.251 	 & $ 8.87\pm0.07$ 	 & 2.65 	 & 2.51 	 & . . . 	 & 1.20 	 & 2.65 	 & 0.555 \\ 
021004 & 2.327 	 & $ 10.20\pm0.18$ 	 & . . . 	 & . . . 	 & . . . 	 & 3.12 	 & 3.12 	 & -0.705 \\ 
021211 & 1.006 	 & $ 10.32\pm0.63$ 	 & . . . 	 & . . . 	 & . . . 	 & 3.01 	 & 3.01 	 & -0.841 \\ 
030328 & 1.520 	 & $ 8.83\pm0.52$ 	 & . . . 	 & . . . 	 & . . . 	 & 3.20 	 & 3.20 	 & 0.680 \\ 
030329 & 0.168 	 & $ 7.74\pm0.06$ 	 & 0.11 	 & 0.09 	 & . . . 	 & 0.21 	 & 0.11 	 & 0.304 \\ 
030528 & 0.782 	 & $ 8.82\pm0.39$ 	 & . . . 	 & 15.07 	 & 9.96 	 & 10.94 	 & 15.07 	 & 1.355 \\ 
031203 & 0.105 	 & $ 8.82\pm0.43$ 	 & 12.68 	 & 4.08 	 & . . . 	 & 9.53 	 & 12.68 	 & 1.287 \\ 
040924 & 0.859 	 & $ 9.20\pm0.37$ 	 & . . . 	 & 1.88 	 & 0.85 	 & 1.73 	 & 1.88 	 & 0.071 \\ 
041006 & 0.712 	 & $ 8.66\pm0.87$ 	 & . . . 	 & 0.34 	 & . . . 	 & 1.21 	 & 0.34 	 & -0.131 \\ 
050223 & 0.584 	 & $ 9.73\pm0.36$ 	 & . . . 	 & 1.44 	 & 1.65 	 & 8.33 	 & 1.44 	 & -0.568 \\ 
050416 & 0.653 	 & $ 9.84\pm0.74$ 	 & . . . 	 & 2.32 	 & 1.85 	 & 2.29 	 & 2.32 	 & -0.476 \\ 
050509B & 0.225  & $ 11.08\pm0.03$ 	 & . . . 	 & . . . 	 & . . . 	 & 16.87 	 & 16.87 	 & -0.853 \\ 
050709 & 0.161 	 & $ 8.66\pm0.07$ 	 & 0.14 	 & . . . 	 & . . . 	 & 0.21 	 & 0.14 	 & -0.512 \\ 
050724 & 0.257 	 & $ 10.64\pm0.05$ 	 & . . . 	 & . . . 	 & . . . 	 & 18.76 	 & 18.76 	 & -0.367 \\ 
050826 & 0.296 	 & $ 9.79\pm0.11$ 	 & . . . 	 & 9.13 	 & . . . 	 & 1.04 	 & 9.13 	 & 0.172 \\ 
%050904 & 6.295 	 & $ 9.73\pm0.31$ 	 & . . . 	 & . . . 	 & . . . & 3.25 & 3.25 & -0.218 \\ 
051022 & 0.807 	 & $ 10.42\pm0.18$ 	 & . . . 	 & 36.46 	 & 58.19 	 & 28.69 	 & 36.46 	 & 0.142 \\ 
051221 & 0.546 	 & $ 8.61\pm0.64$ 	 & . . . 	 & 2.61 	 & 2.11 	 & 4.08 	 & 2.61 	 & 0.804 \\ 
060218 & 0.034 	 & $ 7.78\pm0.08$ 	 & 0.05 	 & 0.06 	 & . . . 	 & 0.08 	 & 0.05 	 & -0.061 \\ 
060505 & 0.089 	 & $ 9.41\pm0.01$ 	 & 0.43 	 & 0.74 	 & . . . 	 & 1.96 	 & 0.43 	 & -0.777 \\ 
060614 & 0.125 	 & $ 7.95\pm0.13$ 	 & 0.01 	 & . . . 	 & . . . 	 & 0.02 	 & 0.01 	 & -0.863 \\ 
061006 & 0.438 	 & $ 10.43\pm0.23$ 	 & . . . 	 & . . . 	 & . . . 	 & 0.17 	 & 0.17 	 & -2.189 \\ 
061126 & 1.159 	 & $ 10.31\pm0.47$ 	 & . . . 	 & 2.38 	 & . . . 	 & 8.89 	 & 2.38 	 & -0.934 \\ 
\enddata
\tablecomments{SFRs, in units of M$_\odot$ yr$^{-1}$,  are corrected for aperture slit loss and dust extinction. }
\tablenotetext{a}{SFR from H$\alpha$ emission lines.}
\tablenotetext{b}{SFR from [OII] emission lines.}
\tablenotetext{c}{SFR from H$\beta$ emission lines.}
\tablenotetext{d}{SFR from UV luminosities at $\lambda =2800$ \AA.}
\tablenotetext{e}{Final adopted SFR.}
\end{deluxetable}

\begin{table}
\small
\caption[t1]{$T_e$ metallicities}
\begin{center} 
\begin{tabular}{lcccccccc} 
\tableline\tableline&&&&&&&&\\[-5pt]
 GRB  & $z$ & [OIII]$\lambda4363$ & [OIII]$\lambda4959$ & [OIII]$\lambda5007$ & $T_e$ & O$^+$/H$^+$ &
O$^{2+}$/H$^+$ & $12+ \log (\rm O/H)$ \\
 & &  & & & ($10^4$ K) & ($10^{-5}$) & ($10^{-5}$) & \\
[5pt]\tableline&&&&&&&&\\[-5pt]
      980425   &    0.0085  &       127   &     3102  & 12280  &     1.20   &   4.31 & 10.23   &    8.16 \\
      020903   &    0.251  &     2.04  &     65.6  &   208.2 &      1.16 &   5.22 &  11.46   &    8.22 \\
      031203   &    0.105  &     164.1  &      3704  & 11240  &     1.33  &   1.06 &    9.50 &      8.02  \\
      060218   &   0.0338 &      38.2\tablenotemark{a} &      234 &   713   &    2.77    &   0.527 &  1.40 & 7.29\tablenotemark{a} \\
[2pt]\tableline
\end{tabular}
\end{center}
\tablecomments{Emission lines fluxes, in units of $10^{-17}$ erg cm$^{-2}$ s$^{-1}$, are corrected for dust extinction in the host.}
\tablenotetext{a}{The detection of [OIII]$\lambda4363$ is dubious, the flux can be considered an upper limit, and the
metallicity a lower limit.}
\label{tte}
\end{table}
 
\begin{table}
\caption[t1]{O3N2 metallicities}
\begin{center}
\begin{tabular}{lccc}
\tableline\tableline&&\\[-5pt]
 GRB  & $z$ & O3N2\tablenotemark{a}  & $12+ \log {\rm (O/H)}$ \\
[5pt]\tableline&&&\\[-5pt] 
      980425 &  0.0085 &    1.94  &     $8.1\pm0.5$ \\
      990712  &  0.434 &    $>1.29$   &   $<8.3$ \\
     020903    &  0.251 &  2.39   &    $8.0\pm0.5$ \\
      030329 &   0.168 &    $>1.81$   &    $<8.2$ \\
      031203 &   0.1055 &   2.07   &    $8.1\pm0.5$ \\
      060218 &   0.03345 &  1.87   &     $8.13\pm0.25$ \\
      060505 &   0.0889 &  0.92  &     $8.44\pm0.25$ \\
[2pt]\tableline
\end{tabular}
\end{center}
\tablecomments{Quantities are all corrected for  dust extinction and stellar absorption. The uncertainties on metallicities are $\pm0.25$ dex for $-1<O3N2<1.9$, and 0.5 for larger O3N2 values (Pettini \& Pagel 2004).}
\tablenotetext{a}{$\rm O3N2 = \log \{([OIII]5007 / H\beta)/([NII]6583 / H\alpha)\}$.}
\label{to3n2}
\end{table}

\begin{deluxetable}{lcccccccccc}
%\rotate
\tablecaption{GRB host metallicities\label{tr23}}
\tablecolumns{11}
\tablewidth{0pc}
\tabletypesize{\footnotesize}
\tablehead{
  \colhead{GRB}  & 
  \colhead{$\log R_{23}$}  & 
  \colhead{$\log O_{32}$}  &
  \colhead{} & 
  \colhead{} & 
  \colhead{$12+\log(\rm O/H)$} &
  \colhead{} & 
  \colhead{} & 
\colhead{} & 
  \colhead{$\log$([NII]/[OII])} & 
  \colhead{$\log$([NII]/H$\alpha$)} \\
  [4pt]\cline{4-9}\\[-4pt] 
 \colhead{} &
 \colhead{} &
  \colhead{} &
  \colhead{Lower KD02\tablenotemark{a} } &
  \colhead{Upper KK04\tablenotemark{b}} &
  \colhead{Lower N06\tablenotemark{c} } &
 \colhead{O3N2\tablenotemark{d}} &
  \colhead{$T_e$\tablenotemark{e}} &
 \colhead{Adopted\tablenotemark{f}} &
 \colhead{} &
 \colhead{}
 }
\startdata 
980425 & 0.960 & $ 0.550 $  & . . .    & . . .    & $\sim8.1$ & 8.1        & 8.16        & 8.16           & $-1.06$ & $-1.21$  \\ 
980703 & 0.840 & $ -0.529 $ & . . .    & 8.14 & 7.6             & . . .        & . . .           & 7.6/8.14   & . . . & . . . \\ 
990712 & 0.932 & $ 0.302 $  & . . .    & . . .    & $\sim8.1$ & $<8.3$ &  . . .          & 8.1              & $<-0.66$ & $<-0.66$ \\ 
991208 & 0.330 & $ 0.491 $  & . . .    & 8.73 & $<7.4$      & . . .         & . . .           &$<7.4$/8.73 & . . . & . . . \\ 
010921 & 0.857 & $ -0.064 $ & . . .    & 8.15 & 8.0            & . . .         & . . .           & 8.0/8.15     & . . . & . . . \\ 
011121 & 0.566 & $ -0.429 $ & 7.50 & 8.64 & . . .             & . . .         & . . .           & 7.50/8.64  & . . . & . . . \\ 
020405 & 0.759 & $ 0.279 $  & 7.78 & 8.44 & . . .             & . . .         & . . .            & 7.78/8.44  & . . . & . . . \\ 
020903 & 0.957 & $ 0.508 $  & . . .    & . . .    & $\sim8.1$ & 8.0        & 8.22         & 8.22           & $-1.55$ & $-1.67$ \\ 
030329 & 0.820 & $ 0.430 $  & 7.97 & 8.33 & . . .             & $<8.2$  & . . .            & 7.97           & $<-1.08$ & $<-1.25$ \\ 
030528 & 0.935 & $ 0.179 $  & . . .    & . . .    & $\sim8.1$ & . . .        &  . . .           & 8.1             & . . . & . . . \\ 
031203 & 0.965 & $ 1.067 $  & 8.25 & . . .    & . . .             & 8.1        & 8.02          & 8.02          & $-0.68$ & $-1.27$ \\ 
050223 & 0.536 & $ -0.135 $ & . . .    & 8.66 & 7.5            & . . .         &  . . .            & 7.5/8.66 & . . . &. . . \\ 
050416 & 0.741 & $ -0.029 $ & 7.97 & 8.44 & . . .            & . . .         & . . .             & 7.97/8.44 & . . . & . . . \\ 
051022 & 0.556 & $ 0.186 $  & . . .    & 8.65 & 7.5            & . . .         & . . .             & 7.5/8.65 & . . . & . . . \\ 
051221 & 0.614 & $ -0.336 $ & . . .    & 8.59 & 7.6            & . . .         & . . .             & 7.6/8.59 & . . . & . . . \\ 
060218 & 0.927 & $ 0.396 $  & . . .    & . . .    & $\sim8.1$ & 8.13      & 7.29\tablenotemark{g} & 8.13          & $-1.15$ & $-1.22$ \\ 
060505 & 0.770 & $ -0.292 $ & . . .    & 8.37 & 7.8            & 8.44      &  . . .            & 8.44          & $-0.88$ & $-0.75$ \\ 
\enddata
\tablecomments{Quantities are corrected for  dust extinction in the host. No solution was found for the host of GRB~040924, so this is not included in the table.}
\tablenotetext{a}{Derived using the $R_{23}$ lower branch solution, as defined by Kewley \& Dopita (2002), 
and converted to O3N2 metallicity, according to Kewley \& Ellison (2008). }
\tablenotetext{b}{Derived using the $R_{23}$  upper branch solution, as defined by Kobulnicky \& Kewley (2004), 
and converted to O3N2 metallicity, according to Kewley \& Ellison (2008).}
\tablenotetext{c}{Derived using the $R_{23}$  lower branch solution, as defined by Nagao et al.\ (2006), 
and converted to O3N2 metallicity, as described in \S~\ref{nagao}.}
\tablenotetext{d}{Derived using the O3N2 prescription of Pettini \& Pagel (2004). Uncertainty is 0.25 or 0.5, for $12+\log (\rm O/H) >8.1$ or $12+\log (\rm O/H) \leq8.1$, respectively.}
\tablenotetext{e}{Derived using the electron temperature $T_e$ prescription of Izotov et al.\ (2006).}
\tablenotetext{f}{Final adopted metallicity. For a subsample of GRB hosts, both the lower and upper branch solutions are considered.}
\tablenotetext{g}{This value is likely a lower limit, due to the dubious detection of the [OIII]$\lambda4363$ line.}
\end{deluxetable}

 \begin{table}
\caption[t1]{Electron gas density}
\begin{center}
\begin{tabular}{lcccccc}
\tableline\tableline&&&&&&\\[-5pt]
 GRB  & redshift & [OII]$\lambda3726$\tablenotemark{a} & [OII]$\lambda3729$\tablenotemark{a} & ratio\tablenotemark{b} & Ref & n$_e$\tablenotemark{c} \\
 	   &		  &   & & & & ($10^2$ cm$^{-3}$) \\
[5pt]\tableline&&&&&&\\[-5pt] 
990506 & 1.31 & $ 1.69\pm0.29$ & $ 3.01\pm0.27$ & 0.56 & [1] & 13.5 \\ 
000418 & 1.118 &  12.23 &  16.59 & 0.74 & [1] & 7.5 \\ 
020405 & 0.691 & $ 15.7\pm1.3$ & $ 18.2\pm1.0$ & 0.86 & [2] & 4.8 \\ 
020903 & 0.251 &  98.8 &  137.7 & 0.72 & [3] & 8.2 \\ 
030329 & 0.168 & $ 10.6\pm2.3$ & $ 15.8\pm2.2$ & 0.67 & [4] & 10.0 \\ 
060218 & 0.03345 & $ 147.1\pm6.7$ & $ 236.9\pm7.8$ & 0.62 & [5] & 12.6 \\ 
[2pt]\tableline
\end{tabular}
\end{center}
\tablenotetext{a}{[OII] fluxes, in units of $10^{-17}$ erg s$^{-1}$ cm$^{-2}$, are corrected for dust extinction and slit-aperture flux loss.}
\tablenotetext{b}{[OII]$\lambda3726$-to-[OII]$\lambda3729$ line ratio.}
\tablenotetext{c}{Electron gas density.}
\tablecomments{References: 
[1] Bloom et al.\ (2003) [2] Price et al.\ (2003) [3] Soderberg et al.\ (2004) [4] Th{\"o}ne et al.\ (2007) [5] Wiersema et al.\ (2007).}
\label{tne}
\end{table}

\clearpage

\begin{deluxetable}{lcccccccc}
%\rotate
\tablecaption{GRB host summary\label{tsummary}}
\tablecolumns{9}
\tablewidth{0pc}
\tabletypesize{\footnotesize}
\tablehead{
  \colhead{GRB}  & 
  \colhead{type\tablenotemark{a}} & 
  \colhead{redshift} & 
  \colhead{$M_K$\tablenotemark{b}}  & 
  \colhead{$\log M_\ast$\tablenotemark{c}} & 
  \colhead{$A_V$\tablenotemark{d}} &
  \colhead{SFR} & 
  \colhead{$\log$ SSFR\tablenotemark{e}} & 
  \colhead{$12+\log (\rm O/H)$} \\
 \colhead{} &
 \colhead{} &
 \colhead{} &
 \colhead{} &
 \colhead{[M$_\odot$]} &
 \colhead{}  & 
 \colhead{(M$_\odot$ yr$^{-1}$)} & 
 \colhead{[Gyr$^{-1}$]}
 }
\startdata 
970228 & long & 0.695 	 & $ -17.78\pm0.09$ 	 & $ 8.65\pm0.05$ 	 & . . . 	 & 0.53 	 & 0.082 	 & . . . \\
970508 & long & 0.835 	 & $ -17.88\pm0.24$ 	 & $ 8.52\pm0.10$ 	 & . . . 	 & 1.14 	 & 0.534 	 & . . . \\
970828 & long & 0.960 	 & $ -20.21\pm0.39$ 	 & $ 9.19\pm0.36$ 	 & . . . 	 & 0.87 	 & -0.246 	 & . . . \\
971214 & long & 3.420 	 & $ -22.20\pm0.72$ 	 & $ 9.59\pm0.40$ 	 & . . . 	 & 11.40 	 & 0.467 	 & . . . \\
980425 & long & 0.009 	 & $ -19.44\pm1.26$ 	 & $ 9.21\pm0.52$ 	 & 1.73 	 & 0.21 	 & -0.883 	 & 8.16 \\
980613 & long & 1.097 	 & $ -19.99\pm0.03$ 	 & $ 8.49\pm0.21$ 	 & . . . 	 & 4.70 	 & 1.184 	 & . . . \\
980703 & long & 0.966 	 & $ -22.04\pm0.15$ 	 & $ 9.33\pm0.36$ 	 & . . . 	 & 16.57 	 & 0.885 	 & 7.6/8.14\tablenotemark{f} \\
990123 & long & 1.600 	 & $ -20.99\pm0.79$ 	 & $ 9.42\pm0.49$ 	 & . . . 	 & 5.72 	 & 0.340 	 & . . . \\
990506 & long & 1.310 	 & $ -21.10\pm0.15$ 	 & $ 9.48\pm0.18$ 	 & . . . 	 & 2.50 	 & -0.081 	 & . . . \\
990705 & long & 0.842 	 & $ -22.37\pm1.06$ 	 & $ 10.20\pm0.76$ 	 & . . . 	 & 6.96 	 & -0.357 	 & . . . \\
990712 & long & 0.433 	 & $ -19.31\pm0.09$ 	 & $ 9.29\pm0.02$ 	 & $0.39\pm0.09$ & 2.39 	 & 0.093 	 & $\sim8.1$ \\
991208 & long & 0.706 	 & $ -18.88\pm0.23$ 	 & $ 8.53\pm0.37$ 	 & . . . 	 & 4.52 	 & 1.121 	 & $<7.4$/8.73\tablenotemark{f} \\
000210 & long & 0.846 	 & $ -19.75\pm0.23$ 	 & $ 9.31\pm0.08$ 	 & . . . 	 & 2.28 	 & 0.049 	 & . . . \\
000418 & long & 1.118 	 & $ -20.47\pm0.21$ 	 & $ 9.26\pm0.14$ 	 & . . . 	 & 10.35 	 & 0.757 	 & . . . \\
000911 & long & 1.058 	 & $ -19.70\pm0.51$ 	 & $ 9.32\pm0.26$ 	 & . . . 	 & 1.57 	 & -0.124 	 & . . . \\
000926 & long & 2.036 	 & $ -21.36\pm1.13$ 	 & $ 9.52\pm0.84$ 	 & . . . 	 & 2.28 	 & -0.165 	 & . . . \\
010222 & long & 1.480 	 & $ -18.42\pm0.65$ 	 & $ 8.82\pm0.26$ 	 & . . . 	 & 0.34 	 & -0.290 	 & . . . \\
010921 & long & 0.451 	 & $ -20.37\pm0.05$ 	 & $ 9.69\pm0.13$ 	 & $1.06\pm0.62$ & 2.50 	 & -0.289 	 & 8.0/8.15\tablenotemark{f} \\
011121 & long & 0.362 	 & $ -21.01\pm0.38$ 	 & $ 9.81\pm0.17$ 	 & 0.38 	 & 2.24 	 & -0.464 	 & 7.50/8.64\tablenotemark{f} \\
011211 & long & 2.141 	 & $ -21.62\pm1.12$ 	 & $ 9.77\pm0.47$ 	 & . . . 	 & 4.90 	 & -0.084 	 & . . . \\
020405 & long & 0.691 	 & $ -21.16\pm1.07$ 	 & $ 9.75\pm0.25$ 	 & . . . 	 & 3.74 	 & -0.174 	 & 7.78/8.44\tablenotemark{f} \\
020813 & long & 1.255 	 & $ -19.32\pm2.36$ 	 & $ 8.66\pm1.41$ 	 & . . . 	 & 6.76 	 & 1.167 	 & . . . \\
020819B & long & 0.410 	 & $ -22.53\pm0.30$ 	 & $ 10.50\pm0.14$ 	 & . . . 	 & 6.86 	 & -0.664 	 & . . . \\
020903 & long & 0.251 	 & $ -18.97\pm0.33$ 	 & $ 8.87\pm0.07$ 	 & 0.59 	 & 2.65 	 & 0.555 	 & 8.22 \\
021004 & long & 2.327 	 & $ -22.06\pm0.38$ 	 & $ 10.20\pm0.18$ 	 & . . . 	 & 3.12 	 & -0.705 	 & . . . \\
021211 & long & 1.006 	 & $ -21.89\pm1.38$ 	 & $ 10.32\pm0.63$ 	 & . . . 	 & 3.01 	 & -0.841 	 & . . . \\
030328 & long & 1.520 	 & $ -20.56\pm0.40$ 	 & $ 8.83\pm0.52$ 	 & . . . 	 & 3.20 	 & 0.680 	 & . . . \\
030329 & long & 0.168 	 & $ -16.69\pm0.12$ 	 & $ 7.74\pm0.06$ 	 & $<0.1$ 	 & 0.11 	 & 0.304 	 & 7.97 \\
030528 & long & 0.782 	 & $ -20.80\pm0.40$ 	 & $ 8.82\pm0.39$ 	 & . . . 	 & 15.07 	 & 1.355 	 & $\sim8.1$ \\
031203 & long & 0.105 	 & $ -19.87\pm0.04$ 	 & $ 8.82\pm0.43$ 	 & $0.03\pm0.05$ & 12.68 	 & 1.287 	 & 8.02 \\
040924 & long & 0.859 	 & $ -20.11\pm0.78$ 	 & $ 9.20\pm0.37$ 	 & . . . 	 & 1.88 	 & 0.071 	 & . . . \\
041006 & long & 0.712 	 & $ -19.04\pm0.72$ 	 & $ 8.66\pm0.87$ 	 & . . . 	 & 0.34 	 & -0.131 	 & . . . \\
050223 & long & 0.584 	 & $ -21.38\pm0.14$ 	 & $ 9.73\pm0.36$ 	 & . . . 	 & 1.44 	 & -0.568 	 & 7.5/8.66\tablenotemark{f} \\
050416 & short & 0.653 	 & $ -21.27\pm1.50$ 	 & $ 9.84\pm0.74$ 	 & . . . 	 & 2.32 	 & -0.476 	 & 7.97/8.44\tablenotemark{f} \\
050509B & short & 0.225  & $ -24.22\pm0.08$ 		 & $ 11.08\pm0.03$ 	 & . . . 	 & 16.87 	 & -0.853 	 & . . . \\
050709 & short & 0.161 	 & $ -18.43\pm0.18$ 	 & $ 8.66\pm0.07$ 	 & $\sim0$ 	 & 0.14 	 & -0.512 	 & . . . \\
050724 & short & 0.257 	 & $ -23.74\pm0.05$ 	 & $ 10.64\pm0.05$ 	 & . . . 	 & 18.76 	 & -0.367 	 & . . . \\
050826 & long & 0.296 	 & $ -21.07\pm0.22$ 	 & $ 9.79\pm0.11$ 	 & . . . 	 & 9.13 	 & 0.172 	 & . . . \\
050904 & long & 6.295 	 & $ >-23.8$ 	                   & $<10.0$                   & . . .  & . . . & . . . & . . . \\
051022 & long & 0.807 	 & $ -22.73\pm0.15$ 	 & $ 10.42\pm0.18$ 	 & . . . 	 & 36.46 	 & 0.142 	 & 7.5/8.65\tablenotemark{f} \\
051221 & short & 0.546 	 & $ -20.06\pm0.80$ 	 & $ 8.61\pm0.64$ 	 & . . . 	 & 2.61 	 & 0.804 	 & 7.6/8.59\tablenotemark{f} \\
060218 & long & 0.034 	 & $ -16.35\pm0.35$ 	 & $ 7.78\pm0.08$ 	 & $0.49\pm0.24$ & 0.05 	 & -0.061 	 & 8.13 \\
060505 & long? & 0.089 	 & $ -20.20\pm0.03$ 	 & $ 9.41\pm0.01$ 	 & $0.63\pm0.01$ & 0.43 	 & -0.777 	 & 8.44 \\
060614 & long & 0.125 	 & $ -16.42\pm0.15$ 	 & $ 7.95\pm0.13$ 	 & . . . 	 & 0.01 	 & -0.863 	 & . . . \\
061006 & short & 0.438 	 & $ -21.57\pm0.56$ 	 & $ 10.43\pm0.23$ 	 & . . . 	 & 0.17 	 & -2.189 	 & . . . \\
061126 & long & 1.159 	 & $ -22.50\pm1.09$ 	 & $ 10.31\pm0.47$ 	 & . . . 	 & 2.38 	 & -0.934 	 & . . . \\
[2pt]\tableline&&&&&&&&\\[-8pt] 
Median\tablenotemark{g} & long & 0.75 &  $-20.5$	& 9.3	& 0.44 & 2.5 & -0.10 &  7.9/8.3\tablenotemark{h}\\	
\enddata
\tablenotetext{a}{GRB type as defined by its duration, long or short (longer or shorter than $\sim2$ seconds). Long or short GRBs are associated with core-collapse SNe or merger of compact objects, respectively.}
\tablenotetext{b}{$K$-band AB absolute magnitude.}
\tablenotetext{c}{GRB host total stellar mass.}
\tablenotetext{d}{Dust extinction in the visual band, as derived from Balmer decrement.}
\tablenotetext{e}{Specific Star Formation Rate SFR/$M_\ast$.}
\tablenotetext{f}{Two values possible, see text and Table~\ref{tr23}.}
\tablenotetext{g}{Median value for each parameter.}
\tablenotetext{h}{Median metallicity in the case that the lower or upper branch solution is considered for a subsample.}
\end{deluxetable}


\begin{thebibliography}{138}

\bibitem[{{}(2004)}]{}
{Abraham}, R.\ G.\ et al.\ 2004, \aj, 127, 2455
\bibitem[{{}(1979)}]{}
Alloin, D., Collin-Souffrin, S., Joly, M., \& Vigroux, L.\ 1979, \aap, 78, 200
\bibitem[Aoki et al.(2006)]{2006astro.ph.12367A} 
Aoki, K., Furusawa, H., Ohta, K., Yamada, T., \& Kawai, N.\ 2007, ArXiv Astrophysics e-prints, arXiv:astro-ph/0612367


\bibitem[Aoki et al.(2006)]{2006NCimB.121.1427A} Aoki, K., Furusawa, H., 
Ohta, K., Yamada, T., \& Kawai, N.\ 2006, Nuovo Cimento B Serie, 121, 1427 

\bibitem[{{}(2005)}]{}  
Asplund, M., Grevesse, N., Sauval, A.~J.\ 2005, ASP  Conf.~Ser.~336: Cosmic Abundances as Records of Stellar Evolution and Nucleosynthesis, 25
\bibitem[{{}()}]{}
Baldry, I.~K.~\& Glazebrook, K.\ 2003, \apj, 593, 258
\bibitem[Baldry et al.(2008)]{2008MNRAS.388..945B} 
Baldry, I.~K., Glazebrook, K., \& Driver, S.~P.\ 2008, \mnras, 388, 945
\bibitem[Barth et al.(2003)]{2003ApJ...584L..47B} 
Barth, A.~J., et al.\  2003, \apjl, 584, L47
\bibitem[Bauer et al.(2005)]{2005ApJ...621L..89B} 
Bauer, A.~E., Drory, N., Hill, G.~J., \& Feulner, G.\ 2005, \apjl, 621, L89
\bibitem[Berger et al.(2003)]{2003ApJ...588...99B} 
Berger, E., Cowie, L.~L., Kulkarni, S.~R., Frail, D.~A., Aussel, H., \& Barger, A.~J.\ 2003, \apj, 588, 99 
\bibitem[Berger et al.(2007)]{2007ApJ...660..504B} 
Berger, E., Fox, D.~B., Kulkarni, S.~R., Frail, D.~A., \& Djorgovski, S.~G.\ 2007a, \apj, 660, 504
\bibitem[Berger et al.(2005)]{2005Natur.438..988B} 
Berger, E., et al.\ 2005, \nat, 438, 988
\bibitem[Berger et al.(2007)]{2007ApJ...665..102B} 
Berger, E., et al.\ 2007b, \apj, 665, 102 
\bibitem[Berger et al.(2007)]{2007ApJ...664.1000B} 
Berger, E., et al.\ 2007c, \apj, 664, 1000
\bibitem[Bersier et al.(2006)]{2006ApJ...643..284B} 
Bersier, D., et al.\ 2006, \apj, 643, 284
\bibitem[Bloom et al.(2003)]{2003AJ....125..999B} Bloom, J.~S., Berger, E., Kulkarni, S.~R., Djorgovski, S.~G., \& Frail, D.~A.\ 2003, \aj, 125, 999 
\bibitem[Bloom et al.(2001)]{2001ApJ...554..678B} 
Bloom, J.~S., Djorgovski, S.~G., \& Kulkarni, S.~R.\ 2001, \apj, 554, 678
\bibitem[Bloom et al.(1998)]{1998ApJ...507L..25B} 
Bloom, J.~S., Djorgovski, S.~G., Kulkarni, S.~R., \& Frail, D.~A.\ 1998, \apjl, 507, L25 
\bibitem[Bloom et al.(2002)]{2002AJ....123.1111B} 
Bloom, J.~S., Kulkarni, S.~R., \& Djorgovski, S.~G.\ 2002, \aj, 123, 1111
\bibitem[Borch et al.(2006)]{2006A&A...453..869B} 
Borch, A., et al.\ 2006, \aap, 453, 869 
\bibitem[Bresolin(2006)]{2006astro.ph..8410B} 
Bresolin, F.\ 2006, ArXiv Astrophysics e-prints, arXiv:astro-ph/0608410 
\bibitem[Bresolin(2007)]{2007ApJ...656..186B} 
Bresolin, F.\ 2007, \apj, 656, 186
\bibitem[{{}()}]{} 
Brinchmann, J., Charlot, S., White, S.~D.~M., Tremonti, C., Kauffmann, G., Heckman, T., \& Brinkmann, J.\ 2004, \mnras, 351, 1151 
\bibitem[{{}()}]{} 
Calzetti, D. 2001, PASP, 113, 1449 
\bibitem[Calzetti et al.(2000)]{2000ApJ...533..682C} 
Calzetti, D., Armus, L., Bohlin, R.~C., Kinney, A.~L., Koornneef, J., \& Storchi-Bergmann, T.\ 2000, \apj, 533, 682
\bibitem[Cardelli et al.(1989)]{1989ApJ...345..245C} 
Cardelli, J.~A., Clayton, G.~C., \& Mathis, J.~S.\ 1989, \apj, 345, 245 
\bibitem[Castro Cer{\'o}n et al.(2006)]{2006ApJ...653L..85C} 
Castro Cer{\'o}n, J.~M., Micha{\l}owski, M.~J., Hjorth, J., Watson, D., Fynbo, J.~P.~U., \& Gorosabel, J.\ 2006, \apjl, 653, L85 
\bibitem[Castro-Tirado et al.(2001)]{2001A&A...370..398C} 
Castro-Tirado, A.~J., et al.\ 2001, \aap, 370, 398
\bibitem[Castro-Tirado et al.(2005)]{2005A&A...439L..15C} 
Castro-Tirado, A.~J., et al.\ 2005, \aap, 439, L15
\bibitem[Castro-Tirado et al.(2007)]{2007A&A...475..101C} 
Castro-Tirado, A.~J., et al.\ 2007, \aap, 475, 101
\bibitem[Chary et al.(2002)]{2002ApJ...566..229C} 
Chary, R., Becklin, E.~E., \& Armus, L.\ 2002, \apj, 566, 229
\bibitem[Chary et al.(2007)]{2007ApJ...671..272C} 
Chary, R., Berger, E., \& Cowie, L.\ 2007, \apj, 671, 272 
\bibitem[{{}()}]{}
Cid Fernandes, R., Mateus, A., Sods, L., Stasi\'nska, G., Gomes, J.\ M.\ 2005, MNRAS, 358, 363
\bibitem[Cimatti et al.(2002)]{2002A&A...392..395C} 
Cimatti, A., et al.\ 2002, \aap, 392, 395
\bibitem[Christensen et al.(2004)]{2004A&A...425..913C} 
Christensen, L., Hjorth, J., \& Gorosabel, J.\ 2004, \aap, 425, 913
\bibitem[Cobb et al.(2004)]{2004ApJ...608L..93C} 
Cobb, B.~E., Bailyn, C.~D., van Dokkum, P.~G., Buxton, M.~M., \& Bloom, J.~S.\ 2004, \apjl, 608, L93
\bibitem[Cobb et al.(2006)]{2006ApJ...645L.113C} 
Cobb, B.~E., Bailyn, C.~D., van Dokkum, P.~G., \& Natarajan, P.\ 2006a, \apjl, 645, L113
\bibitem[Cobb et al.(2006)]{2006ApJ...651L..85C} 
Cobb, B.~E., Bailyn, C.~D., van Dokkum, P.~G., \& Natarajan, P.\ 2006b, \apjl, 651, L85
\bibitem[Conselice et al.(2005)]{2005ApJ...633...29C} 
Conselice, C.~J., et al.\ 2005, \apj, 633, 29
\bibitem[Covino et al.(2006)]{2006A&A...447L...5C} 
Covino, S., et al.\ 2006, \aap, 447, L5
\bibitem[Della Valle et al.(2006)]{2006Natur.444.1050D} 
Della Valle, M., et al.\ 2006, \nat, 444, 1050 
\bibitem[de Ugarte Postigo et al.(2005)]{2005A&A...443..841D} 
de Ugarte Postigo, A., et al.\ 2005, \aap, 443, 841
\bibitem[Djorgovski et al.(2003)]{2003ApJ...591L..13D} 
Djorgovski, S.~G., Bloom, J.~S., \& Kulkarni, S.~R.\ 2003, \apjl, 591, L13
\bibitem[Djorgovski et al.(2001)]{2001ApJ...562..654D} 
Djorgovski, S.~G., Frail, D.~A., Kulkarni, S.~R., Bloom, J.~S., Odewahn, S.~C., \& Diercks, A.\ 2001, \apj, 562, 654 
\bibitem[Djorgovski et al.(1998)]{1998ApJ...508L..17D} 
Djorgovski, S.~G., Kulkarni, S.~R., Bloom, J.~S., Goodrich, R., Frail, D.~A., Piro, L., \& Palazzi, E.\ 1998, \apjl, 508, L17
\bibitem[Erb et al.(2006)]{2006ApJ...647..128E} 
Erb, D.~K., Steidel, C.~C., Shapley, A.~E., Pettini, M., Reddy, N.~A., \& Adelberger, K.~L.\ 2006, 
\apj, 647, 128
\bibitem[{{}()}]{}
Fioc, M., \& Rocca-Volmerange, B.\ 1997, A\&A, 326, 950 
\bibitem[{{}()}]{}
Fioc, M., \& Rocca-Volmerange, B.\ 1999, arXiv:astro-ph/9912179
\bibitem[Frail et al.(2002)]{2002ApJ...565..829F} 
Frail, D.~A., et al.\ 2002, \apj, 565, 829
\bibitem[Fruchter et al.(2001)]{2001GCN..1087....1F} 
Fruchter, A., Burud, I., Rhoads, J., \& Levan, A.\ 2001, GRB Coordinates Netw., 1087
\bibitem[Fruchter et al.(1999)]{1999ApJ...519L..13F} 
Fruchter, A.~S., et al.\ 1999, \apjl, 519, L13
\bibitem[Fruchter et al.(2006)]{2006Natur.441..463F} 
Fruchter, A.~S., et al.\ 2006, \nat, 441, 463
\bibitem[Fynbo et al.(2003)]{2003A&A...406L..63F} 
Fynbo, J.~P.~U., et al.\ 2003, \aap, 406, L63
\bibitem[Fynbo et al.(2006)]{2006Natur.444.1047F} 
Fynbo, J.~P.~U., et al.\ 2006, \nat, 444, 1047 
\bibitem[Galama et al.(1998)]{1998Natur.395..670G} 
Galama, T.~J., et al.\ 1998, \nat, 395, 670 
\bibitem[Galama et al.(2003)]{2003ApJ...587..135G} 
Galama, T.~J., et al.\ 2003, \apj, 587, 135 
\bibitem[Gal-Yam et al.(2006)]{2006Natur.444.1053G} 
Gal-Yam, A., et al.\ 2006, \nat, 444, 1053
\bibitem[Garnavich et al.(2003)]{2003ApJ...582..924G} 
Garnavich, P.~M., et al.\ 2003, \apj, 582, 924
\bibitem[Gehrels et al.(2004)]{2004ApJ...611.1005G} 
Gehrels, N., et al.\ 2004, \apj, 611, 1005 
\bibitem[Gehrels et al.(2005)]{2005Natur.437..851G} 
Gehrels, N., et al.\ 2005, \nat, 437, 851
\bibitem[Glazebrook et al.(1999)]{1999MNRAS.306..843G}
Glazebrook, K., Blake, C., Economou, F., Lilly, S., \& Colless, M.\ 1999, \mnras, 306, 843
\bibitem[{{}()}]{}
Glazebrook, K., et al.\ 2004, \nat, 430, 181
\bibitem[Gorosabel et al.(2005)]{2005NCimC..28..677G} 
Gorosabel, J., Jel{\'{\i}}nek, M., de Ugarte Postigo, A., Guziy, S., \& Castro-Tirado, A.~J.\ 2005a, Nuovo Cimento C, 28, 677 
\bibitem[Gorosabel et al.(2005)]{2005A&A...444..711G} 
Gorosabel, J., et al.\ 2005b, \aap, 444, 711 
\bibitem[Gorosabel et al.(2006)]{2006A&A...450...87G} 
Gorosabel, J., et al.\ 2006, \aap, 450, 87
\bibitem[Hammer et al.(2006)]{2006A&A...454..103H} 
Hammer, F., Flores, H., Schaerer, D., Dessauges-Zavadsky, M., Le Floc'h, E., \& Puech, M.\ 2006, \aap, 454, 103
\bibitem[Hjorth et al.(2002)]{2002ApJ...576..113H} 
Hjorth, J., et al.\ 2002, \apj, 576, 113
\bibitem[Hjorth et al.(2003)]{2003Natur.423..847H} 
Hjorth, J., et al.\ 2003, \nat, 423, 847
\bibitem[Hjorth et al.(2005)]{2005Natur.437..859H} 
Hjorth, J., et al.\ 2005, \nat, 437, 859
\bibitem[{{}()}]{}
Holland, S., et al.\ 2000, GRB Coordinates Netw, 753
\bibitem[Izotov et al.(2006)]{2006A&A...448..955I} 
Izotov, Y.~I.,  Stasi{\'n}ska, G., Meynet, G., Guseva, N.~G., \& Thuan, T.~X.\ 2006, \aap, 448, 955
%\bibitem[Jakobsson et al.(2004)]{2004A&A...427..785J} 
%Jakobsson, P., et al.\ 2004, \aap, 427, 785
\bibitem[Jakobsson et al.(2005)]{2005ApJ...629...45J} 
Jakobsson, P., et al.\ 2005, \apj, 629, 45
\bibitem[Jansen et al.(2000)]{2000ApJS..126..331J} 
Jansen, R.~A., Fabricant, D., Franx, M., \& Caldwell, N.\ 2000, \apjs, 126, 331
\bibitem[{{}()}]{}
Juneau, S.\ et al.\ 2005, \apj, 619, L135 
%\bibitem[Kauffmann et al.(2003a)]{2003MNRAS.341...33K} 
%Kauffmann, G., et al.\ 2003a, \mnras, 341, 33 
\bibitem[Kauffmann et al.(2003)]{2003MNRAS.346.1055K} 
Kauffmann, G., et al.\ 2003, \mnras, 346, 1055
\bibitem[Kelly et al.(2008)]{2008ApJ...687.1201K} 
Kelly, P.~L., Kirshner, R.~P., \& Pahre, M.\ 2008, \apj, 687, 1201 
\bibitem[Kennicutt(1998)]{1998ARA&A..36..189K} 
Kennicutt, R.~C., 1998, \araa, 36, 189 
\bibitem[{{}()}]{}
Kennicutt, R.~C., Bresolin, F., \& Garnett, D.~R.\ 2003, \apj, 591, 801 
\bibitem[Kewley et al.(2007)]{2007AJ....133..882K} 
Kewley, L.~J., Brown, W.~R., Geller, M.~J., Kenyon, S.~J., \& Kurtz, M.~J.\ 2007, \aj, 133, 882
\bibitem[{{}()}]{}
Kewley, L.~J., \& Dopita, M.~A.\ 2002, \apjs, 142, 35
\bibitem[Kewley et al.(2001)]{2001ApJ...556..121K} 
Kewley, L.~J., Dopita, M.~A., Sutherland, R.~S., Heisler, C.~A., \& Trevena, J.\ 2001, \apj, 556, 121
\bibitem[Kewley \& Ellison(2008)]{2008ApJ...681.1183K} 
Kewley, L.~J., \& Ellison, S.~L.\ 2008, \apj, 681, 1183 
\bibitem[{{}()}]{}
Kewley, L.~J., Geller, M.~J., \& Jansen, R.~A.\ 2004, \aj, 127, 2002 
\bibitem[Klebesadel et al.(1973)]{1973ApJ...182L..85K} 
Klebesadel, R.~W., Strong, I.~B., \& Olson, R.~A.\ 1973, \apjl, 182, L85
\bibitem[{{}()}]{}
Kobulnicky, H.~A., Kennicutt, R.~C., \& Pizagno, J.~L.\ 1999, \apj, 514, 544 
\bibitem[{{}()}]{}
Kobulnicky, H.~A., \& Kewley L.~J.\ 2004, ApJ, 617, 240
\bibitem[{{}()}]{}  
Kroupa, P. 2001, MNRAS, 322, 231
\bibitem[Kulkarni et al.(1998)]{1998Natur.393...35K} 
Kulkarni, S.~R., et al.\ 1998, \nat, 393, 35 
\bibitem[K{\"u}pc{\"u} Yolda{\c s} et al.(2006)]{2006A&A...457..115K} 
K{\"u}pc{\"u} Yolda{\c s}, A., Greiner, J., \& Perna, R.\ 2006, \aap, 457, 115
\bibitem[K{\"u}pc{\"u} Yolda{\c s} et al.(2007)]{2007A&A...463..893K} 
K{\"u}pc{\"u} Yolda{\c s}, A., Salvato, M., Greiner, J., Pierini, D., Pian, E., \& Rau, A.\ 2007, \aap, 463, 893
\bibitem[Lamareille et al.(2004)]{2004MNRAS.350..396L} 
Lamareille, F., Mouhcine, M., Contini, T., Lewis, I., \& Maddox, S.\ 2004, \mnras, 350, 396
\bibitem[Le Borgne \& Rocca-Volmerange(2002)]{2002A&A...386..446L} 
Le Borgne, D., \& Rocca-Volmerange, B.\ 2002, \aap, 386, 446
\bibitem[{{}()}]{} 
Le Borgne, D., Rocca-Volmerange, B., Prugniel, P., Lan\c{c}on, A., Fioc, M., \& Soubiran, C. 2004, A\&A, 425, 881
\bibitem[Lee et al.(2006)]{2006ApJ...647..970L} 
Lee, H., Skillman, E.~D., Cannon, J.~M., Jackson, D.~C., Gehrz, R.~D., Polomski, E.~F., \& Woodward, C.~E.\ 2006, \apj, 647, 970
\bibitem[Le Floc'h et al.(2006)]{2006ApJ...642..636L} 
Le Floc'h, E., Charmandaris, V., Forrest, W.~J., Mirabel, I.~F., Armus, L., \& Devost, D.\ 
2006, \apj, 642, 636
\bibitem[Le Floc'h et al.(2002)]{2002ApJ...581L..81L} 
Le Floc'h, E., et al.\ 2002, \apjl, 581, L81
\bibitem[Le Floc'h et al.(2003)]{2003A&A...400..499L} 
Le Floc'h, E., et al.\ 2003, \aap, 400, 499 
\bibitem[Lequeux et al.(1979)]{1979A&A....80..155L} 
Lequeux, J., Peimbert, M., Rayo, J.~F., Serrano, A., \& Torres-Peimbert, S.\ 1979, \aap, 80, 155
\bibitem[{{}()}]{} 
Lilly, S. J., Carollo, C. M., \& Stockton, A. N. 2003, ApJ,  597, 730 
\bibitem[{{}()}]{} 
Madau, P., Pozzetti, L., \& Dickinson, M.\ 1998, \apj, 498, 106
\bibitem[Mangano et al.(2007)]{2007A&A...470..105M} 
Mangano, V., et al.\ 2007, \aap, 470, 105 
\bibitem[Masetti et al.(2005)]{2005A&A...438..841M} 
Masetti, N., et al.\ 2005, \aap, 438, 841
\bibitem[Meurer et al.(1999)]{1999ApJ...521...64M} Meurer, G.~R., Heckman, 
T.~M., \& Calzetti, D.\ 1999, \apj, 521, 64
\bibitem[{{}()}]{}
McCall, M. L., Rybski, P. M., \& Shields, G. A. 1985, ApJS, 57, 1
\bibitem[Metzger et al.(1997)]{1997Natur.387..878M} 
Metzger, M.~R., Djorgovski, S.~G., Kulkarni, S.~R., Steidel, C.~C., Adelberger, K.~L., Frail, D.~A., Costa, E., \& Frontera, F.\ 1997, \nat, 387, 878
\bibitem[Micha{\l}owski et al.(2008)]{2008ApJ...672..817M} 
Micha{\l}owski, M.~J., Hjorth, J., Castro Cer{\'o}n, J.~M., \& Watson, D.\ 2008, \apj, 672, 817
\bibitem[Mirabal et al.(2007)]{2007ApJ...661L.127M} 
Mirabal, N., Halpern, J.~P., \& O'Brien, P.~T.\ 2007, \apjl, 661, L127
\bibitem[Mirabal et al.(2002)]{2002ApJ...578..818M} 
Mirabal, N., et al.\ 2002, \apj, 578, 818
\bibitem[Modjaz et al.(2006)]{2006ApJ...645L..21M} 
Modjaz, M., et al.\ 2006, \apjl, 645, L21
\bibitem[Modjaz et al.(2008)]{2008AJ....135.1136M} 
Modjaz, M., et al.\ 2008, \aj, 135, 1136 
\bibitem[M\o ller et al.(2002)]{2002A&A...396L..21M} 
M{\o}ller, P., et al.\ 2002, \aap, 396, L21
\bibitem[Moustakas et al.(2006)]{2006ApJ...642..775M} 
Moustakas, J., Kennicutt, R.~C., Jr., \& Tremonti, C.~A.\ 2006, \apj, 642, 775
\bibitem[Nagao et al.(2006)]{2006A&A...459...85N} 
Nagao, T., Maiolino, R., \& Marconi, A.\ 2006, \aap, 459, 85 
\bibitem[Noeske et al.(2007)]{2007ApJ...660L..47N} 
Noeske, K.~G., et al.\ 2007, \apjl, 660, L47 
\bibitem[Nuza et al.(2007)]{2007MNRAS.375..665N} 
Nuza, S.~E., Tissera, P.~B., Pellizza, L.~J., Lambas, D.~G., Scannapieco, C., \& de Rossi, M.~E.\ 2007, \mnras, 375, 665 
\bibitem[{{}()}]{} 
Osterbrock, D. E. 1989, Astrophysics of Gaseous Nebulae and Active Galactic Nuclei (Mill Valley, CA: Univ.\ Science Books)
\bibitem[Osterbrock \& Ferland(2006)]{2006agna.book.....O} 
Osterbrock, D.~E., \& Ferland, G.~J.\ 2006, Astrophysics of gaseous nebulae and active galactic nuclei (2nd.~ed.: Sausalito, CA: Univ.\ Science Books)
\bibitem[Ovaldsen et al.(2007)]{2007ApJ...662..294O}
Ovaldsen, J.-E., et al.\ 2007, \apj, 662, 294
\bibitem[{{}()}]{} 
Pagel, B. E. J., Edmunds, M. G., Blackwell, D. E., Chun, M. S., \& Smith, G. 1979, MNRAS, 189, 95
\bibitem[Pellizza et al.(2006)]{2006A&A...459L...5P} 
Pellizza, L.~J., et al.\ 2006, \aap, 459, L5
\bibitem[Perley et al.(2007)]{2007astro.ph..3538P} 
Perley, D.~A., et al.\ 2008, ArXiv Astrophysics e-prints, arXiv:astro-ph/0703538
\bibitem[{{}()}]{} 
Pettini, M., \& Pagel, B.~E.~J.\ 2004, \mnras, 348, L59 
\bibitem[P{\'e}rez-Gonz{\'a}lez et al.(2003)]{2003MNRAS.338..525P} 
P{\'e}rez-Gonz{\'a}lez, P.~G., Gil de Paz, A., Zamorano, J., Gallego, J., Alonso-Herrero, A., \& Arag{\'o}n-Salamanca, A.\ 2003, \mnras, 338, 525
\bibitem[Pian et al.(2006)]{2006Natur.442.1011P} 
Pian, E., et al.\ 2006, \nat, 442, 1011
\bibitem[Pilyugin(2000)]{2000A&A...362..325P} 
Pilyugin, L.~S.\ 2000, \aap, 362, 325
%\bibitem[Pilyugin \& Thuan(2005)]{2005ApJ...631..231P} 
%Pilyugin, L.~S., \&  Thuan, T.~X.\ 2005, \apj, 631, 231
\bibitem[Piro et al.(2002)]{2002ApJ...577..680P} 
Piro, L., et al.\ 2002, \apj, 577, 680
\bibitem[Price et al.(2002)]{2002ApJ...571L.121P} 
Price, P.~A., et al.\ 2002a, \apjl, 571, L121
\bibitem[Price et al.(2002)]{2002ApJ...573...85P} 
Price, P.~A., et al.\ 2002b, \apj, 573, 85
\bibitem[Price et al.(2003)]{2003ApJ...589..838P} 
Price, P.~A., et al.\ 2003, \apj, 589, 838
\bibitem[Priddey et al.(2006)]{2006MNRAS.369.1189P} 
Priddey, R.~S., Tanvir, N.~R., Levan, A.~J., Fruchter, A.~S., Kouveliotou, C., Smith, I.~A., \& Wijers, R.~A.~M.~J.\ 2006, \mnras, 369, 1189
\bibitem[Prieto et al.(2008)]{2008ApJ...673..999P} 
Prieto, J.~L., Stanek, K.~Z., \& Beacom, J.~F.\ 2008, \apj, 673, 999
\bibitem[Prochaska et al.(2004)]{2004ApJ...611..200P} 
Prochaska, J.~X., et al.\ 2004, \apj, 611, 200
\bibitem[Prochaska et al.(2006)]{2006ApJ...642..989P} 
Prochaska, J.~X., et al.\ 2006, \apj, 642, 989
\bibitem[Rau et al.(2005)]{2005A&A...444..425R} 
Rau, A., Salvato, M., \& Greiner, J.\ 2005, \aap, 444, 425
\bibitem[Rau et al.(2004)]{2004A&A...427..815R} 
Rau, A., et al.\ 2004, \aap, 427, 815
\bibitem[Reddy et al.(2006)]{2006ApJ...653.1004R} 
Reddy, N.~A., Steidel, C.~C., Erb, D.~K., Shapley, A.~E., \& Pettini, M.\ 2006, \apj, 653, 1004
\bibitem[Rol et al.(2007)]{2007ApJ...669.1098R} 
Rol, E., et al.\ 2007, \apj, 669, 1098 
\bibitem[{{}()}]{}
Salpeter, E.~E.\ 1955, \apj, 121, 161
\bibitem[{{}()}]{} 
Savaglio, S., 2006, New J.\ Phys., 8, 195
\bibitem[Savaglio et al.(2005)]{2005ApJ...635..260S} 
Savaglio, S., et al.\ 2005, \apj, 635, 260
\bibitem[Schlegel et al.(1998)]{1998ApJ...500..525S} 
Schlegel, D.~J., Finkbeiner, D.~P., \& Davis, M.\ 1998, \apj, 500, 525
\bibitem[Shapley et al.(2005)]{2005ApJ...626..698S} 
Shapley, A.~E., Steidel, C.~C., Erb, D.~K., Reddy, N.~A., Adelberger, K.~L., Pettini, M., Barmby, P., \& Huang, J.\ 2005, \apj, 626, 698 
\bibitem[Soderberg et al.(2004)]{2004ApJ...606..994S} 
Soderberg, A.~M., et al.\ 2004, \apj, 606, 994
\bibitem[Soderberg et al.(2006)]{2006ApJ...636..391S} 
Soderberg, A.~M., et al.\ 2006a, \apj, 636, 391
\bibitem[Soderberg et al.(2006)]{2006ApJ...650..261S} 
Soderberg, A.~M., et al.\ 2006b, \apj, 650, 261
\bibitem[Soderberg et al.(2007)]{2007ApJ...661..982S} 
Soderberg, A.~M., et al.\ 2007, \apj, 661, 982
\bibitem[Sollerman et al.(2005)]{2005NewA...11..103S}
Sollerman, J., {\"O}stlin, G., Fynbo, J.~P.~U., Hjorth, J., Fruchter, A., \& Pedersen, K.\ 2005, New Astron., 11, 103
\bibitem[Sollerman et al.(2006)]{2006A&A...454..503S} 
Sollerman, J., et al.\ 2006, \aap, 454, 503
\bibitem[{{}()}]{} 
Spergel, D.~N., et al.\ 2003, \apjs, 148, 175 
\bibitem[Stanek et al.(2003)]{2003ApJ...591L..17S} 
Stanek, K.~Z., et al.\ 2003, \apjl, 591, L17 
\bibitem[Stanek et al.(2006)]{2006AcA....56..333S} 
Stanek, K.~Z., et al.\ 2006, Acta Astron., 56, 333 
\bibitem[Stasi{\'n}ska(2005)]{2005A&A...434..507S} 
Stasi{\'n}ska, G.\ 2005, \aap, 434, 507 
\bibitem[Stasi{\'n}ska(2006)]{2006A&A...454L.127S} 
Stasi{\'n}ska, G.\ 2006, \aap, 454, L127
\bibitem[Stasi{\'n}ska et al.(2006)]{2006MNRAS.371..972S} 
Stasi{\'n}ska, G., Cid Fernandes, R., Mateus, A., Sodr{\'e}, L., \& Asari, N.~V.\ 2006, \mnras, 371, 972
\bibitem[Tanvir et al.(2004)]{2004MNRAS.352.1073T} 
Tanvir, N.~R., et al.\ 2004, \mnras, 352, 1073
\bibitem[Th{\"o}ne et al.(2007)]{2007ApJ...671..628T} 
Th{\"o}ne, C.~C., Greiner, J., Savaglio, S., \& Jehin, E.\ 2007, \apj, 671, 628
\bibitem[Th{\"o}ne et al.(2008)]{2008ApJ...676.1151T} 
Th{\"o}ne, C.~C., et al.\ 2008, \apj, 676, 1151
\bibitem[Totani et al.(2006)]{2006PASJ...58..485T} 
Totani, T., Kawai, N., Kosugi, G., Aoki, K., Yamada, T., Iye, M., Ohta, K., \& Hattori, T.\ 2006, \pasj, 58, 485
\bibitem[Tremonti et al.(2004)]{2004ApJ...613..898T} 
Tremonti, C.~A., et al.\ 2004, \apj, 613, 898 
\bibitem[Wainwright et al.(2007)]{2007ApJ...657..367W} 
Wainwright, C., Berger, E., \& Penprase, B.~E.\ 2007, \apj, 657, 367
\bibitem[Wiersema et al.(2007)]{2007A&A...464..529W} 
Wiersema, K., et al.\ 2007, \aap, 464, 529
\bibitem[Wiersema et al.(2008)]{2008A&A...481..319W} 
Wiersema, K., et al.\ 2008, \aap, 481, 319
\bibitem[Wolf \& Podsiadlowski(2007)]{2007MNRAS.375.1049W} 
Wolf, C., \& Podsiadlowski, P.\ 2007, \mnras, 375, 1049
\bibitem[Woosley(1993)]{1993ApJ...405..273W} 
Woosley, S.~E.\ 1993, \apj, 405, 273
\bibitem[{{}()}]{}
Worthey G., 1994, ApJS, 95, 107 
\bibitem[Yan et al.(2006)]{2006ApJ...651...24Y} 
Yan, H., Dickinson, M., Giavalisco, M., Stern, D., Eisenhardt, P.~R.~M., \& Ferguson, H.~C.\ 2006, \apj, 651, 24
\bibitem[Y{\"u}ksel et al.(2008)]{2008ApJ...683L...5Y} Y{\"u}ksel, H., Kistler, M.~D., Beacom, J.~F., \& Hopkins, A.~M.\ 2008, \apjl, 683, L5 

\end{thebibliography}
\end{document}